\documentclass[aip,superscriptaddress,nofootinbib]{revtex4-1}

\usepackage[english]{babel}
\usepackage[utf8]{inputenc}
\usepackage{pstricks}
\usepackage{graphics}
\usepackage{graphicx}
\usepackage[pdftex,pdfpagelabels,plainpages=false,linktocpage=true]{hyperref}
\hypersetup{colorlinks=true,citecolor=blue,urlcolor=blue,filecolor=green,
	linkcolor=red}
\hypersetup{colorlinks}
\usepackage{latexsym}
\usepackage{fancyhdr}
\usepackage[bf]{caption}
\usepackage{verbatim}
\usepackage{color}
\usepackage{amstext,amssymb}
\usepackage[intlimits,fleqn]{amsmath}
\usepackage{bm}
\usepackage{picinpar}
\usepackage{calc}
\usepackage{afterpage}
\usepackage{color}
\usepackage{multirow}
\usepackage{booktabs}
\usepackage{afterpage}
\usepackage{rotating}
\usepackage{array}

\usepackage{natbib}
\usepackage[off]{auto-pst-pdf}
\usepackage{psfrag}
\usepackage{float}
\usepackage[caption=false,labelformat=simple]{subfig}

\usepackage{numprint}
\usepackage{upgreek}
\usepackage{threeparttable} 
\makeatletter
\newcommand*{\Romannum}[1]{\textit{\expandafter\@slowromancap\romannumeral
		#1@}}
\makeatother
\usepackage{booktabs}
\usepackage{tabularx}
\newcommand{\biborder}[1]{}

\newcommand{\uproman}[1]{\uppercase\expandafter{\romannumeral#1}}

\graphicspath{{./fig/}}

\newcommand{\eq}[1]{\hyperref[#1]{Eq.\,\eqref{#1}}}
\newcommand{\eqs}[2]{\hyperref[#1]{Eqs.\,\eqref{#1}} and
	\hyperref[#2]{\eqref{#2}}}
\newcommand{\eqqs}[3]{\hyperref[#1]{Eqs.\,\eqref{#1}},
	\hyperref[#2]{\eqref{#2}} and \hyperref[#3]{\eqref{#3}}}
\newcommand{\tabs}[2]{\hyperref[#1]{Tables\;\ref{#1}} and
	\hyperref[#2]{\ref{#2}}}
\newcommand{\sct}[1]{\hyperref[#1]{Section~\ref{#1}}}
\newcommand{\cha}[1]{\hyperref[#1]{Chapter~\ref{#1}}}
\newcommand{\app}[1]{\hyperref[#1]{Appendix~\ref{#1}}}

\newcommand{\Figs}[1]{\hyperref[#1]{Figures\;\ref{#1}}}

\newcommand{\abs}[1]{\lvert#1\rvert}

{

	\definecolor{dgreen}{rgb}{0.2,0.6,0.2}

\begin{document}


\title{Investigation of cohesive particle deagglomeration in 
	homogeneous isotropic turbulence using particle-resolved 
        direct numerical simulation}

\affiliation{Professur f\"ur Str\"omungsmechanik, 
	Helmut-Schmidt-Universit\"at Hamburg, 
	D-22043 Hamburg, Germany}

\author{A.\ Khalifa}
\author{M.\ Breuer}
\email{breuer@hsu-hh.de}

\date{\today}
	                 
%
\begin{abstract}
In this study, agglomerate breakage in homogeneous isotropic
turbulence is investigated using particle-resolved direct numerical
simulations. Single agglomerates composed of 500 monodisperse
spherical particles are considered, and their interaction with the
turbulent flow is resolved through an immersed boundary method coupled
with a soft-sphere discrete element model. A range of Reynolds numbers
and cohesion levels is examined to assess their influence on the
breakup behavior.  Detailed insights into the underlying breakage
mechanisms are provided through the analysis of local flow structures
and fluid stresses. Strain-dominated regions are identified as the
primary contributors to the onset and propagation of particle
erosion. The benefits of the particle-resolved simulation framework in
capturing these physical processes in detail are demonstrated. The
predicted fragment size distributions and breakup modes are analyzed
leading to the outcome that erosion-driven breakage is the dominating
mechanism. The time evolution of the fragment number and the main
agglomerate structure is quantified. The breakage rate is evaluated
and its dependence on the modified adhesion number is established,
showing a power-law decay that agrees with general trends reported in
the literature. In addition, the analysis of the fragment ejection
direction reveals a strong alignment with the local deformation plane
spanned by the most extensional and compressive strain-rate
eigenvectors, indicating that breakage results from the interplay
between flow stretching and compression. The results contribute to the
development of physics-informed breakup kernels for use in efficient
but less-detailed simulation approaches such as point-particle
Euler--Lagrange predictions with agglomerates represented by effective
spheres or Euler--Euler simulations.
\end{abstract}


\maketitle 


%
\clearpage
\section{Introduction}
Micrometer-sized particles suspended in gas flows are a common
occurrence in natural and industrial systems. At this scale, cohesive
forces between particles often surpass gravitational forces, promoting
the formation of agglomerates. These agglomerate structures are
dynamic in time, since they are constantly exposed to different kinds
of stresses including fluid stresses that can trigger their
fragmentation. Understanding how the agglomerate size evolves under
such conditions is crucial, as it influences a wide spectrum of
processes and applications across both natural and engineered
systems. For example, particle-laden gas flows play a crucial role in
natural processes such as wind-driven particle erosion and deposition
\citep{shao1993efficiencies} and the atmospheric transport of
aerosols, which influence weather patterns and climate dynamics.  In
the pharmaceutical industry, understanding particle behavior in gas
flows is critical for applications like dry powder inhalers, where
precise particle dispersion and delivery are essential for effective
drug application and targeted therapeutic outcomes
\citep{darquenne2012aerosol}.

Despite numerous experimental studies in related fields, research
specifically addressing particle deagglomeration remains limited,
often focusing on application-driven objectives (see, e.g.,
\citet{kousaka1993dispersion,weiler2008generierung, saha2015breakup,
  du2024experimental}). Most existing studies emphasize integral
quantities, such as particle size distribution at the outlet of
experimental setups or devices, rather than providing detailed
insights into the nature of individual breakup events. This is
primarily due to the inherent complexity of fluid-particle and
particle-particle interactions, as well as the technical challenges
associated with experimentally capturing these dynamics. Observing and
quantifying agglomeration and deagglomeration processes require
advanced measurement equipment, including high-resolution imaging
systems, accurate particle tracking methods, and precise environmental
controls, all of which demand substantial technical expertise and
financial resources.

Numerical simulations have become indispensable for studying
particle-laden flows, offering detailed insights into complex
phenomena. However, they face significant challenges, particularly in
turbulent flows containing agglomerates. These challenges stem from
the wide range of length and time scales involved
\citep{balachandar2009scaling}, which result in prohibitive
computational costs. Fully resolved simulations, known as
particle-resolved direct numerical simulations (PR-DNS, see, e.g.,
\citet{bagchi2003effect,burton2005fully,kuerten2016point}), capture
both fluid-particle interfaces and all turbulent scales but remain
infeasible for many practical applications.

A computationally more manageable alternative is the Euler--Lagrange
framework, employing point-particle approximations to model
fluid-particle interactions and the discrete element method (DEM) for
inter-particle interactions
\citep{zhou2010discrete,kuerten2016point}. These methods are
computationally less demanding but face limitations in dense
agglomerate systems, where particles are closely packed. Challenges
include accounting for drag shielding effects and the particle-induced
feedback on the fluid. Point-particle approaches typically rely on
constitutive models to compute forces and torques on the particles,
assuming an undisturbed flow and using empirical constants validated
for isolated particles. Consequently, their reliability diminishes in
particle-dense environments with complex interactions.

Within the context of the point-particle Euler--Lagrange framework
coupled with DEM, numerous studies have explored the restructuring and
deagglomeration of agglomerates in various flows, including laminar
shear flows, uniform flows, and homogeneous isotropic turbulence. Both
one-way (see, e.g., \citet{higashitani2001, fanelli2006prediction,
  fanelli2006prediction2, becker2009restructuring,
  eggersdorfer2010fragmentation, calvert2011mechanistic,
  ruan2020structural, zhao2021flocculation, yu2025shielding}) and two-way 
  coupling
approaches (see, e.g., \citet{dizaji2019collision,
  yao2021deagglomeration}) have been employed. These studies primarily
focus on the interplay between fluid forces, characterized by the
shear rate in shear flows or the root-mean-square velocity
fluctuations in turbulence, and cohesive forces between particles. Key
outcomes include quantifying breakage dynamics, such as the
development of the agglomerate size and shape, and the determination
of the time required for breakup initiation, and the breakage rate
under varying fluid conditions and agglomerate properties.

The Stokesian Dynamics approach offers another alternative, enabling
detailed force calculations by accounting for neighboring particle
influences while assuming a linear flow field around the
agglomerate. This method has been used to study restructuring and
fragmentation in a variety of cases including simple shear flows
\citep{harada2006dependence, frungieri2021aggregation} and homogeneous
isotropic turbulence \citep{debona2014}.

Despite their inherent limitations, insights from these investigations
have significantly contributed to the development of deagglomeration
models for computationally efficient coarse-grained frameworks, such
as Euler--Euler and Euler--Lagrange methods, where agglomerates are
often simplified as effective spheres. Traditionally, these
mechanistic models heavily rely on basic theoretical arguments and
experimental observations to approximate complex interactions
\citep{rumpf1932desagglomeration,tomi1978behaviour,kousaka1979dispersion,kusters1991,weiler2010new,breuer2019revisiting,breuer2019refinement}.

Recent advancements in computational power and numerical methods have
also opened new opportunities for detailed analyses of
agglomerate-laden flows through particle-resolved simulations combined
with DEM. Although such techniques remain constrained to simplified
setups involving basic geometries, low Reynolds numbers, and small
agglomerates consisting of a limited number of particles, they are
invaluable for studying fluid-particle interactions in densely packed
particle arrangements, such as agglomerates. Fully resolved
simulations provide deeper insights into agglomerate dynamics and
their interactions with complex flow fields, capturing the influence
of neighboring particles in detail. For instance, fully-resolved
lattice-Boltzmann simulations have been applied to study restructuring
and deagglomeration in steady \citep{saxena2022numerical,
  saxena2023role} and accelerating shear flows
\citep{saxena2025exposure}.  These studies investigated the influence
of shear rate and inter-particle forces on the breakage rate and the
evolution of the agglomerate's radius of gyration, shedding light on
the factors affecting deagglomeration under various flow conditions.

The present study employs a particle-resolved direct numerical
simulation (PR-DNS) approach combined with DEM to investigate the
deagglomeration of particle clusters in a forced homogeneous isotropic
turbulence, marking, to the best of the authors' knowledge, a first of
its kind. By combining the immersed boundary method with the discrete
element method, the study focuses on the breakup of nearly-spherical
agglomerates composed of solid, dry, electrostatically neutral
spherical particles bound by van-der-Waals forces. The analysis is
conducted at three different flow Reynolds numbers and for three
levels of particle cohesion, characterized by varying Hamaker
constants. The objective is to provide novel insights into the breakup
behavior under these conditions and to compare the results with those
from methods that do not resolve the fluid-particle interface.

The remainder of the paper is organized as follows:
Section~\ref{sec:method} explains the applied PR-DNS / DEM
methodology. Section~\ref{sec:comp_setup} presents the computational
setup for agglomerate-laden homogeneous isotropic turbulence.
Section~\ref{sec:results} discusses the results, and
Section~\ref{sec:conclusions} concludes the study.

\clearpage
\section{Particle-resolved DNS methodology
	\label{sec:method}}
An immersed boundary DNS-DEM approach that resolves the particle-fluid
interface is employed, utilizing a customized version of the
open-source framework "CFDEMcoupling-PUBLIC"
\citep{hager2014parallel,kloss2012models}. It couples the CFD solver
OpenFOAM with the DEM solver LIGGGHTS. The customized version
incorporates several key modifications, including the addition of a
source term in the fluid-phase momentum equation to enforce
homogeneous isotropic turbulence (see
Section~\ref{subsubsec:forcingterm}). Additionally, further algorithms
have been introduced for particle detection, along with critical bug
fixes to improve the determination of the fluid velocity field in the
presence of particles.

While the original methodology is comprehensively described in the
referenced literature, a brief overview is provided below for
completeness, with a particular focus on the enhancements introduced
in this study to extend the framework's capabilities.

\subsection{Description of the continuous phase \label{subsect:desc_contin}}

\subsubsection{Immersed boundary (fictitious domain) method}

The fluid phase within the coupled CFD-DEM framework is described
using an incompressible flow formulation governed by the
three-dimensional Navier-Stokes equations. The OpenFOAM solver
"pisoFoam" serves as the computational foundation, ensuring accurate
and efficient resolution of the flow field. The mass and momentum
conservation equations are given as:
\begin{align}
	\frac{\partial u_i}{\partial x_i} & = 0 \; , \\ \frac{\partial
          u_i}{\partial t} + u_j \; \frac{\partial u_i}{\partial x_j}
        & = - \frac{1}{\rho_\text{f}}\;\frac{\partial p}{\partial x_i}
        + \nu_\text{f} \; \frac{\partial^2 u_i}{\partial x_j \partial
          x_j} + S_{t, i} \;, \label{eq:mom}
\end{align}
where $x_i$ are the spatial coordinates, $u_i$ are the fluid velocity
components, $p$ denotes the pressure, $\nu_\text{f}$ stands for the
kinematic viscosity of the fluid, $\rho_\text{f}$ is the density, and
$S_{t, i}$ is a linear source term for enforcing homogeneous isotropic
turbulence (see Section~\ref{subsubsec:forcingterm}).

Initially, the fluid phase is treated independently, solving the
Navier-Stokes equations across the entire domain while temporarily
neglecting the influence of particles. The presence of particles is
then incorporated using an immersed boundary method based on the
fictitious domain approach
\citep{glowinski1998fictitious,glowinski2001fictitious}. In this
method, the particle-fluid interaction is accounted for by imposing
the particle velocities predicted by DEM (see
Section~\ref{subsec:descr_disperse}) within the regions occupied by
the particles. This approach adjusts the fluid velocity in solid
regions to align with the particle motion while considering the degree
of cell occupancy by particles. The velocity update is performed by
iterating over cells partially or fully occupied by the particle p:
\begin{equation}
	\label{eq:vel_corr}
	u_i = \phi_\text{p} \: U_{\text{p},i} \; + \; \alpha_\text{c}
        \: u_i \; .
\end{equation} 
Here, $\phi_\text{p}$ represents the solid volume fraction within the
cell c contributed exclusively by the particle p.  Two cases are
distinguished for the coefficient $\alpha_\text{c}$.  If the cell has
not been previously updated (i.e., not associated with another
particle during the same time step), the coefficient $\alpha_\text{c}$
corresponds to the overall fluid volume fraction $\phi_\text{c}$ of
the cell, which accounts for the presence of all particles within the
cell. Otherwise, $\alpha_\text{c}$ is set to unity. This approach
differs from the standard implementation in the original program,
which does not accurately handle cases where multiple particles
intersect a single cell. The original method considered only the
overall fluid volume fraction $\phi_\text{c}$ of the cell, neglecting
the individual contributions $\phi_\text{p}$ of each particle within
it. Note that the velocity of the particle $U_{\text{p}}$ in the cell
is computed relying on the DEM calculations for the position
$r_\text{p}$, the translational velocity $u_{\text{p}}$, and the
rotational velocity $\omega_{\text{p}}$ of the particle as:
\begin{equation}
	\label{eq:vel_incell} 
	U_{\text{p},i} = u_{\text{p},i} \: + \: \varepsilon_{ijk} \;
        \left( r_{\text{c}} - r_{\text{p}} \right)_k \:
        \omega_{\text{p},j}\; .
\end{equation}
Here, $\varepsilon_{ijk}$ is the Levi-Civita symbol for the cross
product and $\left( r_{\text{c}} - r_{\text{p}} \right)_k$ represents
the position vector relative to the particle center.

To maintain the incompressibility condition, the corrected velocity
field is made divergence-free through a pressure correction step,
adhering to the PISO algorithm.

To ensure accurate coupling, a dynamic local mesh refinement strategy
is employed to effectively capture the fluid-particle interface. In
the present study, the local mesh refinement guarantees that the
particle diameter is resolved by eight cells in each direction in all
considered Reynolds number cases, while avoiding unnecessary
refinement in regions far from the particles. Such a resolution was
found to be sufficient for achieving grid independent results
\citep{hager2014thesis}.  The refinement is performed based on a field
variable denoted the "interFace", which has a non-zero value in a
spherical region around the particle possessing twice its diameter.

The IB method is fully compatible with parallel computing algorithms,
as it enables the detection of particles across parallel blocks and
periodic boundaries. However, the original code had a limitation in
calculating the position vector $r_\text{p}$ of the particle within
cells near periodic boundaries in Eq.~(\ref{eq:vel_incell}) used to
compute the particle velocity. Specifically, it only considered the
particle's physical position without adjusting for periodic boundary
conditions, unlike the present customized version which correctly
accounts for this adjustment.

The refined velocity and pressure fields are subsequently utilized to
resolve the fluid-particle interactions. Specifically, the fluid force
and torque acting on the particle are computed as follows:
\begin{align}
	\bm{F}^\text{f-p} &= \int_{V_p} \left( \rho_\text{f} \:\nu_\text{f} \: 
	\frac{\partial^2 u_i}{\partial x_j \partial x_j} - \frac{\partial 
	p}{\partial 
		x_i} \right) \,  dV_\text{p} \;,\\
	\bm{M}^\text{f-p} &= \int_{V_\text{p}} \varepsilon_{ijk} \: 
	(r_{\text{c}} - 
	r_{\text{p}})_j \left( \rho_\text{f} \: \nu_\text{f} 
	\:\frac{\partial^2 u_k}{\partial x_m \partial x_m} - \frac{\partial 
	p}{\partial 
		x_k} \right) \, dV_\text{p}\;,
\end{align}
where $V_\text{p}$ denotes the volume of the particle. Given the
discrete nature of the numerical solution, the force and torque
expressions are reformulated as:
\begin{align}
	\bm{F}^\text{f-p} &= \sum_{c=1}^{N_{\text{cells}}} \left( 
	\rho_\text{f} 
	\, \nu_\text{f} \, \frac{\partial^2 u_i}{\partial x_j \partial x_j} - 
	\frac{\partial p}{\partial x_i} \right) \: V_c \;,  \label{eq:drag}\\
	\bm{M}^\text{f-p} &= \sum_{c=1}^{N_{\text{cells}}} 
	\varepsilon_{ijk} \left( r_{\text{c}} - 
	r_{\text{p}} \right)_j \left( \rho_\text{f} \, \nu_\text{f} 
	\, \frac{\partial^2 u_k}{\partial x_m \partial x_m} - \frac{\partial 
		p}{\partial x_k} \right) \: V_c \;, \label{eq:fluidtorq}
\end{align}
where $c$ is an index running over all cells ($N_\text{cells}$) that
are completely or partially occupied by the particle, and $V_c$
represents the total volume of each cell.  The use of the total cell
volume $V_c$, even for partially covered cells, is justified since the
fluid velocity $u_i$ and the pressure $p$ are volume-averaged
quantities over the cell, accounting for the particle volume fraction
in partially occupied cells. This ensures that the contributions of
partially occupied cells are appropriately weighted in the force and
torque calculations.


The finite-volume method used in this study is second-order accurate
in space, employing a Gauss linear scheme for gradient and divergence
discretizations.  For Laplacian terms, the Gauss linear corrected
scheme is applied to account for mesh non-orthogonality. Surface
normal gradients are handled using the corrected scheme, while field
variable interpolation utilizes the linear scheme for consistency and
accuracy. In addition, a second-order accurate implicit backward
differentiation scheme is used for time integration, ensuring
high-fidelity simulations of fluid-particle interactions.
\subsubsection{Forcing term for homogeneous isotropic turbulence
	\label{subsubsec:forcingterm}}

To achieve and sustain a statistically stationary homogeneous
isotropic turbulence under prescribed conditions, the inclusion of a
forcing term $S_{t,i}$ in the momentum equation~(\ref{eq:mom}) of the
fluid is necessary.  In this study, a linear forcing approach based on
maintaining constant turbulent kinetic energy, as proposed by
\citet{bassenne2016constant}, is adopted. Its formulation in physical
space is given by:
\begin{equation}
	S_{t,i} = A(t) \; [u_i - \overline{u_i}] \; ,
\end{equation}
where $\overline{u_i}$ represents the spatially averaged velocity
within the triply periodic domain, which should tend to zero, and
$A(t)$ is the forcing coefficient defined as:
\begin{equation}
	\label{eq:for_coeff}
	A(t) = \dfrac{\epsilon(t) - 3\: A_\infty \: G_k \: [\overline{k}(t) - 
		k_\infty]}{2 
		\: \overline{k}(t)} \; . 
\end{equation}
Here, $\overline{k}(t)$ and $\overline{\epsilon}(t)$ denote the
spatially averaged turbulent kinetic energy and its dissipation rate,
respectively. The parameter $G_k$ is a dimensionless constant
representing the proportional gain of the turbulent kinetic energy,
while the forcing parameter $A_\infty$ and the steady-state turbulent
kinetic energy $k_\infty$ are computed for a given Taylor-microscale
Reynolds number $Re_\lambda =$ $(15\:u_\text{rms} \:L_\text{c} /
\nu_\text{f})^{1/2}$ following the procedure outlined by
\citet{carroll2013proposed}, summarized as follows:
\begin{align}
	A_\infty &= \dfrac{Re^2_\lambda \; \nu_\text{f}}{45 \; L_\text{c}^2} 
	\; , 
	\label{eq:a_forcing}\\
	k_\infty &= \dfrac{27}{2} \; A^2_\infty \: L_\text{c}^2 \; 
	,\label{eq:k_infty}
\end{align}
where $L_\text{c}$ is the integral length-scale, which is
approximately 20~\% of the length of the computational domain
\citep{rosales2005linear,carroll2013proposed}. Note that the forcing
parameter $A_\infty$ represents a time scale corresponding to the eddy
turn-over time (i.e., $A_\infty =\epsilon_\infty/ \; 2\:k_\infty$).
In addition to $A_\infty$ and $k_\infty$, the corresponding
steady-state dissipation rate $\epsilon_\infty$ and the
root-mean-square velocity $u_\text{rms}$ are of interest and can be
computed as:
\begin{align}
	\epsilon_\infty &= 27 \; A^3_\infty \: L_\text{c}^2 \; , 
	\label{eq:eps_infty}\\
	u_\text{rms} &= \sqrt{\frac{2}{3} \: k_\infty} \; \; . \label{eq:urms}
\end{align}
%

\subsection{Description of the disperse phase}
\label{subsec:descr_disperse}

As is customary in DEM, particles are tracked by determining their
trajectories in a Lagrangian frame of reference. The deformation of
particles during their interactions with other particles is modeled by
allowing the interacting surfaces to overlap \citep{cundall1979}.
Accordingly, the contact forces are computed as continuous functions
of the deformation (overlap). For practical reasons, vector notation
is adopted in this section. The equations governing the translatory
and rotary motions of particle $i$ interacting with particle $j$ are
given by:
\begin{align}
	m_{\text{p},i}\dfrac{d\bm{u}_{\text{p},i}}{dt} &= 
	\bm{F}^\text{f-p}_{i} + \sum_{j} 
	\bm{F}_{ij} \;, \\
	I_{\text{p},i}\dfrac{d\bm{\omega}_{\text{p},i}}{dt} &= 
	\bm{M}^\text{f-p} \: + \: \sum_{j} 
	(\bm{M}_{ij}+\bm{M}_{ij,r}) \;,
\end{align}
respectively. Here, $m_{p,i}$ and $\bm{u}_{p,i}$ denote the mass and
velocity of particle $i$, respectively. Similarly, $I_{p,i}$ and
$\bm{\omega}_{p,i}$ represent the moment of inertia and angular
velocity of the particle. The fluid-particle interaction is included
through the force $\bm{F}^\text{f-p}_{i}$ and torque
$\bm{M}^\text{f-p}$, as described in
Eqs.~(\ref{eq:drag})~and~(\ref{eq:fluidtorq}), respectively. The
contact force $\bm{F}_{ij}$ is decomposed into a normal component
$\bm{F}_{n,ij}$ and a tangential component $\bm{F}_{t,ij}$. The latter
generates a torque $\bm{M}_{ij}$ at the contact point, causing the
particle to rotate. Furthermore, the relative rotational motion
between the two particles induces a rolling friction torque
$\bm{M}_{ij,r}$, which resists the rolling motion. The models used to
describe these contact forces and torques are summarized in the
following.

Based on a spring-dashpot analogy, the normal force is generally
expressed as an elastic and a damping force:
\begin{equation}
	\label{eq:normal_contact}
	\bm{F}_{n,ij} = \bm{F}^{El}_{n,ij} + \bm{F}^{D}_{n,ij} \; = \; -k_n 
	\;\bm{\delta}_n  + c_n  \; \Delta \bm{u}_n \;, 
\end{equation}
where $k_n$ is the normal elastic coefficient and $\bm{\delta}_n$ is
the normal overlap vector computed based on the positions of the
particles' centers and the size of their diameters.  In addition,
$c_n$ and $\Delta \bm{u}_n$ are the normal damping coefficient and
normal relative velocity vector, respectively.  In the present
notation, the negative sign in Eq.~(\ref{eq:normal_contact}) stands
for the repulsive nature of the elastic force. The normal elastic
coefficient is given based on the theory of \citet{hertz1882}:
\begin{equation}
	\label{eq:normal_elastic}
	k_n =  \frac{4}{3} \; E^*_s 
	\; \sqrt{R^* \; \delta_n} \;,
\end{equation}
where $\delta_n$ is the magnitude of the normal overlap vector.
Moreover, the normal damping coefficient is given as a function of the
normal restitution coefficient $e$ (\citet{tsuji1992}):
\begin{equation}
	\label{eq:norm_damp}
	c_n =- 2 \:\sqrt{\frac{5}{6}}\; \dfrac{\ln(e)}{\ln(e)^2+\pi^2} \; 
	\sqrt{S_n \; m^*}  \;,
\end{equation}
with:
\begin{equation}
	S_n = 2 \; E^*_s 
	\;\sqrt{R^* \; \delta_n} \;.
\end{equation}
Here, $R^*$ is the effective radius, $E^*_s$ is the effective Young's
modulus of elasticity and $m^*$ is the effective mass
\citep{khalifa2020data}.

The tangential force is accounted for based on a simplified version of
the model by \citet{mindlin1953elastic}, which is proposed by
\citet{tsuji1992}:
\begin{equation}
	\label{eq:tang_force}
	\bm{F}_{t,ij} = 
	\begin{cases}
		-  k_t \; \bm{\delta}_t+ c_t \; \Delta \bm{u}_t  &  
		\abs{k_t \; \bm{\delta}_t } \leq \mu_s \;\abs{ 
			\bm{F}_{n,ij}}\;, \\
		-\mu_s \;\abs{ 
			\bm{F}_{n,ij}} \cdot \dfrac{\bm{\delta}_t}{\abs{\bm{\delta}_t}} 
		& 
		\abs{k_t \; \bm{\delta}_t } > \mu_s \;\abs{ 
			\bm{F}_{n,ij}} \; ,  
	\end{cases}
\end{equation}
suggesting that in case of sliding, the tangential force is replaced
by the force of friction computed based on the coefficient of static
friction $\mu_s$ assuming that it is valid also for kinetic conditions
\citep{di2005improved}.  The tangential elastic coefficient $k_t$ in
Eq.~(\ref{eq:tang_force}) is given
as~(\citet{mindlin1949compliance}):
\begin{equation}
	\label{eq:spring_const_hertz}
	k_t = 8 \; G^* \; \sqrt{R^* \; \delta_n} \;,
\end{equation}
where $G^*$ is the effective shear modulus
\citep{khalifa2020data}. The tangential overlap vector $\bm{\delta}_t$
is computed by integrating the relative tangential velocity $\Delta
\bm{u}_t$ over the contact time $\Delta
t_c$~(\citet{cleary1998well}):
\begin{equation}
	\bm{\delta}_t=\int_{\Delta t_c} \Delta \bm{u}_t \; \text{d}t \;,
\end{equation}
which is practically achieved by summing $\Delta \bm{u}_t \: \Delta t$
over the time steps during the contact period.

The tangential dissipation coefficient $c_t$ is computed based on a
relation analogous to that of the normal counterpart
(\ref{eq:norm_damp}):
\begin{equation}
	\label{eq:c_t}
	c_t = -2 \:\sqrt{\frac{5}{6}}\; \dfrac{\ln(e)}{\ln(e)^2+\pi^2} \; 
	\sqrt{S_t \; m^*} \;,
\end{equation} 
with: 
\begin{equation}
	S_t = 8 \; G^* \; \sqrt{R^* \; \delta_n} \; .
\end{equation} 
%
%
Due to the relative rotational motion of the surfaces in contact, the
rolling friction torque resisting the rotation of the particle
$i$ reads:
\begin{equation}
	\label{eq:rollin_fric}
	\bm{M}_{ij,r} =  - \mu_r \: \abs{\bm{F}^{El}_{n,ij}} \; R^*\;
	\dfrac{\bm{\omega}_{rel}}{\abs{\bm{\omega}_{rel}}} \; ,
\end{equation}
where $\mu_r$ is the rolling friction coefficient, and
$\bm{\omega}_{rel}$ is the relative angular velocity between particles
$i$ and $j$.

The cohesion between primary particles is taken into account relying
on the van-der-Waals force model by \citet{hamaker1937}, which
guarantees a smooth transition between the touch and non-touch state
\citep{dong2012settling,parteli2014attractive}:
\begin{equation}
	\label{eq:fvdw_hamaker}
	\bm{F}^{\text{vdW}}_{n,ij} = 
	\begin{cases}
		\dfrac{H \: R^*}{6 \: \delta^2_0} \cdot 
		\dfrac{\bm{\delta}_n}{\abs{\bm{\delta}_n}}  & \delta_n \geq 0,
		\vspace{10pt}\\ 
		\dfrac{H \: R^*}{6 \: (\delta_n-\delta_0)^2} \cdot 
		\dfrac{\bm{\delta}_n}{\abs{\bm{\delta}_n}}  & -l_{\text{max}} \leq 
		\delta_n < 0, \vspace{10pt} \\
		0 & \delta_n < -l_{\text{max}} \;.
	\end{cases}
\end{equation}
When particles are in physical contact ($\delta_n \geq 0$), the
cohesive force is calculated based on the Hamaker constant $H$ and the
minimum surface separation distance $\delta_0$, which accounts for
surface roughness. As particles move apart, the cohesive force
gradually diminishes to zero. To reduce computational cost, a
van-der-Waals force cut-off distance $l_{\text{max}} = 5\cdot10^{-2}
\; d_p $ is introduced. This parameter defines the maximum separation
distance beyond which the cohesive force between particles effectively
vanishes. The resulting van-der-Waals force contribution,
$\bm{F}^{\text{vdW}}_{n,ij}$, is then directly added to the normal
contact force described in Eq.~(\ref{eq:normal_contact}).

The motivation for the proposed van-der-Waals cohesion model is
twofold. First, the model accounts for the attractive forces between
nearby, non-contacting particles, which can significantly influence
the behavior of micron-sized particles \cite[]{parteli2014attractive},
as considered in this work. Second, given the material properties and
particle size used (see Table~\ref{tab:sio2prop}), the dimensionless
Tabor parameter $\mu_{\text{Ta}} = (4 R^* \gamma_s^2 / E_s^{*^{2}}
\delta_0 )^{1/3}$ is about 0.03.  This range supports the
applicability of models based on the Derjaguin-Muller-Toporov (DMT)
theory \cite[]{derjaguin1975effect}, such as the present van-der-Waals
force model \citep{dong2012settling}. The Tabor parameter is a widely
used metric to characterize the impact of cohesive forces, expressed
via the surface energy $\gamma_s$, on the contact area predicted by
the Hertzian theory \citep{hertz1882}. For small particles with low
surface energy and elastic modulus (i.e., $\mu_{\text{Ta}} < 1.0$, see
\citet{liu2010applicability,li2011adhesive}), the deviation of the
contact area from the Hertzian solution due to cohesion is
negligible. Therefore, the current contact model, where the normal
elastic force remains independent of the cohesive force, is
appropriate. Hence, the DEM code LIGGGHTS was extended in a previous
work \citep{khalifa2020data} to incorporate the van der Waals force
model.

Furthermore, the equation of motion of the single particle is
integrated based on the velocity Verlet scheme
\citep{swope1982computer}, which is second-order accurate in velocity
and fourth-order accurate in position.

Note that the DEM framework discussed here has been described and used
in previous pure DEM investigations (without fluid flow) on the
breakage of agglomerates due to wall impacts
\citep{khalifa2020data,khalifa2021efficient,khalifa2021les} and
particle-particle collisions
\citep{khalifa2022neural,khalifa2023data,khalifa2024eval}.


\section{Computational setup for homogeneous isotropic 
	turbulence  \label{sec:comp_setup}}

A homogeneous isotropic turbulent flow in a periodic environment
provides an ideal setup for studying the deagglomeration of particles
by turbulence in isolation from other competing mechanisms, such as
wall impact. Consequently, homogeneous isotropic turbulence has been
frequently utilized for the study of agglomerate breakup in turbulent
flows (e.g., \citet{yao2021deagglomeration,zhao2021flocculation}).
\subsection{Flow configuration and agglomerate properties 
	\label{subsec:descr_flow_agg}}

The simulations are carried out in a 3D triply periodic box with an
edge length of $ L_\text{box} =2 \cdot 10^{-4}$~m according to about
$200~d_\text{p}$, where $d_\text{p}$ is the diameter of the primary
particle that makes up the agglomerate under considerations. This
length is chosen as a compromise between required computational
resources and achievable accuracy. It ensures that the flow acting on
a particle is not influenced by its periodic images and that the
breakage behavior remains unaffected by collisions with already
separated particles crossing the periodic boundaries.

The fluid properties correspond to those of air, with a density of $
\rho_\text{f} = 1.196 $ kg/m$^3$ and a kinematic viscosity of $
\nu_\text{f} = 1.57 \cdot 10^{-5} $ m$^2$/s. To investigate the
effects of the turbulence intensity on breakage, three flow cases are
considered. The corresponding Taylor-microscale Reynolds numbers are
set to $ Re_\lambda =$ 30, 43, and 64, inspired by the study of
\citet{yao2021deagglomeration}, denoted hereafter as $Re_{30}$,
$Re_{43}$, and $Re_{64}$, respectively. To realize these flow cases
the parameters of the forcing coefficient $A(t)$ in
Eq.~(\ref{eq:for_coeff}) are adjusted for each case. Specifically, the
forcing parameter $ A_\infty $ and the steady-state turbulent kinetic
energy $ k_\infty $ are set according to the corresponding
$Re_\lambda$ value, as predicted by Eqs.~(\ref{eq:a_forcing}) and
(\ref{eq:k_infty}). Furthermore, the proportional gain of the kinetic
energy $G_k$ is set to a value of 300 for all cases, which ensures a
rapid convergence to a statistically steady state.

Spherical, equally sized, dry, electrostatically neutral, and cohesive
particles with a diameter of $d_\text{p} = 0.97~\mu\text{m}$ are
considered.  The particle properties correspond to those of silica, as
listed in Table~\ref{tab:sio2prop}~\citep{khalifa2020data,
  khalifa2021efficient, weiler2008generierung}. To investigate the
effect of cohesive forces on the< deagglomeration behavior, the
Hamaker constant of silica ($2.418 \cdot 10^{-20}~\text{J}$) is
artificially increased twice by a factor of about 10, resulting in
values of $2 \cdot 10^{-19}~\text{J}$ and $2 \cdot
10^{-18}~\text{J}$~(\citet{yao2021deagglomeration}). Based
on the flow conditions and particle properties, several characteristic
non-dimensional numbers are used to describe the nature of the
fluid-particle interactions, as summarized in
Table~\ref{tab:par_parameter}. These include the ratio of primary
particle diameter to the Kolmogorov length scale
$d_\mathrm{p}/l_\mathrm{k}$, the particle Stokes number
$St_\mathrm{p}$, the particle Reynolds number $Re_\mathrm{p}$, and the
ratio of the initial agglomerate diameter to the Kolmogorov length
scale $d_\mathrm{agg}/l_\mathrm{k}$. Here, the agglomerate diameter
$d_\mathrm{agg}$ refers to the diameter of a sphere possessing the
same volume as the convex hull enclosing the agglomerate structure
(see, e.g., Fig.~\ref{fig:break_steps}). The Kolmogorov length scale
is expressed as $l_\text{k} = (\nu^3_\text{f}/\epsilon_\infty)^{1/4}
$.  The Stokes number is defined as $St_\mathrm{p} =
\tau_\mathrm{p}/\tau_\mathrm{k}$, where the particle response time is
given by $\tau_\mathrm{p} = \rho_\mathrm{p} \: d_\mathrm{p}^2 / (18 \:
\rho_\mathrm{f} \: \nu_\mathrm{f})$, and the Kolmogorov time scale is
defined as $\tau_\mathrm{k} = \sqrt{\nu_\mathrm{f} /
  \epsilon_\infty}$. The characteristic particle Reynolds number is
computed based on the fluid root-mean-velocity $u_\mathrm{rms}$ as:
\begin{equation}
	\label{eq:reyp}
	Re_\mathrm{p} 
	= \dfrac{u_\mathrm{rms} \: d_\mathrm{p}}{\nu_\mathrm{f}} \: .
\end{equation}
\begin{table}[b!]
	\centering
		\caption{Dimensionless parameters for the different cases.}
	\begin{tabular}{lccc}
		\toprule[1.5pt]
		Parameter ~~~~~& ~~$Re_{30}$~~~ & $~~Re_{43}$~~~ & $Re_{64}$  \\
		\midrule[1.5pt]
		$d_\mathrm{p}/l_\mathrm{k}$ & $0.52$ & 0.9 & 
		1.63
		\\
		$St_\mathrm{p}$ & $25.4$ & $74.8$ & $247.0$\\
		$Re_\mathrm{p}$ & $1.46$ & $2.99$ & 6.62\\
		$d_\mathrm{agg}/l_\mathrm{k}$ &  6.4 & 11.0 
		& 20.0 \\
		\bottomrule[1.5pt]
	\end{tabular}
	\label{tab:par_parameter}
\end{table}

By combining three different flow Reynolds numbers with three distinct
Hamaker constants, a total of nine independent simulation cases are
considered. Each case involves a single agglomerate composed of 500
primary particles arranged in a nearly spherical structure. The
agglomerate is generated using a two-step procedure. In the first
step, a pure DEM simulation is carried out: 500 particles are randomly
distributed within a spherical domain and then driven toward the
center by a centripetal force until a mechanically stable
configuration is achieved \citep{khalifa2020data,
  yang2008agglomeration}.  During this step, the cohesive van der
Waals interactions are modeled using the Hamaker constant for silica
($H_1$). In the second step, the DEM simulation is restarted without
the centripetal force, allowing the agglomerate to relax to its
natural network of cohesive contacts before the deagglomeration
simulations begin. To generate agglomerates corresponding to higher
cohesion levels, i.e., about 10 times ($H_{10}$) and 100 times
($H_{100}$) the Hamaker constant of silica, the Hamaker constant is
adjusted only in this relaxation step.

\subsection{Computational aspects}
\label{subsec:num_config}

To achieve particle-resolved DNS, the numerical grid must satisfy both
the general DNS resolution requirements in regions far from particles
and the specific requirement of resolving the particle interface.

Given the relatively small ratio $d_\text{p}/l_\text{k}$ (see
Table~\ref{tab:par_parameter}) across all cases, resolving the
particle interface using an equidistant Cartesian grid, without
applying dynamic mesh refinement, would require an extremely fine
mesh, far exceeding standard DNS requirements.

%
\begin{figure}[!t]
	\centering
	{\includegraphics[width=0.45\textwidth,trim= 0.cm 0.cm
		0.cm 0.cm,clip]
		{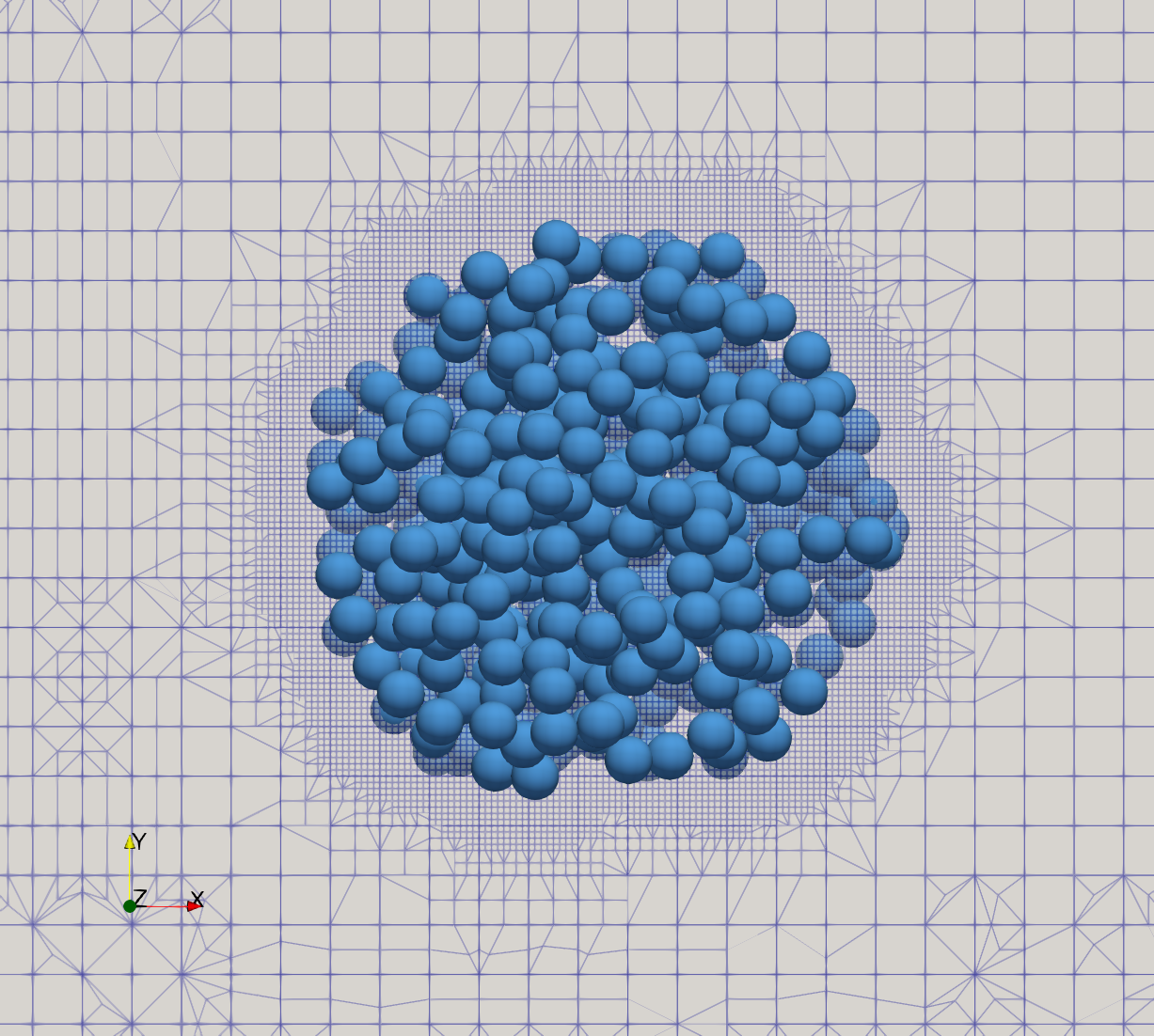}} \\
	\caption{
		A zoom-in of a slice at the center of the
                computational domain, showing the grid resolution
                around the particles. The mesh artifacts outside the
                spherical shell surrounding the particles are
                rendering issues and can be ignored. }
	\label{fig:agg_grid_resol}
\end{figure}

%
On the other hand, if the initial grid is too coarse, the particle
volume may not be accurately represented, which can hinder the
effectiveness of adaptive mesh refinement and prevent the required
resolution from being achieved around the particles. To address this
issue, the initial grid spacing is chosen to be comparable to the
particle diameter, while a dynamic mesh refinement strategy is applied
within a spherical region surrounding the particle, extending to twice
its diameter. The cells in this region are refined three times,
resulting in a final resolution of eight cells per particle diameter
in each spatial direction as depicted in
Fig.~\ref{fig:agg_grid_resol}.

For the considered simulation box with an edge length of about
$200~d_\text{p}$, an initial grid spacing comparable to the particle
diameter requires 200 control volumes per direction, regardless of the
flow Reynolds number $Re_\lambda$. The resulting resolution, defined
as the highest resolved wavenumber $(\pi / \Delta x)$ multiplied by
$l_\text{k}$, is 5.83, 3.39, and 1.87 for $Re_\lambda =$ 30, 43, and
64, respectively. All values are above the recommended minimum DNS
threshold of 1.5~(\citet{bassenne2016constant}), ensuring
sufficient resolution for turbulent structures even for the case with
the highest $Re_\lambda$. It is worth mentioning that utilizing a
unique initial grid for all $Re_\lambda$ cases eliminates grid
resolution effects from the analysis, providing consistency in the
results despite the not necessarily increased computational costs for
$Re_\lambda =$ 30 and 43.

For each $Re_\lambda$ case, the computational domain is divided into
216 parallel blocks (6 blocks per direction) and distributed across
three nodes of HPC system, each equipped with 72 cores (256 GB RAM and
two Intel(R) Xeon(R) Platinum 8360Y processors).

A critical aspect of particle-resolved CFD-DEM simulations is the
choice of the time-step size, which must be sufficiently small to
accurately capture both particle collisions and the smallest turbulent
eddies while maintaining numerical stability, particularly in the
presence of dynamic mesh refinement.  On the DEM side, the
elastic-response (Hertz) time scale \citep{li2011adhesive} and the
Rayleigh-wave time scale \citep{li2005comparison} are commonly used to
determine appropriate time-step sizes. In this study, the DEM time
step is set to $10^{-10}$~s ensuring that it remains below 10~\% of
either characteristic time scale, thereby providing reliable collision
detection and resolution. Such a time-step size is naturally smaller
than what is required for DNS and also satisfies the stability
constraints of the CFD solver. However, to maintain consistency
between the fluid and particle solvers and to eliminate the need for
separate time-stepping schemes, the time steps of both the fluid
solver and DEM are set to be identical, enabling fully synchronized
time advancement.

\subsection{Simulation procedure}
\label{subsec:procedure}
In the first step, single-phase flow simulations are performed to
establish the statistically stationary state of homogeneous isotropic
turbulence for each Reynolds number. To initialize these simulations,
a velocity field is generated using random numbers following a
Gaussian distribution with zero mean and a standard deviation equal to
the corresponding $u_\text{rms}$ according to
Eq.~(\ref{eq:urms}). Once the statistically stationary state is
reached, the agglomerates consisting of 500 particles are introduced
at the center of the domain with zero velocity, while the solution of
the translational and rotational equations of motion for the particles
remains disabled. During this phase, the flow is allowed to develop
around the particles before enabling their motion and interaction with
the fluid. This step is important to mitigate the influence of
pressure peaks and other numerical artifacts that could arise from the
direct insertion of particles, which might otherwise affect the
deagglomeration behavior. After the flow has stabilized again,
indicated by the spatially averaged turbulent kinetic energy and
dissipation rate reaching a new steady state, the solution of the
equations of motion for the particles is activated, and the monitoring
of results begins.

\newpage
\section{Results and discussion
	\label{sec:results}}
In this section, the unladen fluid flow is first examined, with focus
on its convergence toward the intended homogeneous isotropic
turbulence for each Reynolds number. Subsequently, a detailed
discussion of the results for the agglomerate-laden flow is provided.
\subsection{Single-phase simulations \label{subsec:res_sing}}
\begin{figure}[!t]
	\centering
	\subfloat[$Re_\lambda$ = 30]{%
		\includegraphics[width=0.32\textwidth]{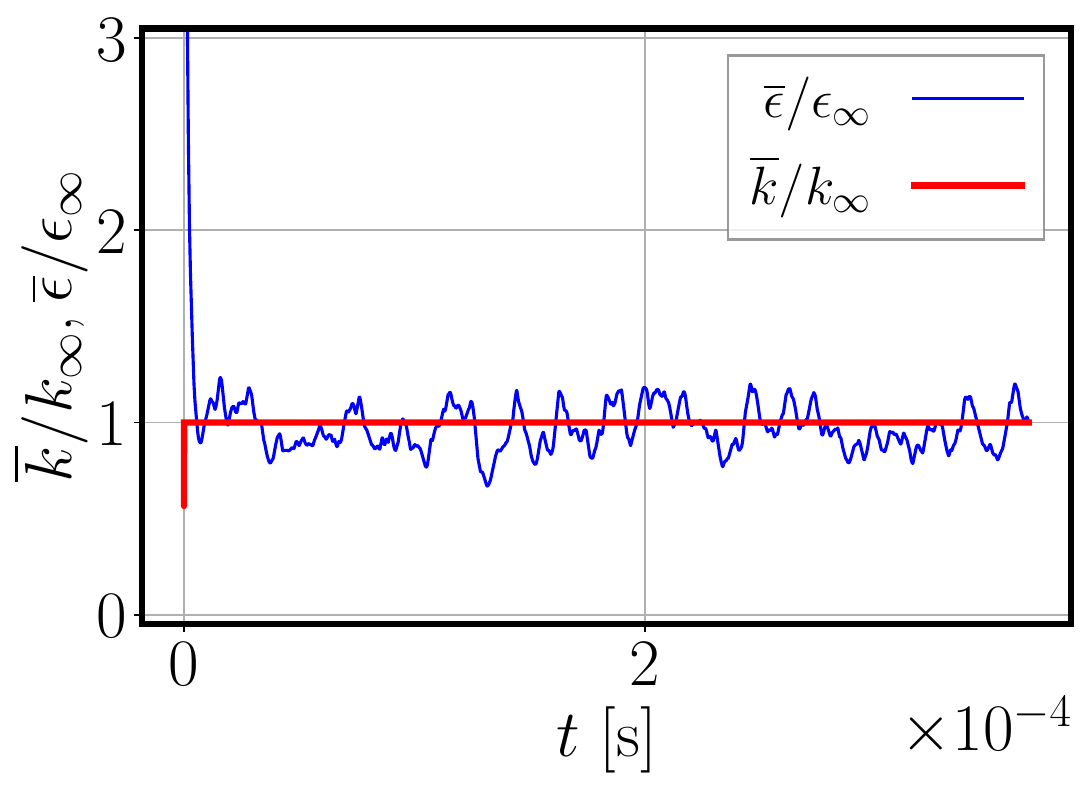}
		\label{subfig:k_eps_re30}
	}
	\hfill
	\subfloat[$Re_\lambda$ = 43]{%
		\includegraphics[width=0.32\textwidth]{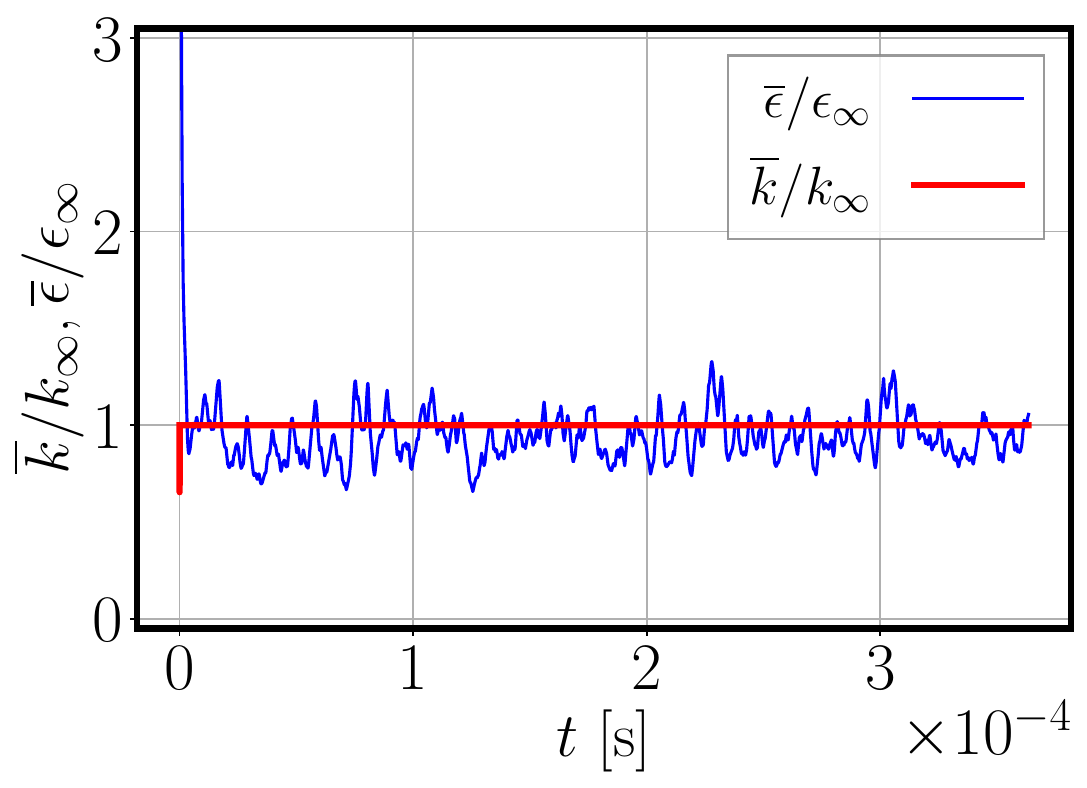}
		\label{subfig:k_eps_re43}
	}
	\hfill
	\subfloat[$Re_\lambda$ = 64]{%
		\includegraphics[width=0.32\textwidth]{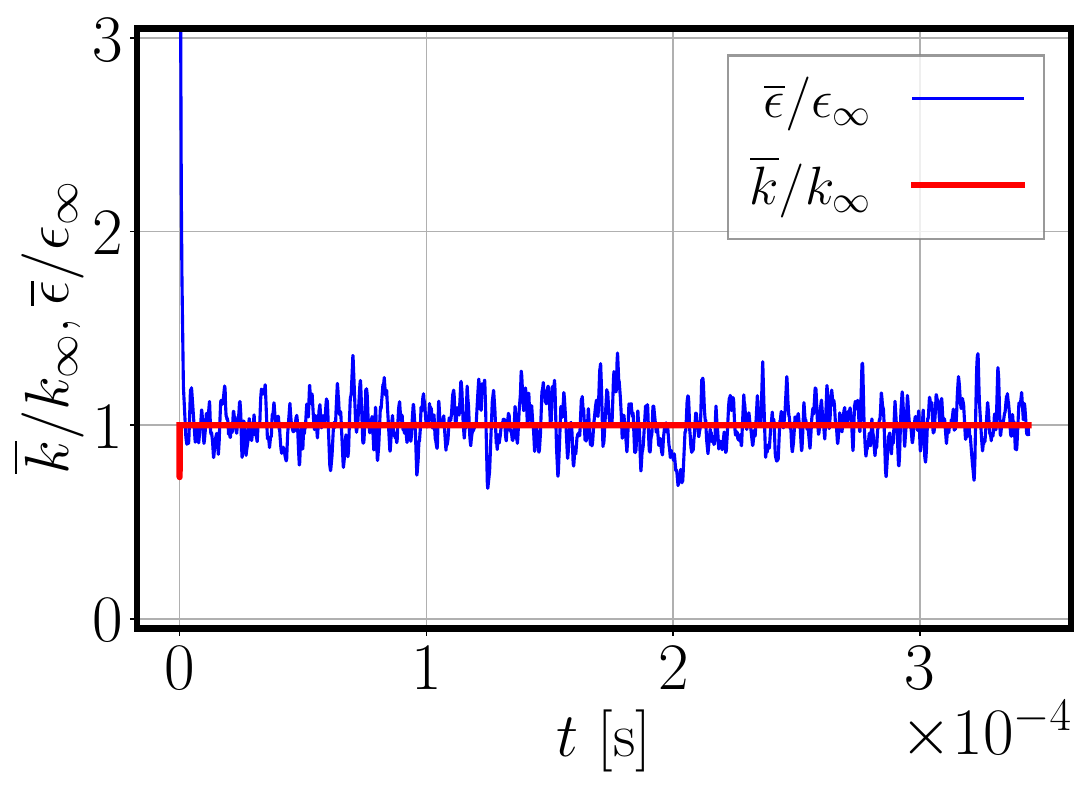}
		\label{subfig:k_eps_re64}
	}
	
	\caption{Time histories of the spatially averaged turbulent kinetic energy 
		$\overline{k}/k_\infty$ and turbulent dissipation rate 
		$\overline{\epsilon}/\epsilon_\infty$.}
	\label{fig:k_eps}
\end{figure}

As discussed earlier, the applied forcing approach (see
Eq.~(\ref{eq:for_coeff})) ensures that the turbulent kinetic energy
remains constant at $k_\infty$. Figure~\ref{fig:k_eps} shows that, for
all $Re_\lambda$ cases, the spatially averaged turbulent kinetic
energy $\overline{k}$ rapidly reaches its corresponding value
$k_\infty$ and remains stable throughout the simulation. Additionally,
while the theoretically predicted dissipation rate $\epsilon_\infty$
is not explicitly imposed by the forcing term, the results indicate
that the spatially averaged dissipation rate $\overline{\epsilon}$
fluctuates closely around the expected value $\epsilon_\infty$ as also
visible in Fig.~\ref{fig:k_eps}. This is consistent with the findings
of \citet{bassenne2016constant}, who originally proposed the
source-term formulation. Overall, the results in Fig.~\ref{fig:k_eps}
confirm that the integral quantities, such as the spatially averaged
$\overline{k}$ and $\overline{\epsilon}$, align well with the target
values for the given $Re_\lambda$.

Furthermore, the evolution of the spatially averaged velocity vector
$\overline{u_i}$ (not depicted here) reveals that the ratios
$\abs{\overline{u_i}}/u_\text{rms}$ are in the order of $10^{-9}$,
$10^{-7}$, and $10^{-6}$ for $Re_\lambda = 30$, 43, and 64,
respectively. In addition, the three components of $\overline{u_i}$
contribute equally to $u_\text{rms}$ in each case, indicating an
overall negligibly small mean velocity that is unique across all
spatial directions. This behavior is consistent with the expectations
for homogeneous isotropic turbulence. However, to further demonstrate
the isotropy and homogeneity properties of the generated fields, a
more detailed analysis follows.

\begin{figure}[!t]
	\centering
	\subfloat[$Re_\lambda$ = 30]{%
		\includegraphics[width=0.32\textwidth]{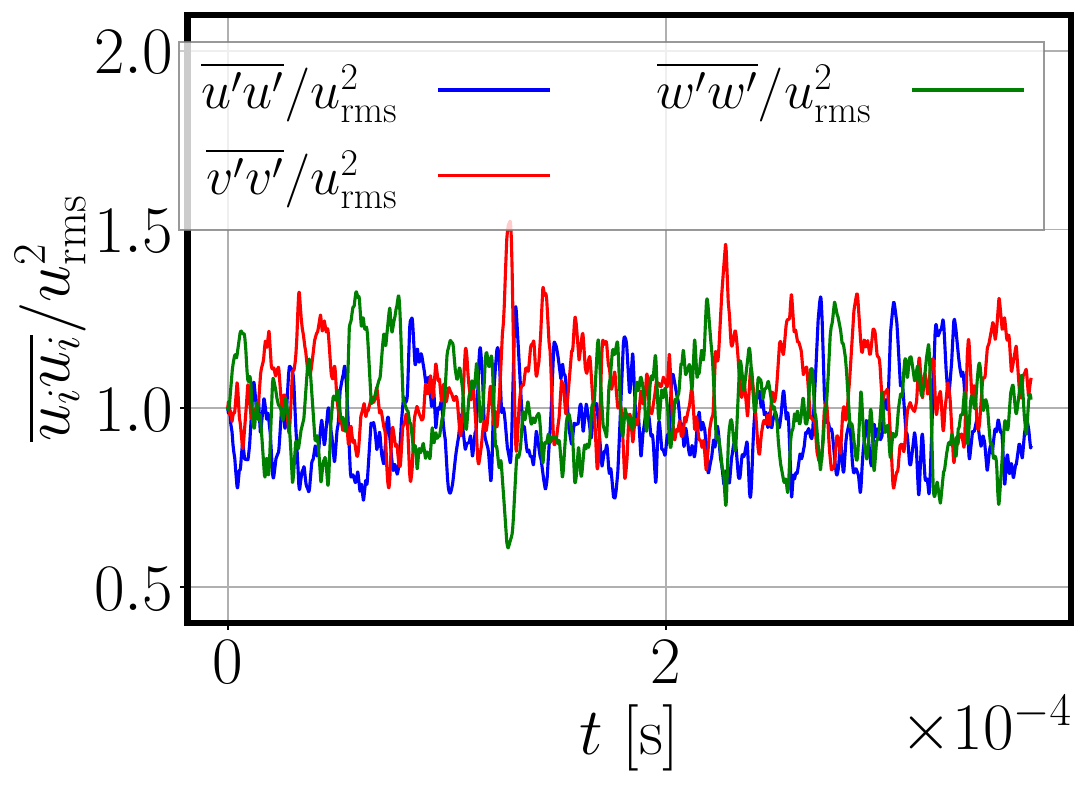}
		\label{subfig:stress_re30}
	}
	\hfill
	\subfloat[$Re_\lambda$ = 43]{%
		\includegraphics[width=0.32\textwidth]{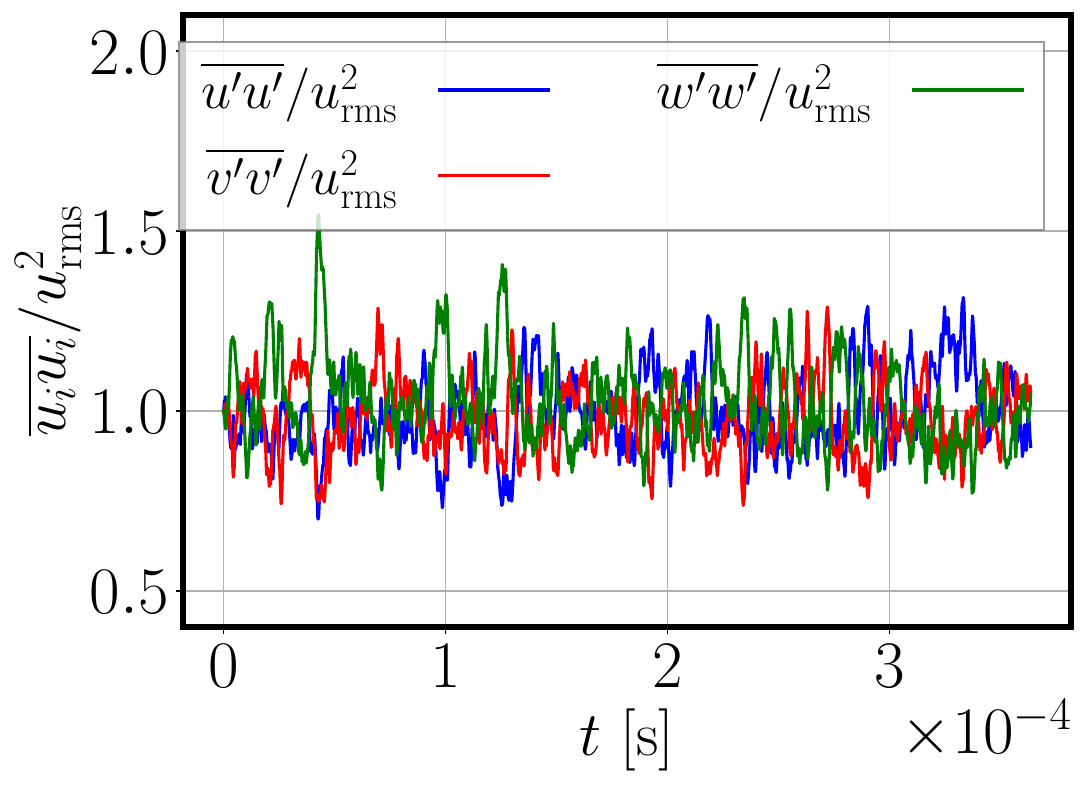}
		\label{subfig:stress_re43}
	}
	\hfill
	\subfloat[$Re_\lambda$ = 64]{%
		\includegraphics[width=0.32\textwidth]{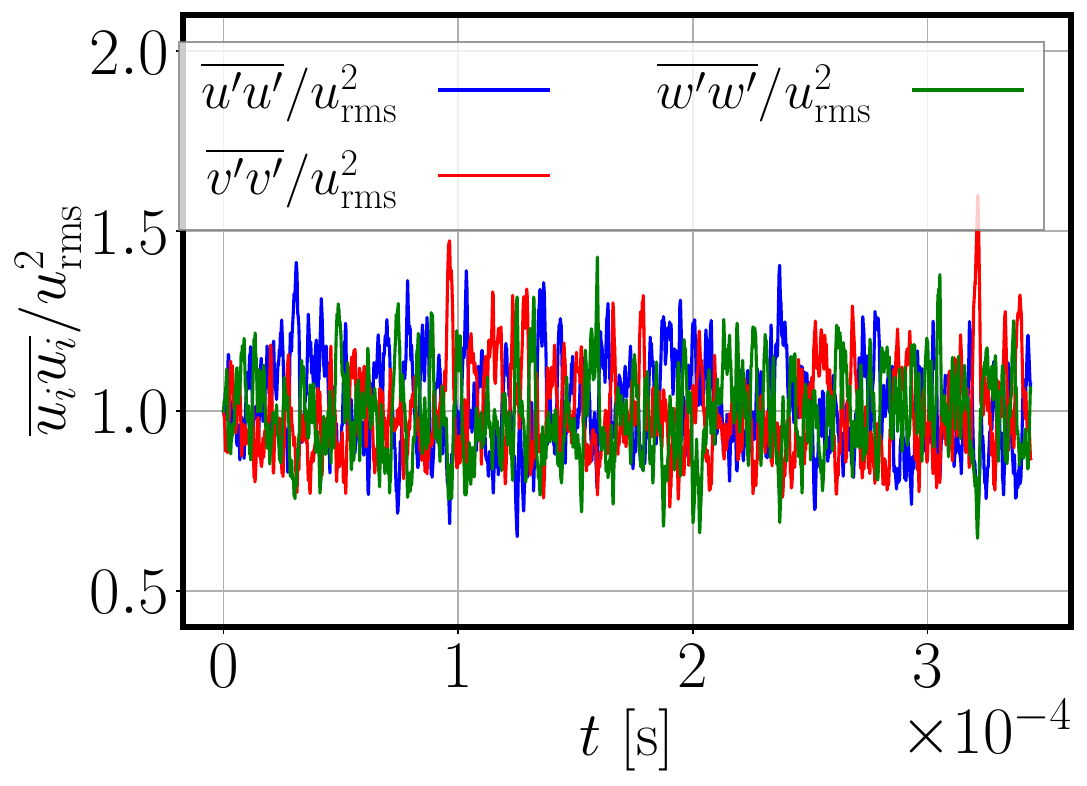}
		\label{subfig:stress_re64}
	}
	\caption{Time histories of the spatially averaged normal Reynolds stresses 
		$\overline{u_i^\prime 
			u_i^\prime}/u^2_\mathrm{rms}$.}
	\label{fig:flow_stresses}
\end{figure}

Isotropic turbulence implies that the turbulent characteristics are
independent of the spatial direction. Figure~\ref{fig:flow_stresses}
depicts the time histories of the spatially averaged normal components
of the Reynolds stress tensor $\overline{u_i^\prime
  u_i^\prime}/u^2_\mathrm{rms}$. The time histories show that for all
three $Re_\lambda$ cases, the components in all three directions
fluctuate around similar values. This suggests that the flow is
isotropic to a good extent. Note that the sum of the normal Reynolds
stresses is always nearly constant as shown by $\overline{k}$ in
Fig.~\ref{fig:k_eps}.

\begin{figure}[h]
	\centering
	\subfloat[$R_{uu}$, $Re_\lambda$ = 30]{%
		\includegraphics[width=0.32\textwidth]{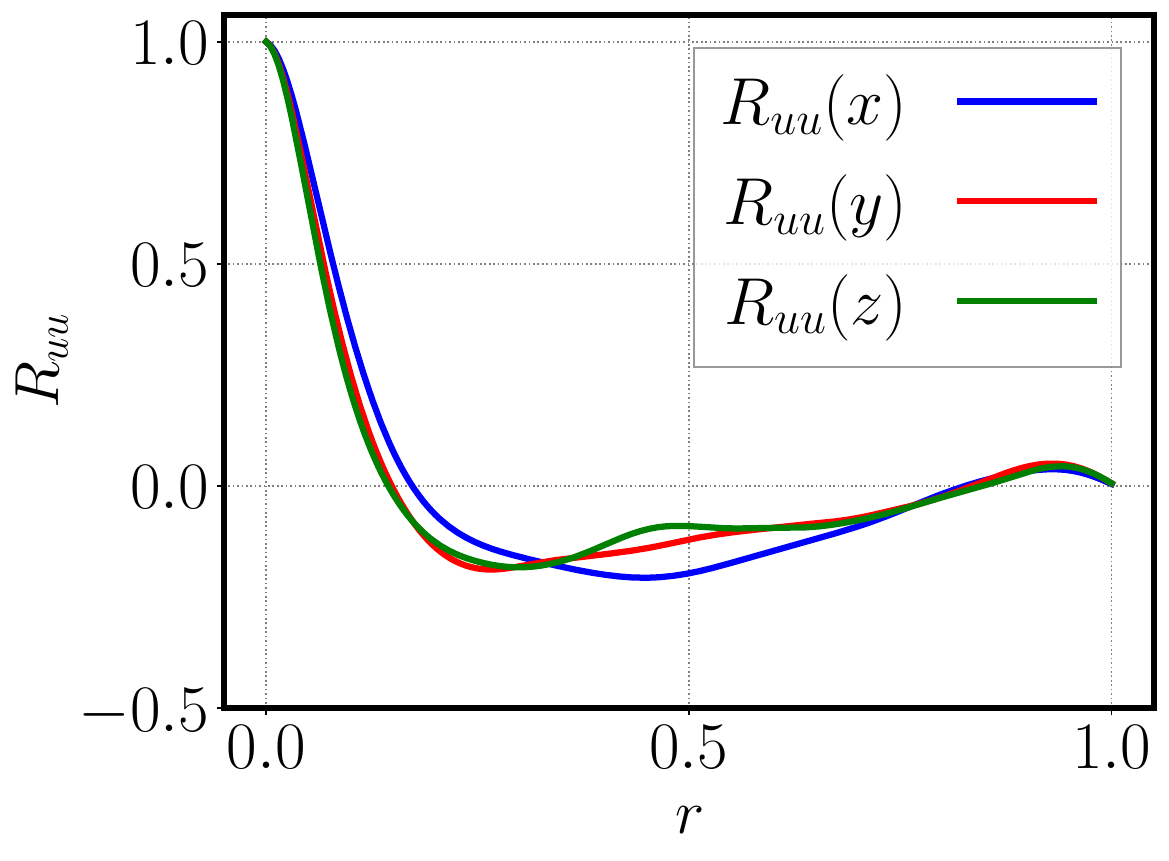}
		\label{subfig:Ruu_30}
	}
	\hfill
	\subfloat[$R_{vv}$, $Re_\lambda$ = 30]{%
		\includegraphics[width=0.32\textwidth]{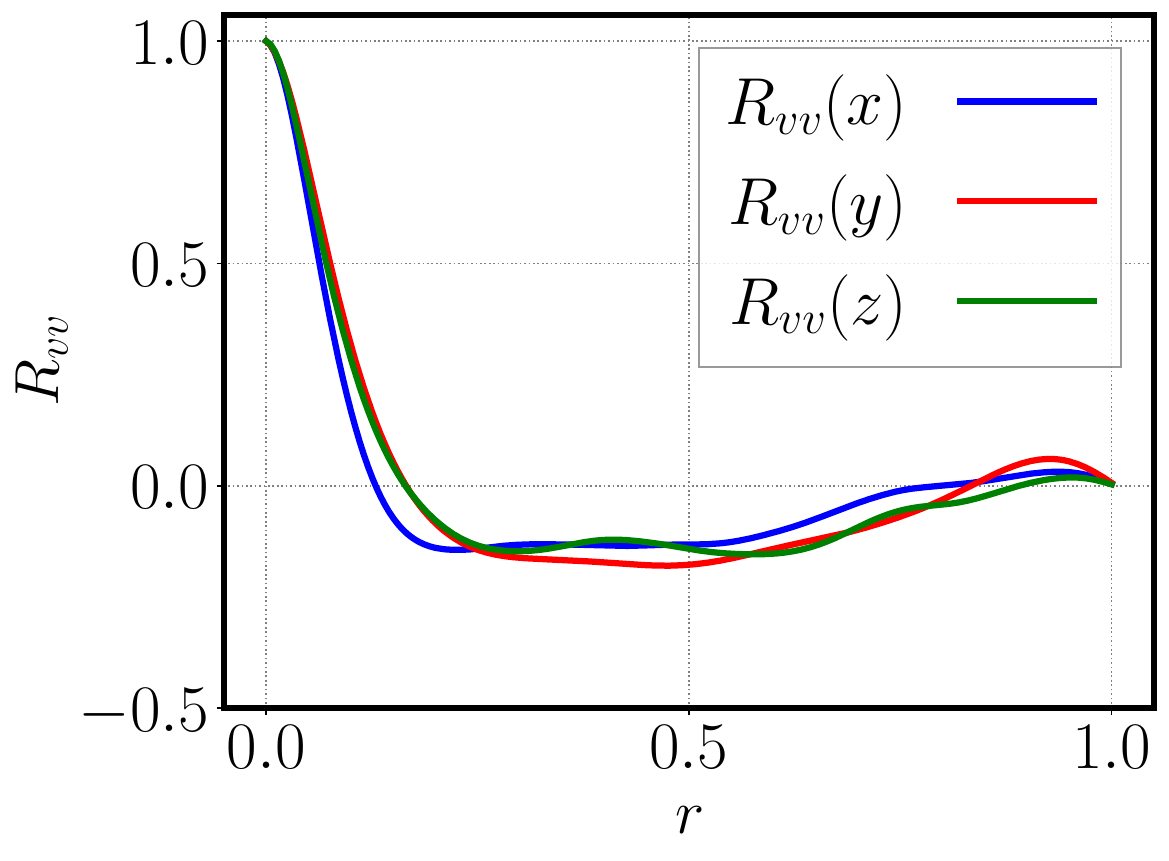}
		\label{subfig:Rvv_30}
	}
	\hfill
	\subfloat[$R_{ww}$, $Re_\lambda$ = 30]{%
		\includegraphics[width=0.32\textwidth]{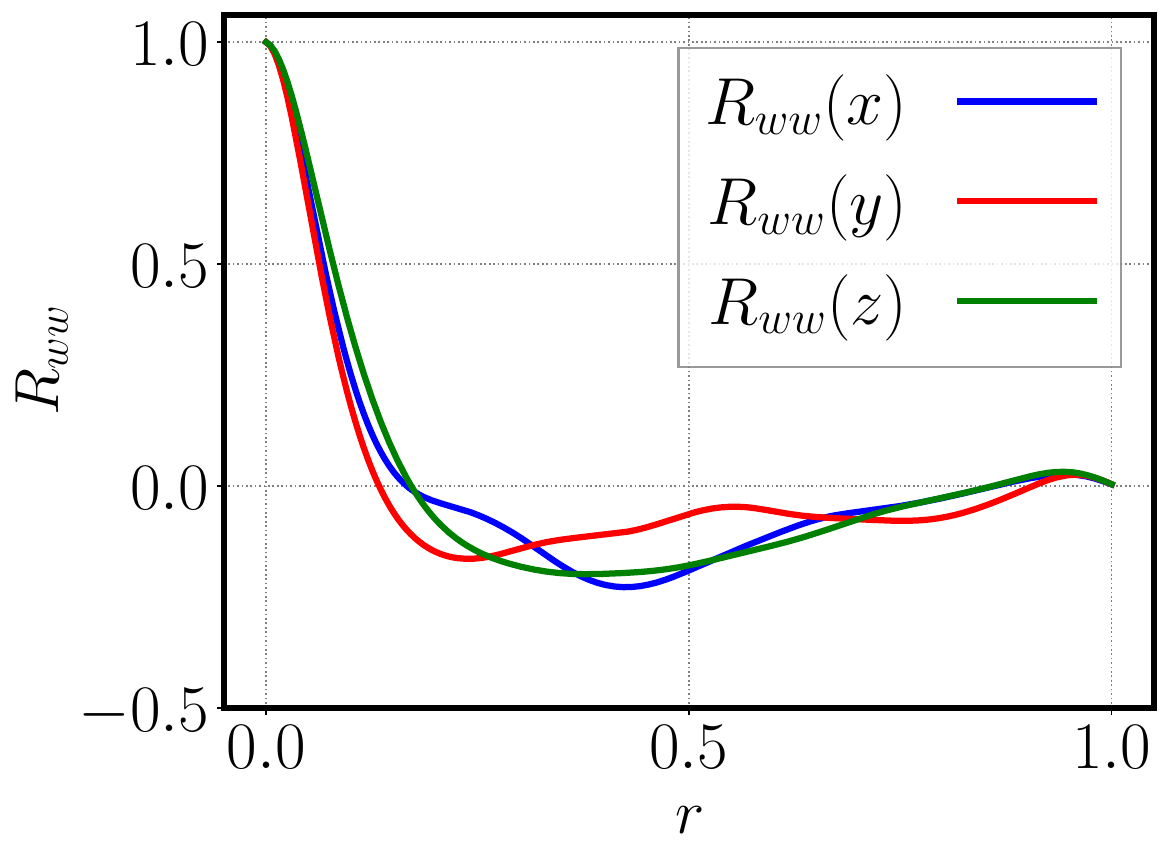}
		\label{subfig:Rww_30}
	} \\
	\subfloat[$R_{uu}$, $Re_\lambda$ = 43]{%
		\includegraphics[width=0.32\textwidth]{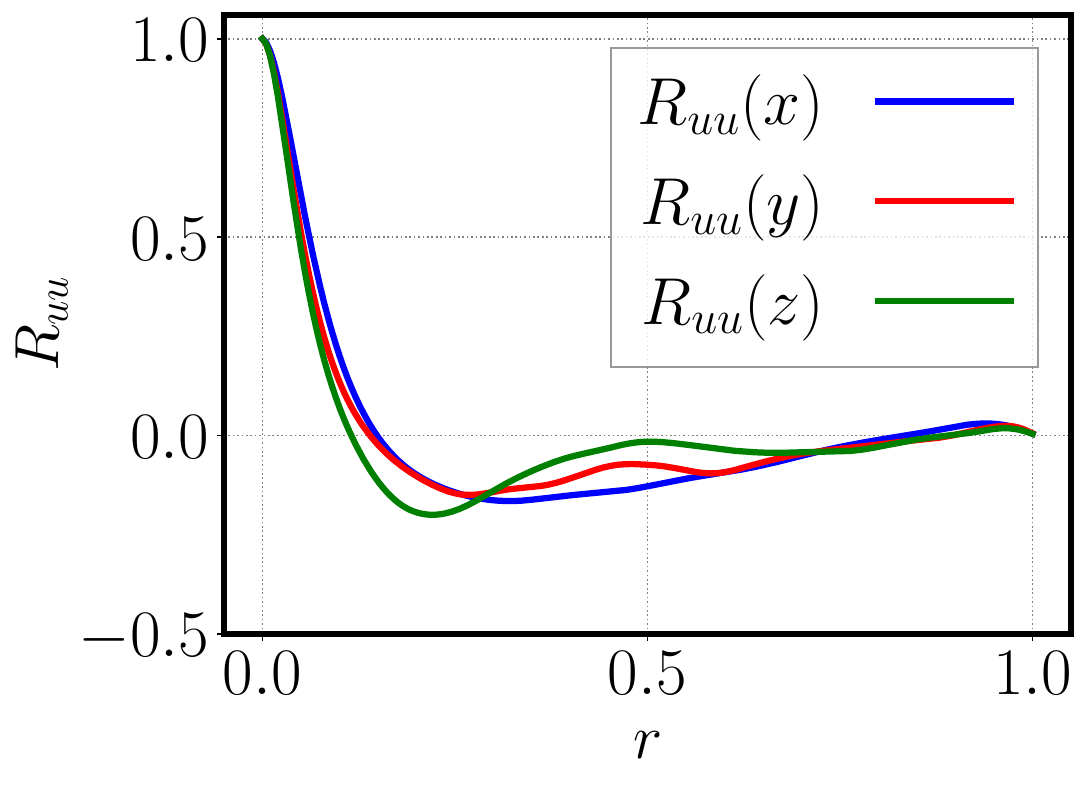}
		\label{subfig:Ruu_43}
	}
	\hfill
	\subfloat[$R_{vv}$, $Re_\lambda$ = 43]{%
		\includegraphics[width=0.32\textwidth]{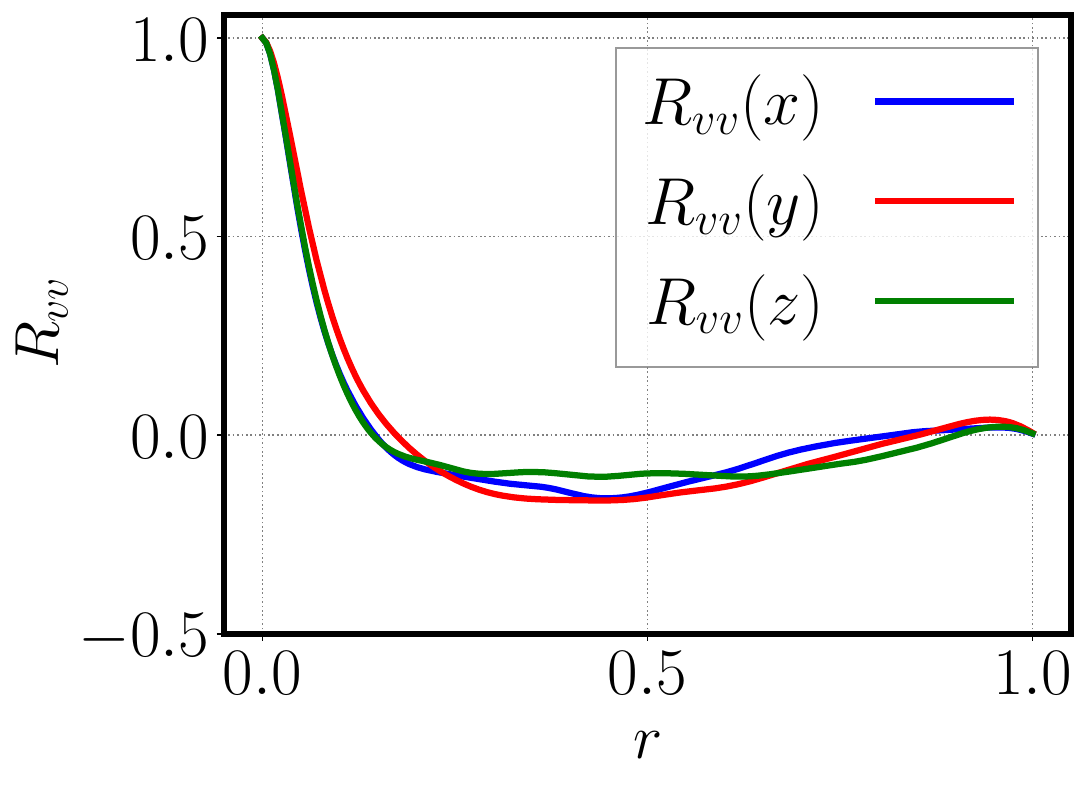}
		\label{subfig:Rvv_43}
	}
	\hfill
	\subfloat[$R_{ww}$, $Re_\lambda$ = 43]{%
		\includegraphics[width=0.32\textwidth]{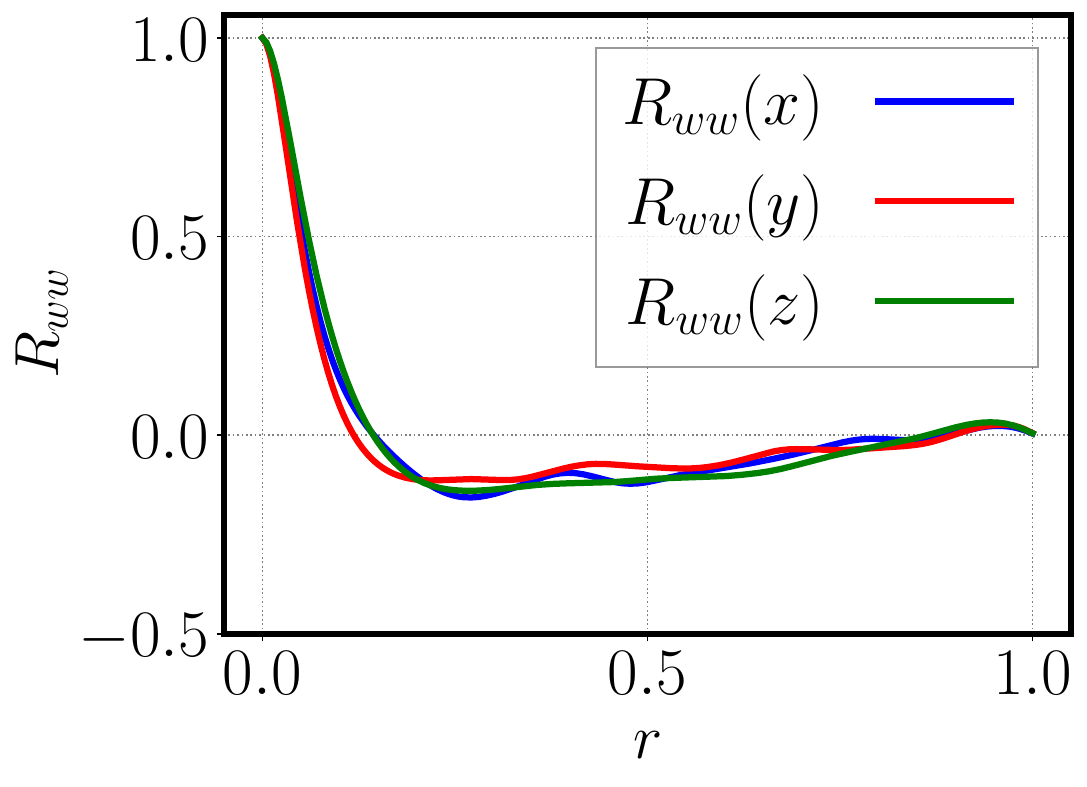}
		\label{subfig:Rww_43}
	}\\
	
	\subfloat[$R_{uu}$, $Re_\lambda$ = 64]{%
		\includegraphics[width=0.32\textwidth]{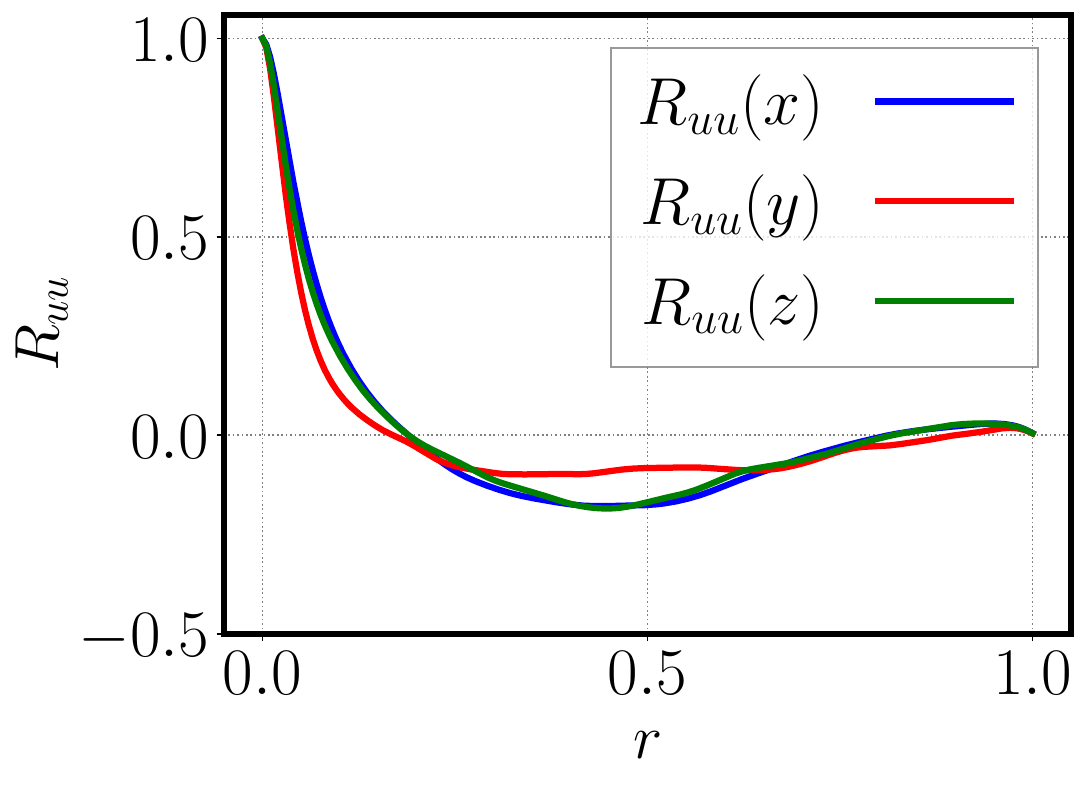}
		\label{subfig:Ruu_64}
	} 
	\hfill
	\subfloat[$R_{vv}$, $Re_\lambda$ = 64]{%
		\includegraphics[width=0.32\textwidth]{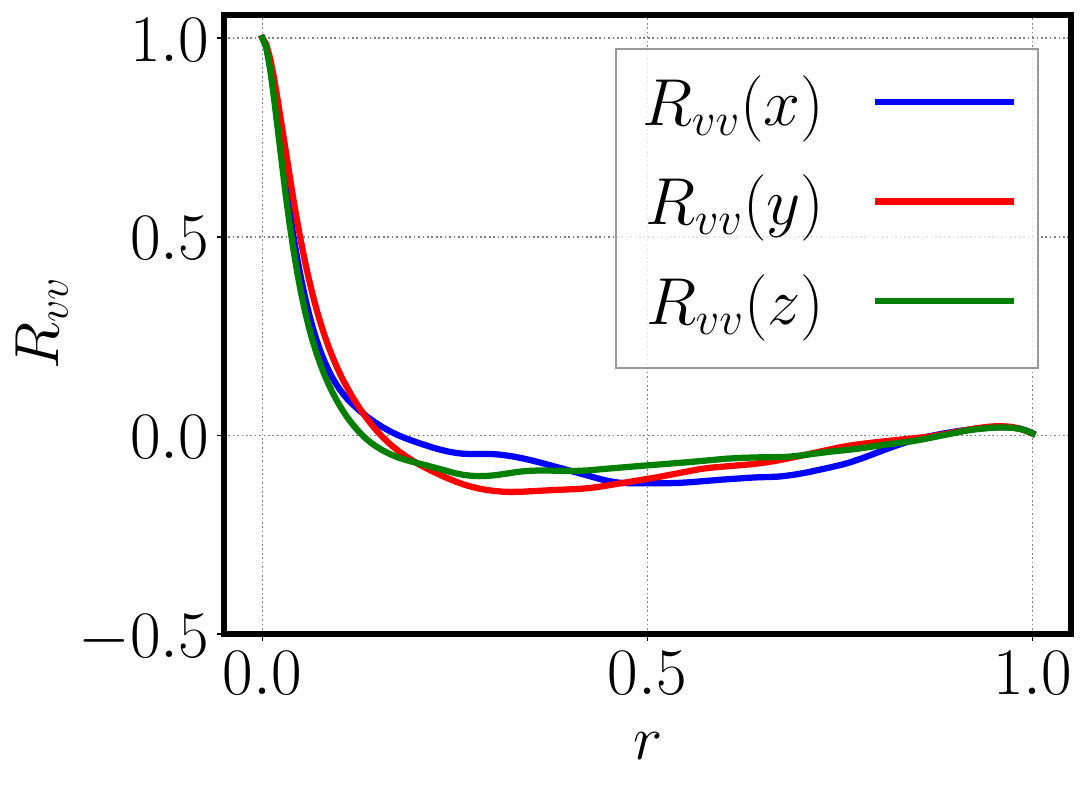}
		\label{subfig:Rvv_64}
	}
	\hfill
	\subfloat[$R_{ww}$, $Re_\lambda$ = 64]{%
		\includegraphics[width=0.32\textwidth]{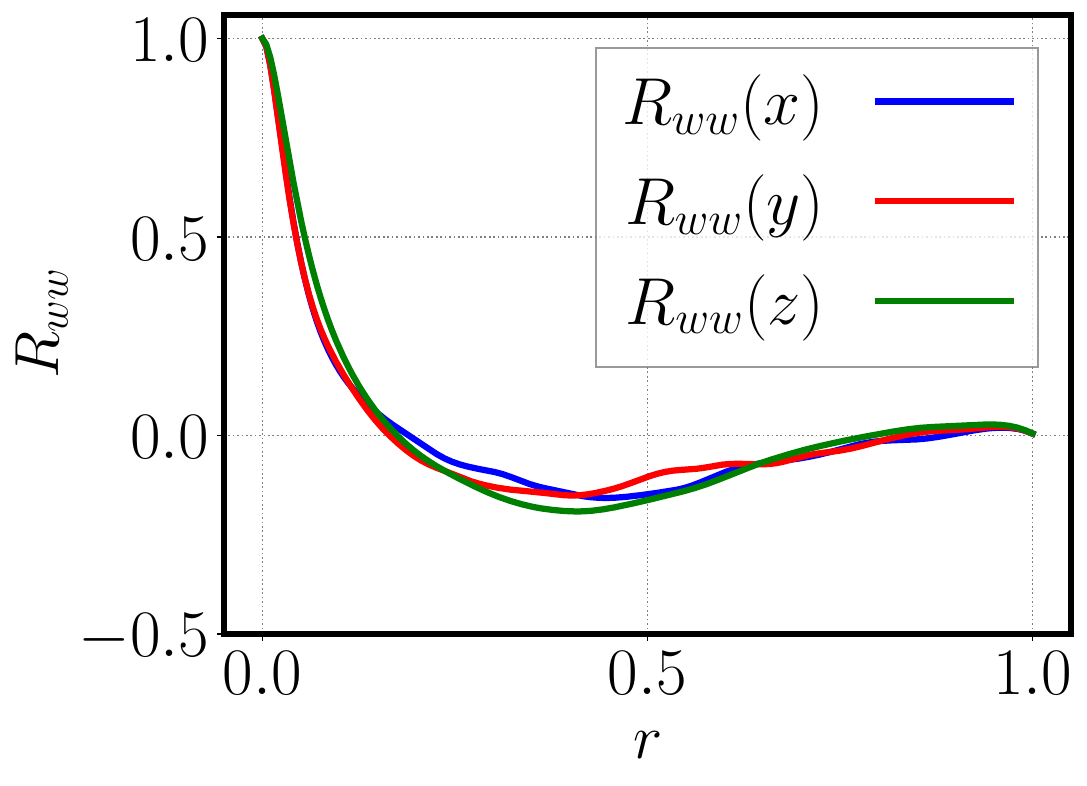}
		\label{subfig:Rww_64}
	}
	\caption{Two-point correlation functions $R_{u_i u_i}$ averaged along all 
		possible lines 
		and shown for all three spatial directions.}
	\label{fig:twopoint}
\end{figure}

Homogeneous turbulence on the other hand implies that statistical
properties remain uniform in space and are independent of the physical
position. To assess the homogeneity of the generated fields, velocity
data are extracted at an arbitrary time when the flow is fully
developed. The two-point correlation function $R_{u_i u_i}$ is
computed along all possible lines $r$ in a given direction, followed
by spatial averaging to obtain a representative measure of velocity
correlations. The results in Fig.~\ref{fig:twopoint} show that for
each $Re_\lambda$, the two-point correlation function $R_{u_i u_i}$ of
the velocity fluctuations in a given direction are nearly identical
across all spatial directions ($x,y,z$). This indicates that $R_{u_i
  u_i}$ depends solely on the distance $r$ between the two points,
which further supports the assumption of homogeneous turbulence.

\clearpage
\subsection{Agglomerate-laden flow simulations\label{subsec:res_coup}}

\subsubsection{Advantages of the particle-resolved simulation framework}

\begin{figure}[!t]
	\centering
	\subfloat[Full agglomerate]{%
		\includegraphics[width=0.32\textwidth]{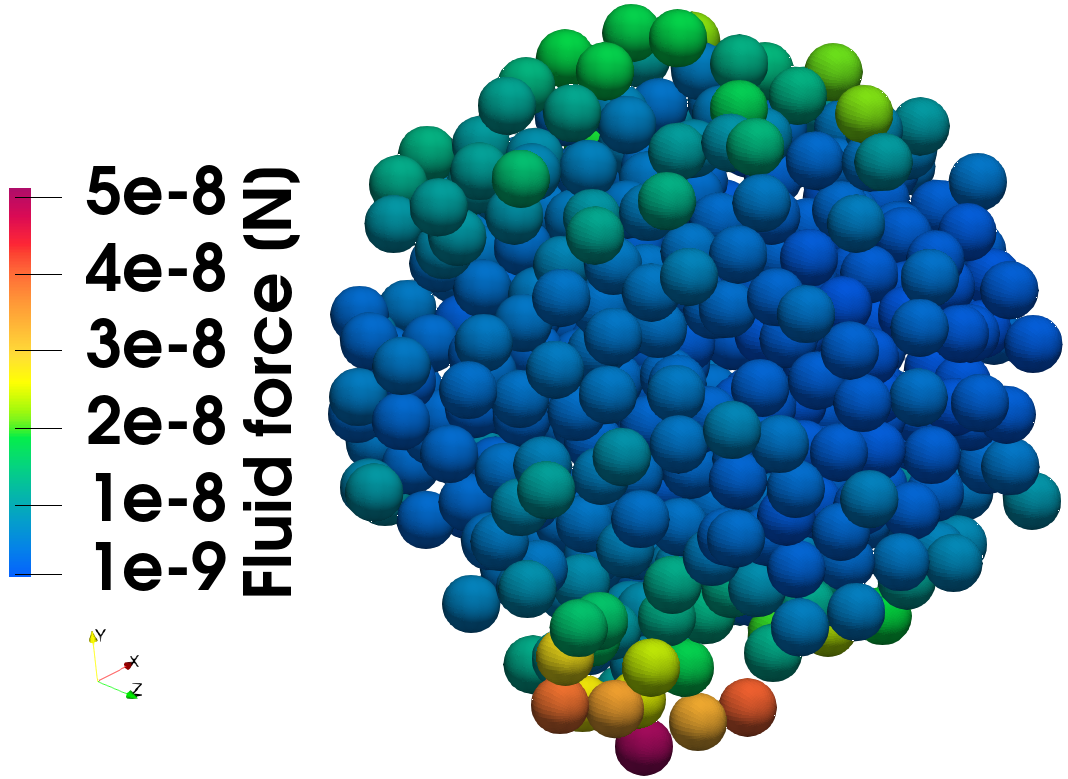}
		\label{subfig:fullagg}
	}
	\hfill
	\subfloat[Zoom of the lower region]{%
		\includegraphics[width=0.32\textwidth]{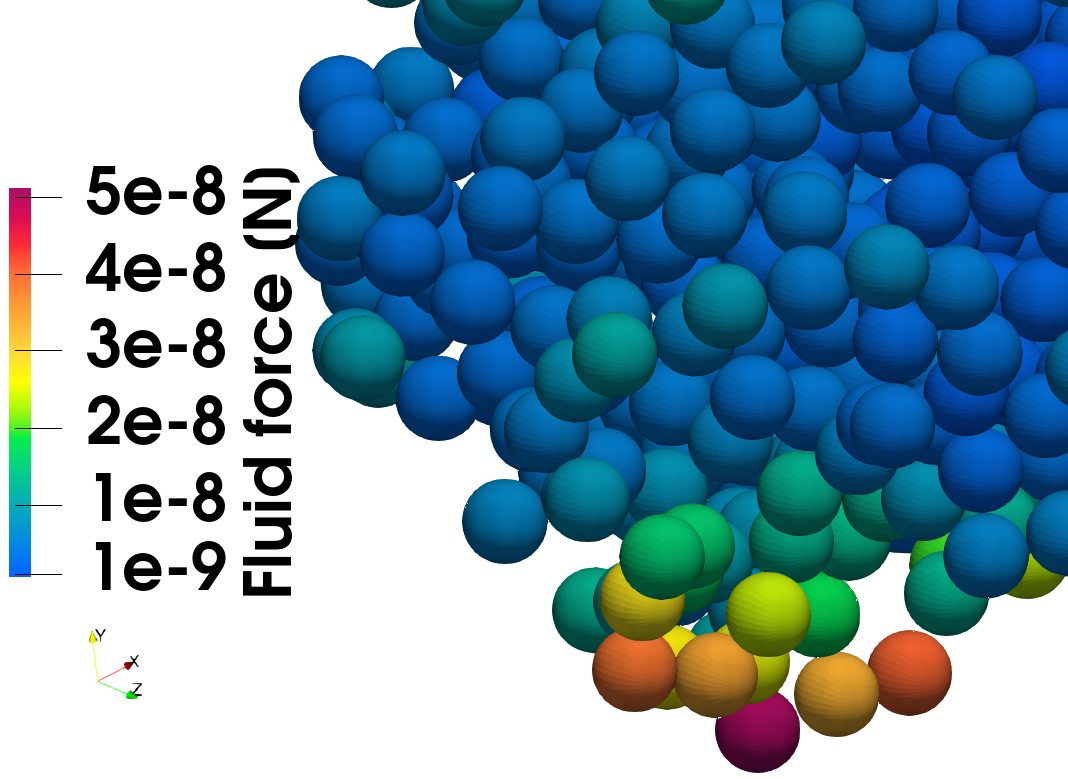}
		\label{subfig:zoomagg}
	}
	\hfill
	\subfloat[Slice near the center]{%
		\includegraphics[width=0.32\textwidth]{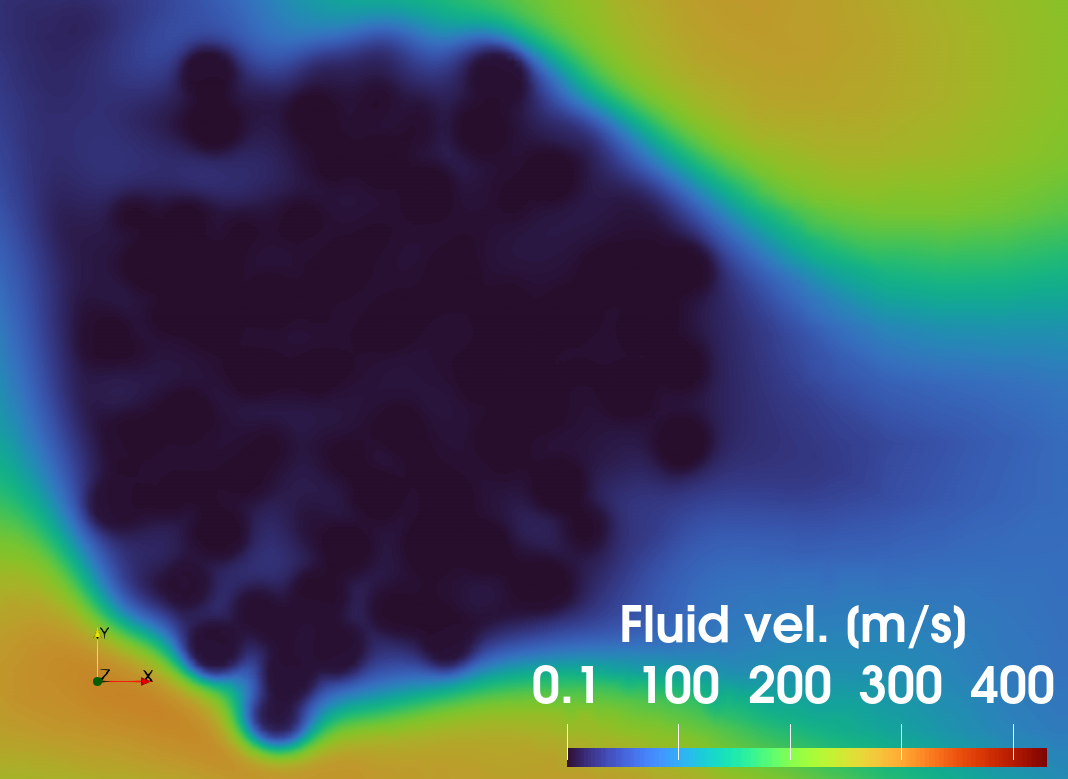}
		\label{subfig:sliceagg}
	}
	\caption{Visualization of the fluid forces acting on the particles ((a) and 
		(b)) and of the magnitude of the fluid velocity within a slice near the 
		center of the 
		agglomerate for the representative case [$Re_{64}, H_{100}$].}
	\label{fig:repres_case}
\end{figure}

The primary objective of the present study is the analysis of highly
resolved deagglomeration data, with the aim of enabling the subsequent
derivation of mechanistic models that can be employed in more
efficient and less-detailed methods. These are the Euler--Euler or the
Euler--Lagrange approach, where agglomerates are represented by
effective spheres. Such models are expected to predict the breakage
rate, the particle size distribution, and the separation axis.
Although attempts have been made in the past to derive such models,
they have been primarily based on unresolved point-particle approaches
limiting their accuracy.

To illustrate the capabilities of the present particle-resolved
simulation framework, a representative case is visualized in
Fig.~\ref{fig:repres_case}, involving the flow at $Re_{64}$ and the
particle cohesion characterized by a Hamaker constant $H_{100}$, which
corresponds to 100 times the Hamaker constant of silica. In
Fig.~\ref{subfig:fullagg}, the full agglomerate is shown, with
particles colored by the magnitude of the fluid force acting upon
them. A magnified view of the region with strong fluid forces is
provided in Fig.~\ref{subfig:zoomagg}. Both visualizations reveal a
highly non-uniform and heterogeneous distribution of fluid forces
throughout the agglomerate. These spatial variations are attributed to
the complex local flow structures interacting with the irregular
geometry of the agglomerate. This heterogeneity plays a critical role
in the deformation and eventual breakup processes and is inherently
resolved in the fully coupled simulations employed here.

In contrast, conventional point-particle models either completely
neglect the effect of neighboring particles or approximate shadowing
using spatially averaged expressions based on empirically determined
crowding factors. However, these models fall short in accurately
capturing the shadowing effect, wake interactions, and localized
accelerations that arise in dense particle configurations. For
instance, the widely used model by \citet{beetstra2007drag} introduces
a volume-fraction-based correction to the drag force in static
particle arrays, but does not account for transient or spatial
variations in the drag force within agglomerates.

Further insights are provided in Fig.~\ref{subfig:sliceagg}, which
depicts a slice of the fluid velocity magnitude near the agglomerate
center, with particles removed for clarity. The influence of the
agglomerate geometry on the flow structure is clearly
visible. Non-negligible velocity gradients can be observed in the
narrow interstitial regions between particles, i.e., regions where
local strain and shear significantly contribute to agglomerate
deformation and breakup. Such flow details cannot be captured by
two-way coupled Euler--Lagrange simulations, which typically represent
particles as point sources of momentum and do not resolve the actual
flow morphology around and between particles.

\begin{figure}[!t]
	\centering
	\subfloat[$t^*$ = 2300]{%
		\includegraphics[width=0.4\textwidth]{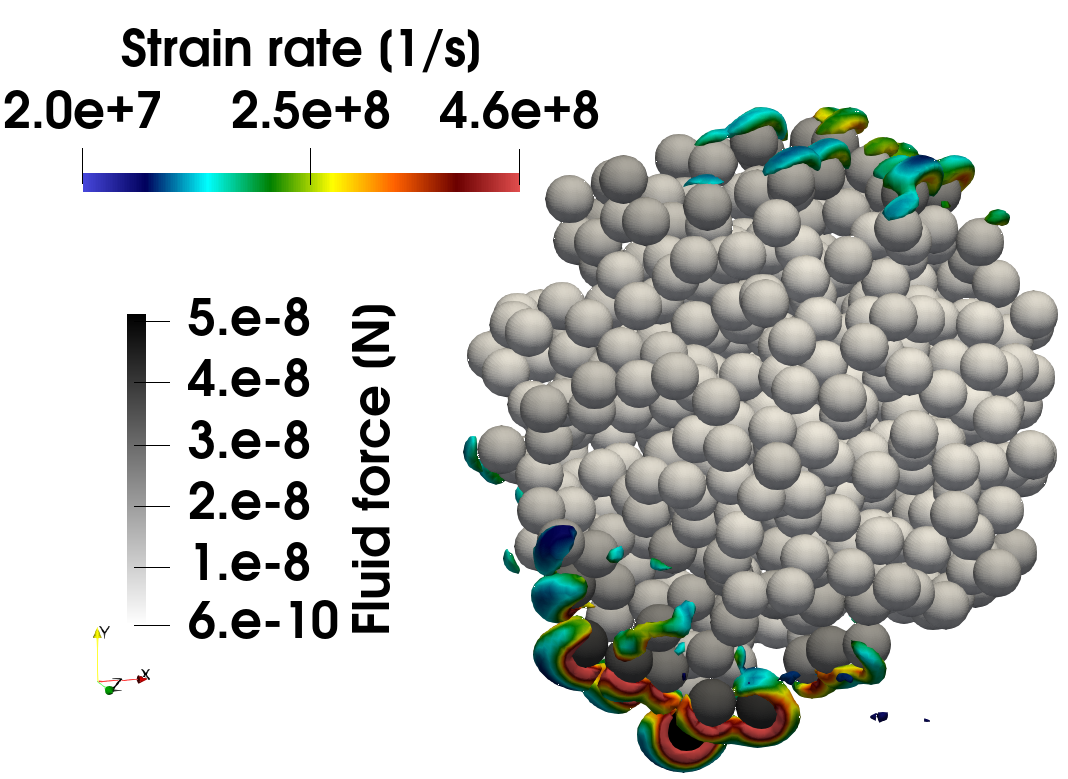}
		\label{subfig:break_step_1}
	}
	\hfil
	\subfloat[$t^*$ = 3200]{%
		\includegraphics[width=0.37\textwidth]{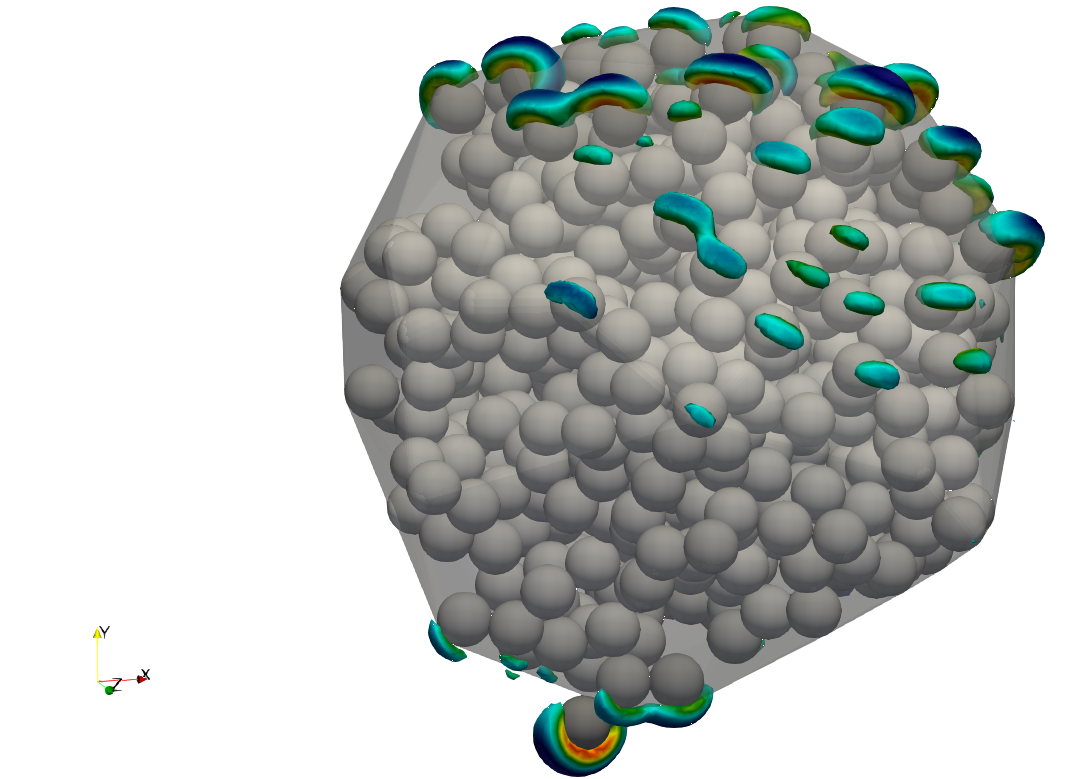}
		\label{subfig:break_step_2} 			
	} \\	
	
	\subfloat[$t^*$ = 34,400]{%
		\includegraphics[width=0.37\textwidth]{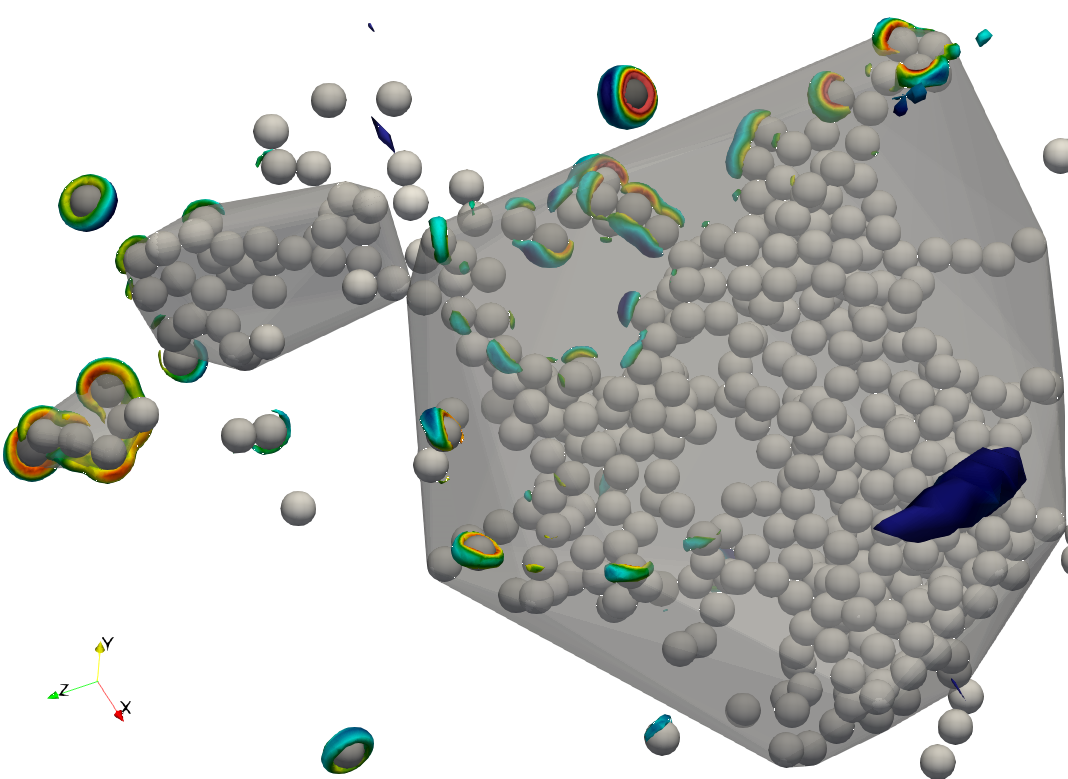}
		\label{subfig:break_step_3} 
	}
	\hfil
	\subfloat[$t^*$ = 57,400]{%
		\includegraphics[width=0.37\textwidth]{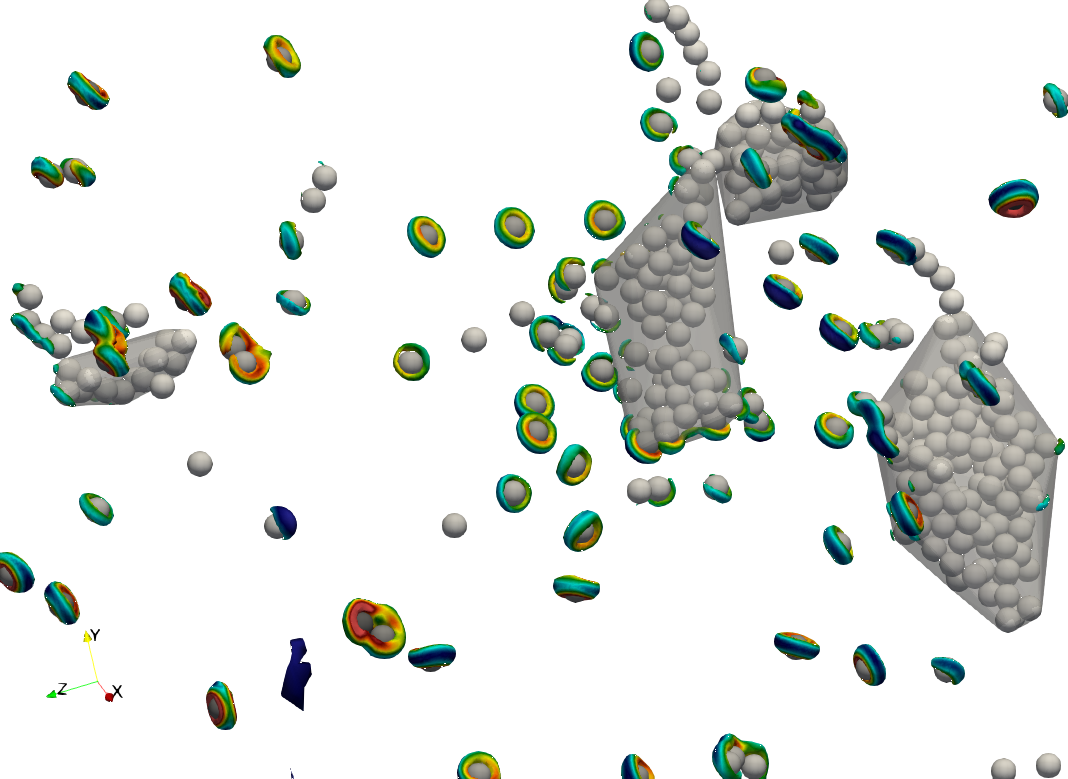}
		\label{subfig:break_step_4} 								
	} 
	\caption{Snapshots of the agglomerate and vortical flow structures 
		($\lambda_2$) at characteristic time steps for the 
		case [$Re_{64}, H_{100}$], illustrating the progression of the breakup 
		process. In figures (b)--(d) the four largest fragments are enclosed by 
		convex hulls to visually distinguish them in the illustrations.}
	\label{fig:break_steps}
\end{figure}

In Fig.~\ref{fig:break_steps}, the influence of local flow structures
on agglomerate breakage is examined using snapshots from the
representative case [$Re_{64}, H_{100}$]. In
Fig.~\ref{subfig:break_step_1}, the intact agglomerate is shown after
2300 simulation time steps, just prior to the onset of
breakup. Different gray scales are assigned to the particles according
to the magnitude of the fluid forces acting on them. To highlight the
local vortical structures, isosurfaces of a suitably chosen negative
$\lambda_2$ value are displayed and colored based on the local strain
rate magnitude. The $\lambda_2$ criterion identifies vortices as
regions where a symmetric tensor derived from the velocity gradient
indicates a local pressure minimum due to rotational motion.

As shown in Fig.~\ref{subfig:break_step_2}, particle erosion begins
after about 3200 time steps in regions exhibiting high strain
rates. The convex hull of the largest remaining fragment is also
visualized to emphasize that the eroded particles no longer belong to
the main structure, confirming a breakage event. In the subsequent
snapshots depicted in
Figs.~\ref{subfig:break_step_3}~and~\ref{subfig:break_step_4}, it can
be observed that breakage continues to occur in areas subjected to
elevated fluid stresses, and that the detached fragments are
surrounded by regions of high strain.

These depictions underscore the advantages of particle-resolved
simulations in delivering detailed and physically consistent insights
into the mechanisms governing agglomerate breakage. Although
computationally demanding, this approach enables a comprehensive
resolution of the fluid--particle interactions that drive
deagglomeration in turbulent flows.

\subsubsection{Breakage dynamics}

Since all investigated cases begin with a single intact agglomerate
composed of 500 particles, it is of interest to compare the time
evolution of the number of resulting fragments under varying flow and
cohesion conditions. To this end, the Fragmentation Ratio (FR) is
introduced:
\begin{equation} \mathrm{FR} = \dfrac{N_\mathrm{frag} - 
		1}{N^\mathrm{max}_\mathrm{frag} - 1} \;\;,
\end{equation} 
where the numerator describes the number of fragments excluding the
original agglomerate and the denominator corresponds to the maximum
attainable number of fragments, again excluding the original
agglomerate, assuming complete deagglomeration.

\begin{figure}[!t]
	\centering
	\subfloat[]{%
		\includegraphics[width=0.49\textwidth]{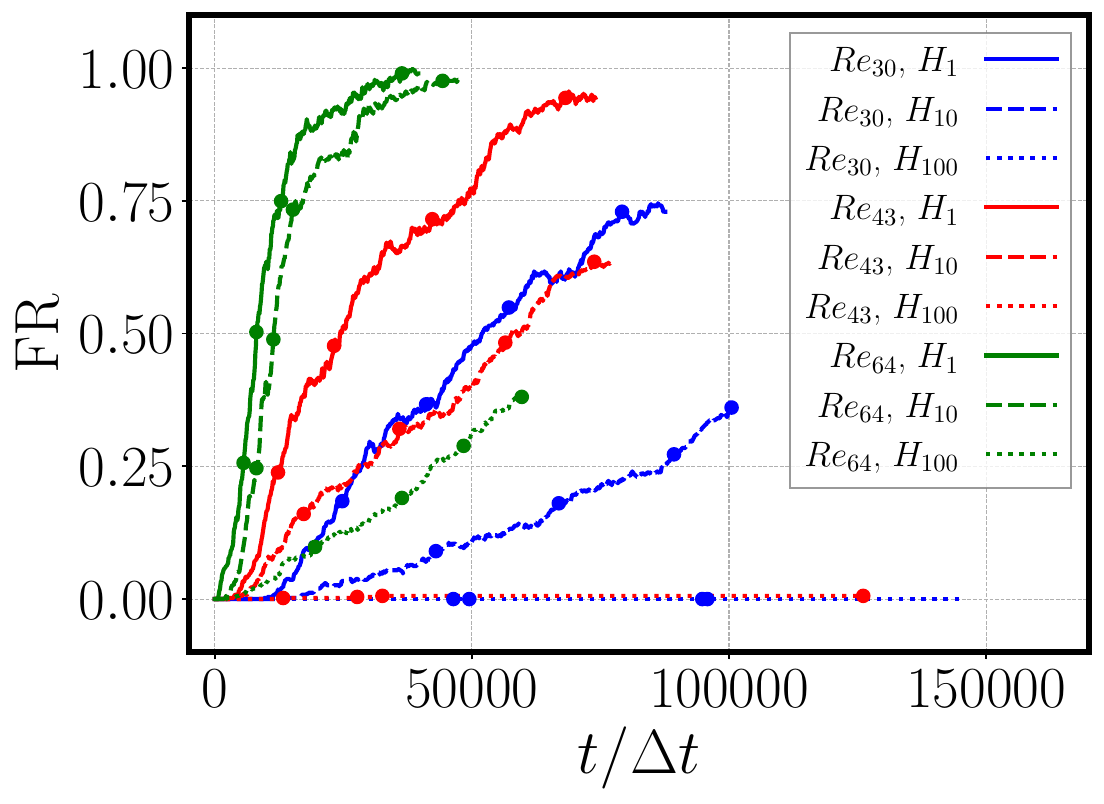}
		\label{subfig:fr_t}
	}	\hfill
	\subfloat[]{%
		\includegraphics[width=0.49\textwidth]{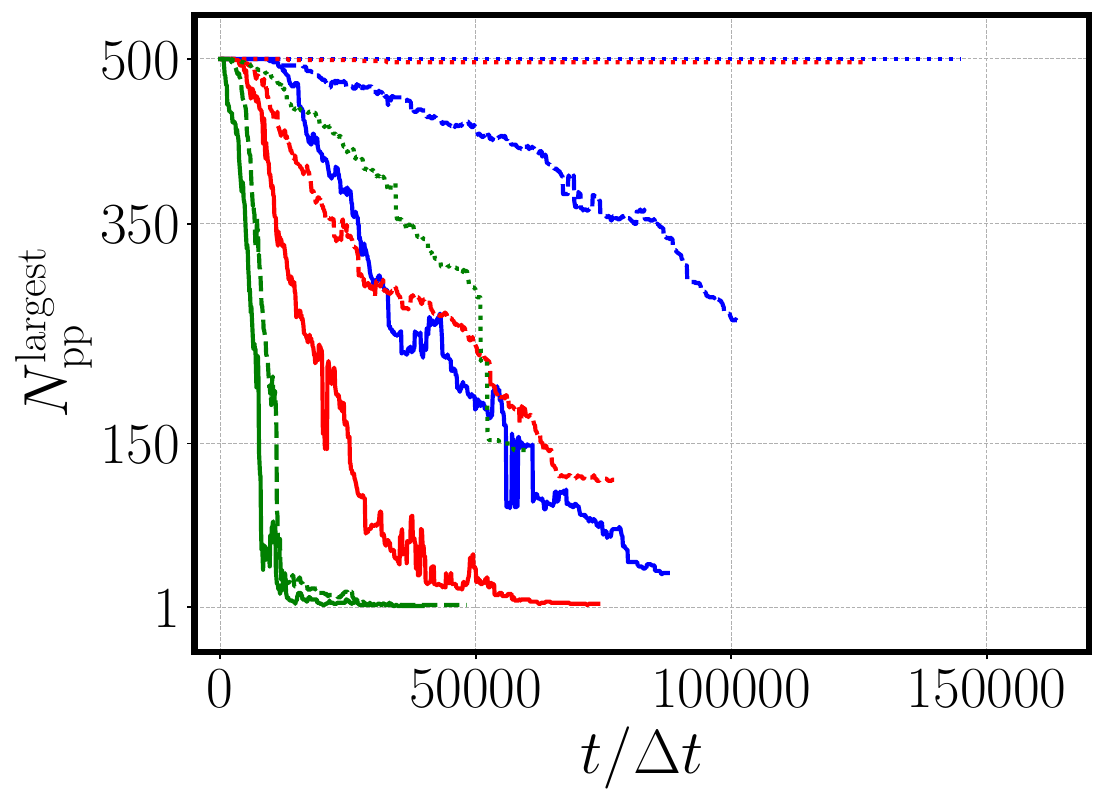}
		\label{subfig:npp_t}
	}\\
	\subfloat[]{%
		\includegraphics[width=0.49\textwidth]{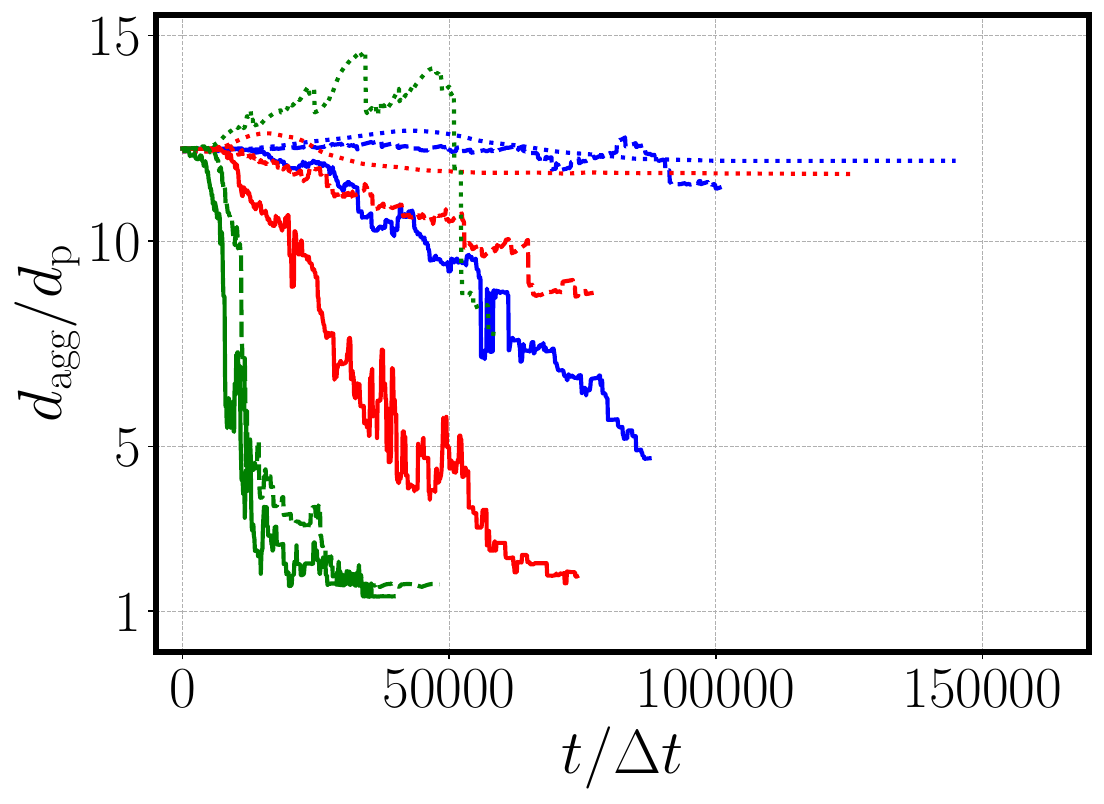}
		\label{subfig:dconv_t}
	} \\ 
	\caption{Time evolution of key fragmentation metrics for different cases: 
		(a) fragmentation ratio (FR), indicating the extent of breakage; (b) 
		number 
		of primary particles contained in the largest remaining fragment; and 
		(c) 
		equivalent diameter of the largest fragment.}
	\label{fig:deagg_dyn}
\end{figure}

Figure~\ref{subfig:fr_t} presents the temporal evolution of the
fragmentation ratio for the different cases. It is evident that, for a
fixed Hamaker constant, increasing the flow Reynolds number leads to
faster and more extensive breakage. Likewise, a reduction in the
Hamaker constant at a constant $Re_\lambda$ results in increased
fragmentation. These trends are physically consistent and align with
expectations: Elevated Reynolds numbers induce stronger fluid
stresses, while reduced Hamaker constants correspond to weaker
cohesive forces between the particles. Figure~\ref{subfig:fr_t} also
demonstrates that the chosen parameter ranges of $Re_\lambda$ and $H$
gurantee the full range of breakup scenarios between $\mathrm{FR}
\approx 0$ and $\mathrm{FR} \approx 1$. The apparent fluctuations in
FR in some cases are attributed to the re-agglomeration between the
eroded particles and fragments.  It is worth mentioning that for each
curve, four representative time instants are highlighted by
markers. These points are approximately evenly spaced along the
respective curves and will serve as reference time instants for the
size distribution analysis presented in the following section.

%
%
Besides the fragmentation ratio, the shape evolution of the largest
fragment, which represents the main remaining part of the original
agglomerate, is of particular interest, as it offers insight into the
underlying mechanisms of the deagglomeration process.
Figures~\ref{subfig:npp_t}~and~\ref{subfig:dconv_t} show the time
evolution of the number of primary particles in the largest fragment,
$N^\mathrm{largest}_\mathrm{pp}$, and the ratio of the agglomerate
diameter to the primary particle diameter, $d_\mathrm{agg} /
d_\mathrm{p}$.  The results reveal that both
$N^\mathrm{largest}_\mathrm{pp}$ and $d_\mathrm{agg} / d_\mathrm{p}$
drop more rapidly with increasing Reynolds number and decreasing
Hamaker constant, indicating a more abrupt disintegration of the
agglomerate as expected based on Fig.~\ref{subfig:fr_t}.

Interestingly, for cases with strong cohesion (specifically,
$H_{100}$), the agglomerate initially stretches under the action of
the turbulent flow.  This is reflected in a temporary increase in the
diameter ratio above its initial value, despite ongoing particle
erosion, as seen in the dotted curves for the [$Re_{64}, H_{100}$]
case in Figs.~\ref{subfig:npp_t}~and~\ref{subfig:dconv_t}.  However,
the slight decrease in the agglomerate diameter ratio in cases with
little or no significant breakup (specifically, $Re_{30}$ and
$Re_{43}$ for $H_{100}$) indicate that the interaction with the
turbulent flow can also compress the agglomerate structure over time.

Moreover, Fig.~\ref{subfig:npp_t} demonstrates that in cases where
breakup does occur, the reduction in the number of primary particles
within the largest fragment proceeds gradually. Even in the strongest
breakage scenarios, i.e., $Re_{64}$ combined with $H_1$ and $H_{10}$,
the decrease in size occurs in discrete steps, albeit at a faster
rate. This behavior suggests that under the present flow and cohesion
conditions, abrupt deagglomeration events, such as instantaneous full
breakup or large-scale fragmentation into a few equally sized pieces,
can be ruled out. A more detailed analysis of the resulting fragment
size distributions is presented in the next section.

\subsubsection{Breakup mode and fragment size distribution}

\begin{figure}[!t]
	\centering
	\subfloat[{[$Re_{30}, H_{1}$]}]{%
		\includegraphics[width=0.49\textwidth]{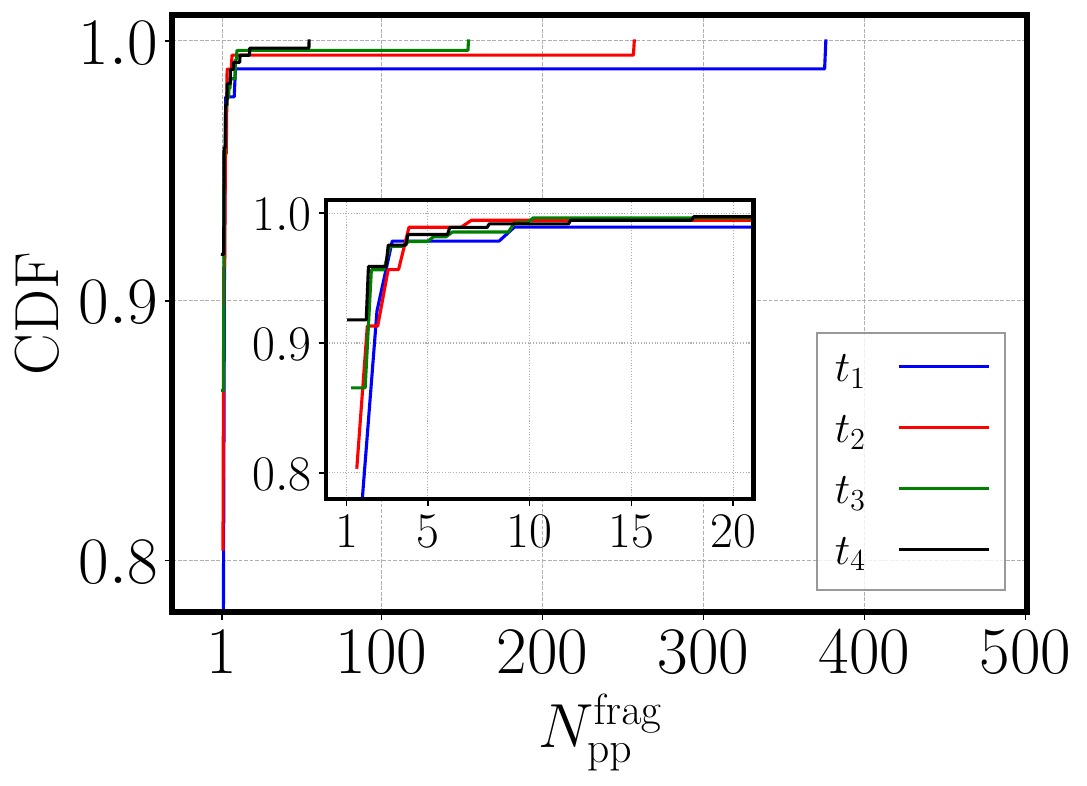}
		\label{subfig:cdf_30H1}
	}
	\hfill
	\subfloat[{[$Re_{30}, H_{10}$]}]{%
		\includegraphics[width=0.49\textwidth]{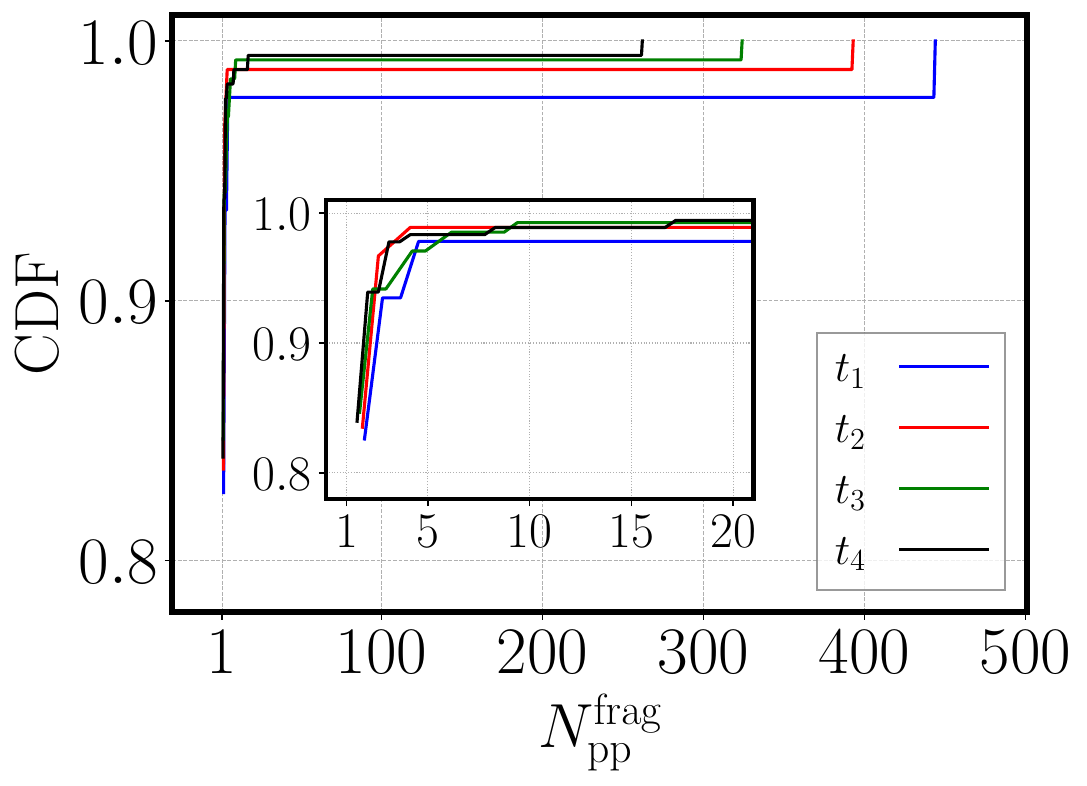}
		\label{subfig:cdf_30H10}
	}
	\\
	\subfloat[{[$Re_{43}, H_{1}$]}]{%
		\includegraphics[width=0.49\textwidth]{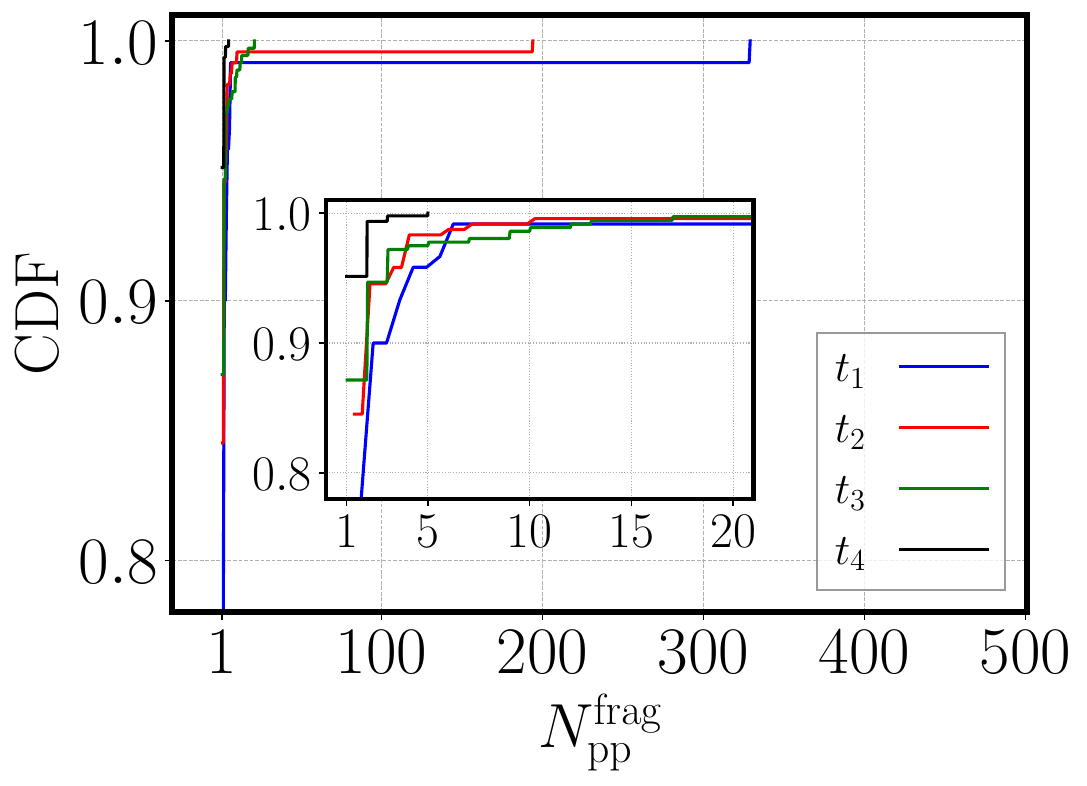}
		\label{subfig:cdf_43H1}
	}
	\hfill
	\subfloat[{[$Re_{43}, H_{10}$]}]{%
		\includegraphics[width=0.49\textwidth]{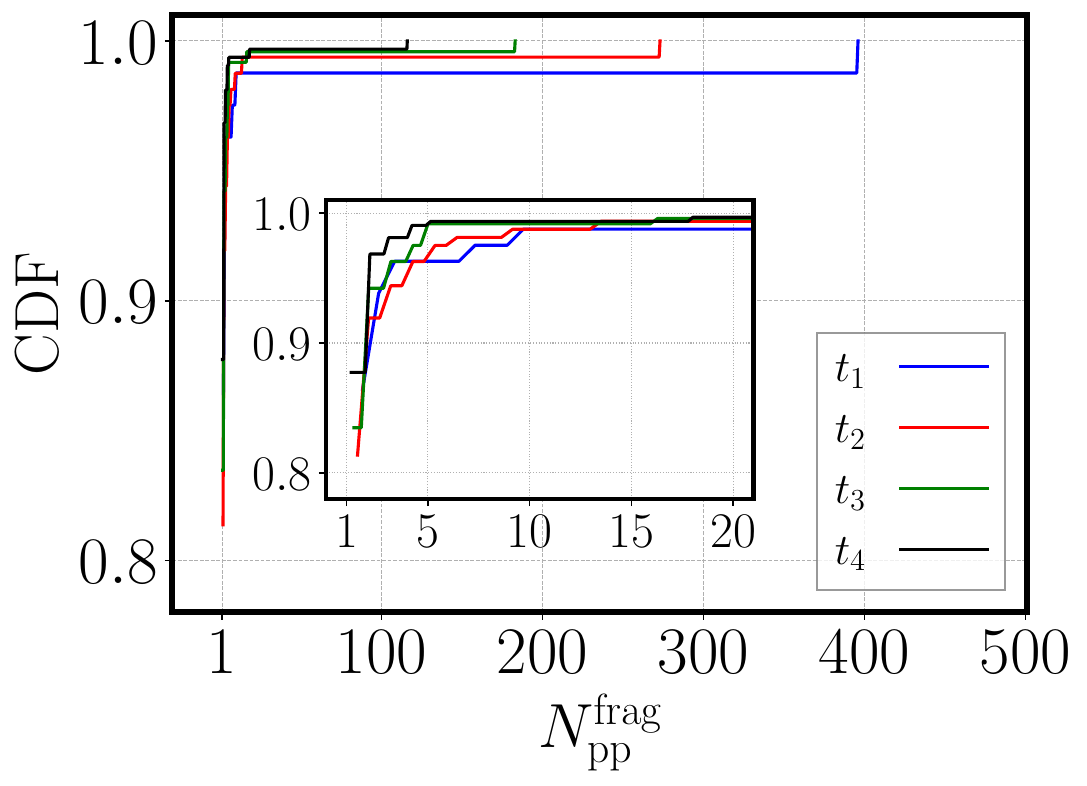}
		\label{subfig:cdf_43H10}
	}
	
	\caption{Cumulative distribution function (CDF) of the number of particles 
		per fragment, $N^\mathrm{frag}_\mathrm{pp}$ for the cases corresponding 
		to 
		$Re_{30}$ and $Re_{43}$ excluding cases for $H_{100}$, shown at four 
		distinct time 
		instants ($t_1$ to $t_4$) during the simulation.}
	\label{fig:CDF_Npp_re3043}
\end{figure}

\begin{figure}[!t]
	\centering
	\subfloat[{[$Re_{64}, H_{1}$]}]{%
		\includegraphics[width=0.49\textwidth]{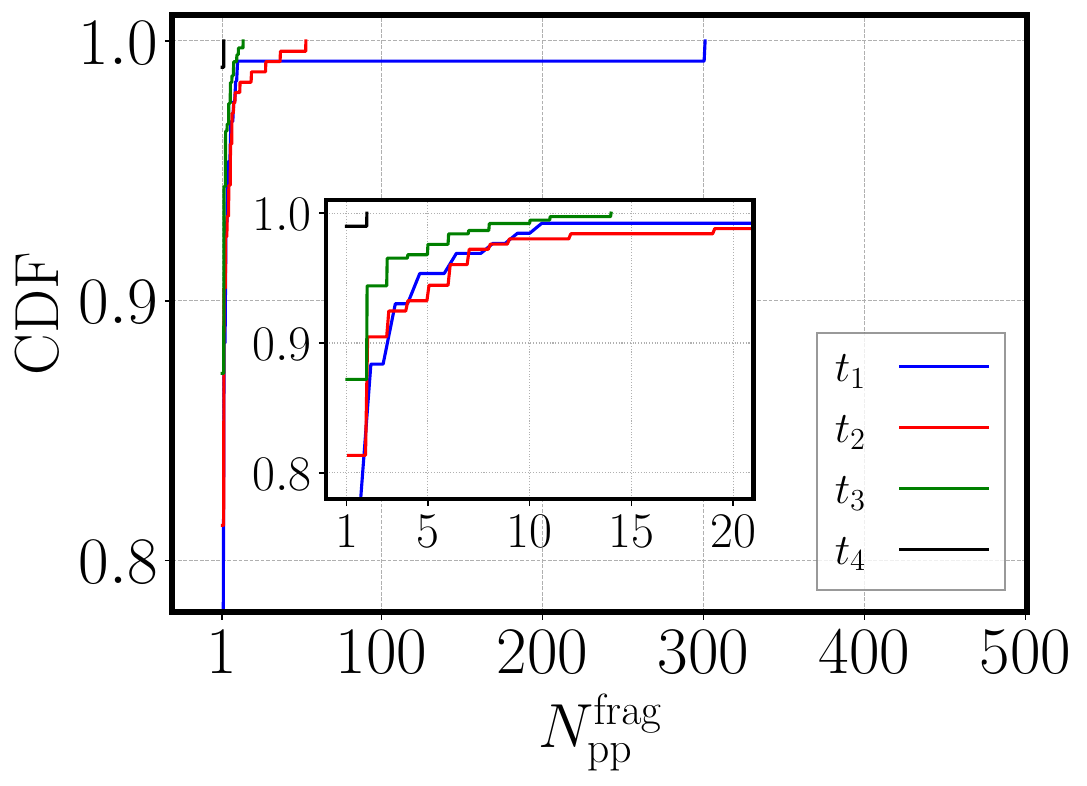}
		\label{subfig:cdf_64H1}
	} 
	\hfill
	\subfloat[{[$Re_{64}, H_{10}$]}]{%
		\includegraphics[width=0.49\textwidth]{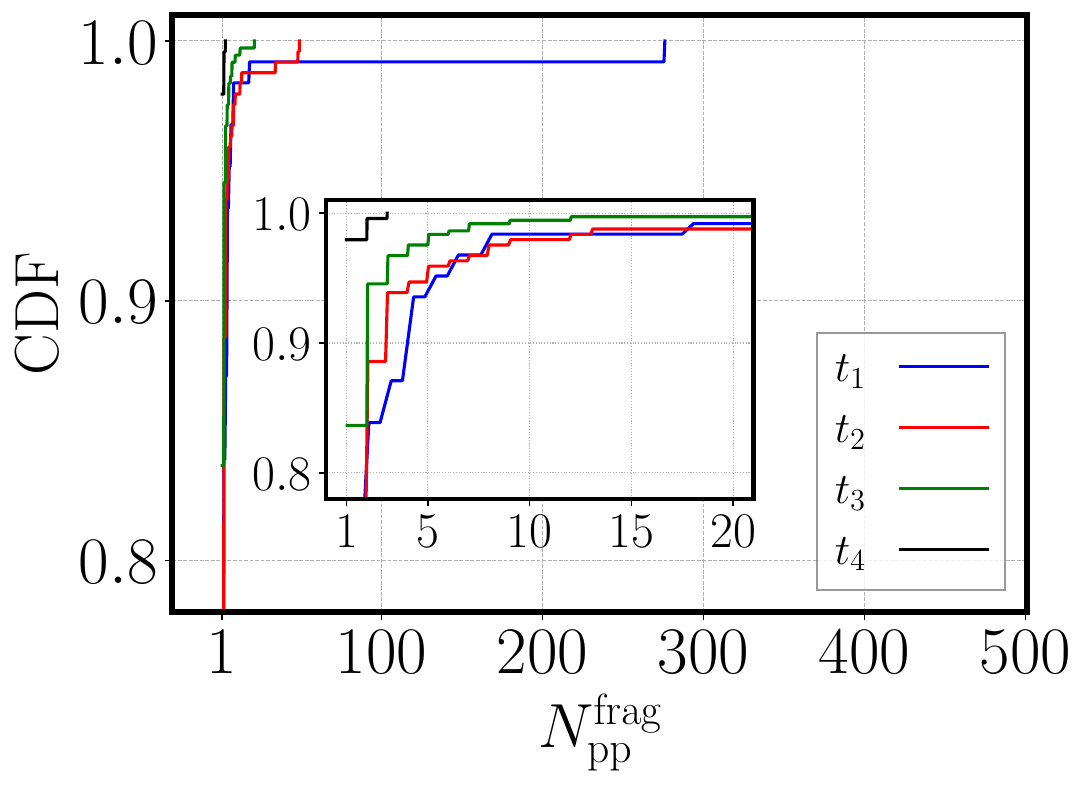}
		\label{subfig:cdf_64H10}
	}
	\\
	\subfloat[{[$Re_{64}, H_{100}$]}]{%
		\includegraphics[width=0.49\textwidth]{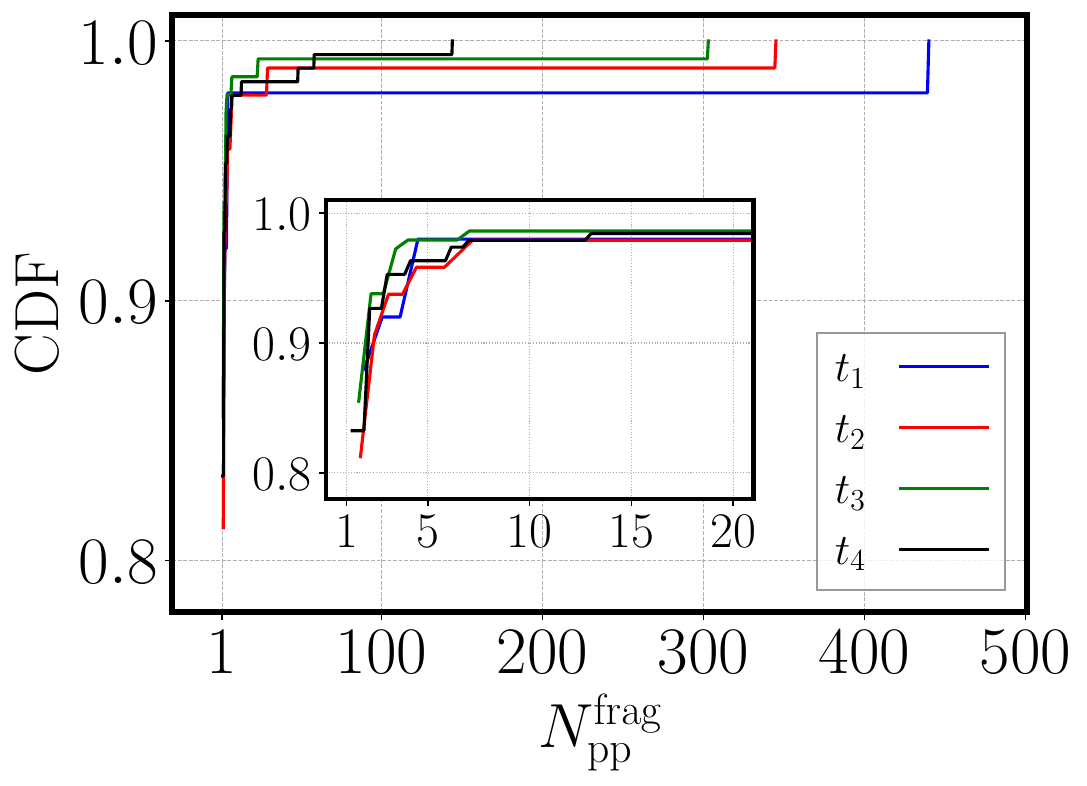}
		\label{subfig:cdf_64H100}
	}
	\caption{Cumulative distribution function (CDF) of the number of particles 
		per fragment, $N^\mathrm{frag}_\mathrm{pp}$, for the highest Reynolds 
		number $Re_{64}$ at all three Hamaker constants shown at four distinct 
		time 
		instants ($t_1$ to $t_4$) during the simulation.}
	\label{fig:CDF_Npp_re64}
\end{figure}

The fragment size distribution is a key parameter for modeling the
breakage process, as it for example directly influences the
formulation of breakup kernels in coarse-grained simulations.
Figure~\ref{fig:CDF_Npp_re3043} presents the cumulative distribution
functions (CDFs) of the number of primary particles per fragment,
$N^\mathrm{frag}_\mathrm{pp}$, at four representative time instants
($t_1$ to $t_4$), for the cases corresponding to $Re_{30}$ and
$Re_{43}$.  Cases with the strongest Hamaker constant ($H_{100}$) are
excluded, as they exhibit negligible or no significant fragmentation.
Similarly, Fig.~\ref{fig:CDF_Npp_re64} displays the CDFs for the
$Re_{64}$ cases but now for all three Hamaker constants. Note that the
chosen four time instants of the analysis corresponds to those marked
on the fragmentation ratio curves in Fig.~\ref{subfig:fr_t}.

The CDFs generally indicate the presence of three distinct fragment
categories.  First, the largest fragment, which typically corresponds
to the original agglomerate or its main remnant. Second, medium-sized
fragments consisting of approximately 2 to 20 primary
particles. Third, individual primary particles, representing fully
separated units. As time progresses, the largest fragment gradually
diminishes in size, while medium-sized fragments tend to either break
down into smaller clusters or fully disintegrate into single
particles.  This trend reflects the continuous erosion and
restructuring of the agglomerate under the influence of homogeneous
isotropic turbulence. Except for the speed and extent of the
deagglomeration, no significant differences in the temporal
development of the fragment size distribution are found between the
seven cases where significant breakage occurs.

\begin{figure}[!t]
	\centering
	\subfloat[{[$Re_{30}, H_{1}$]}, $t^*=$248]{%
		\includegraphics[width=0.24\textwidth]{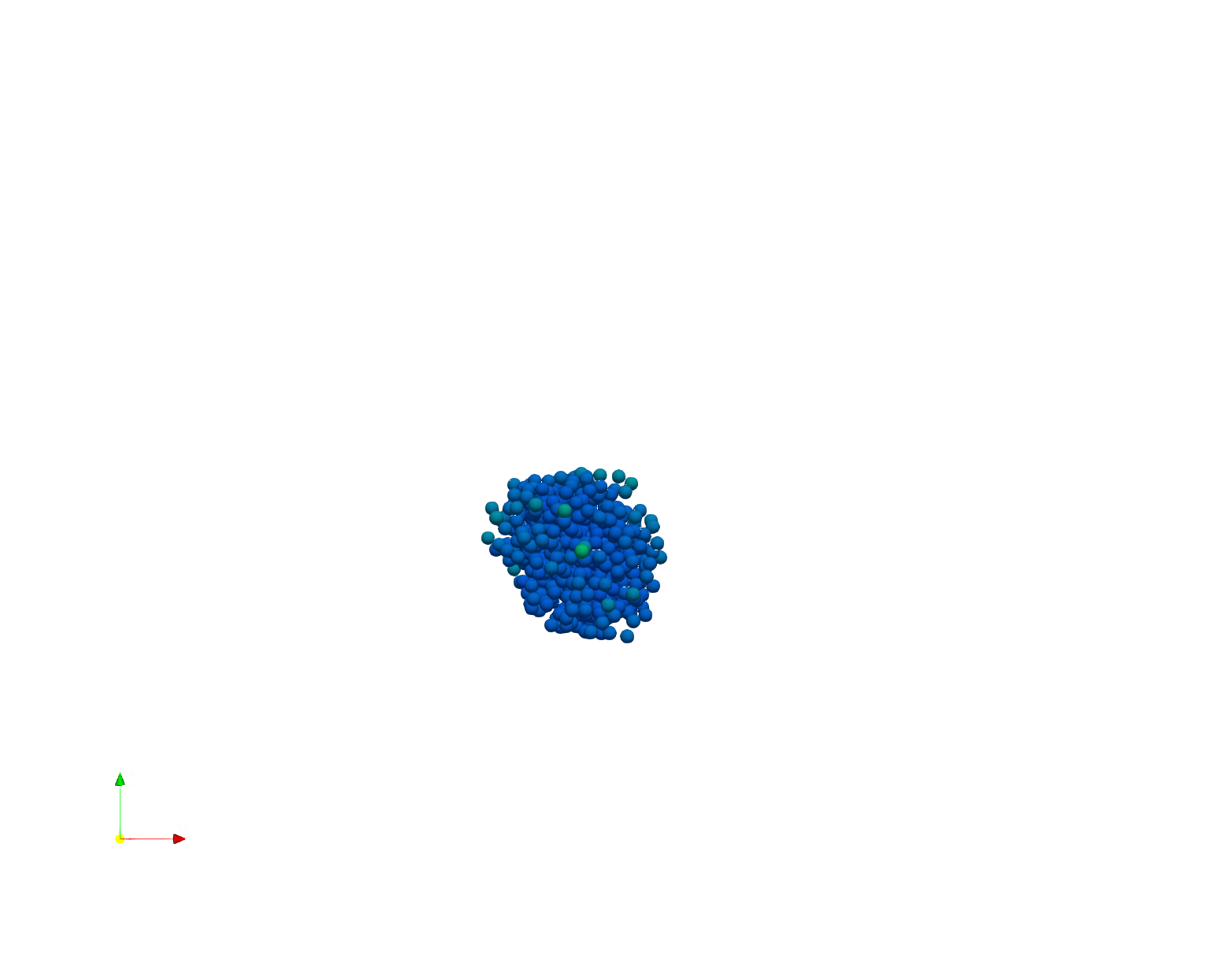}
		\label{subfig:re30_h1_t1}
	}
	\hfill
	\subfloat[$t^*=$411]{%
		\includegraphics[width=0.24\textwidth]{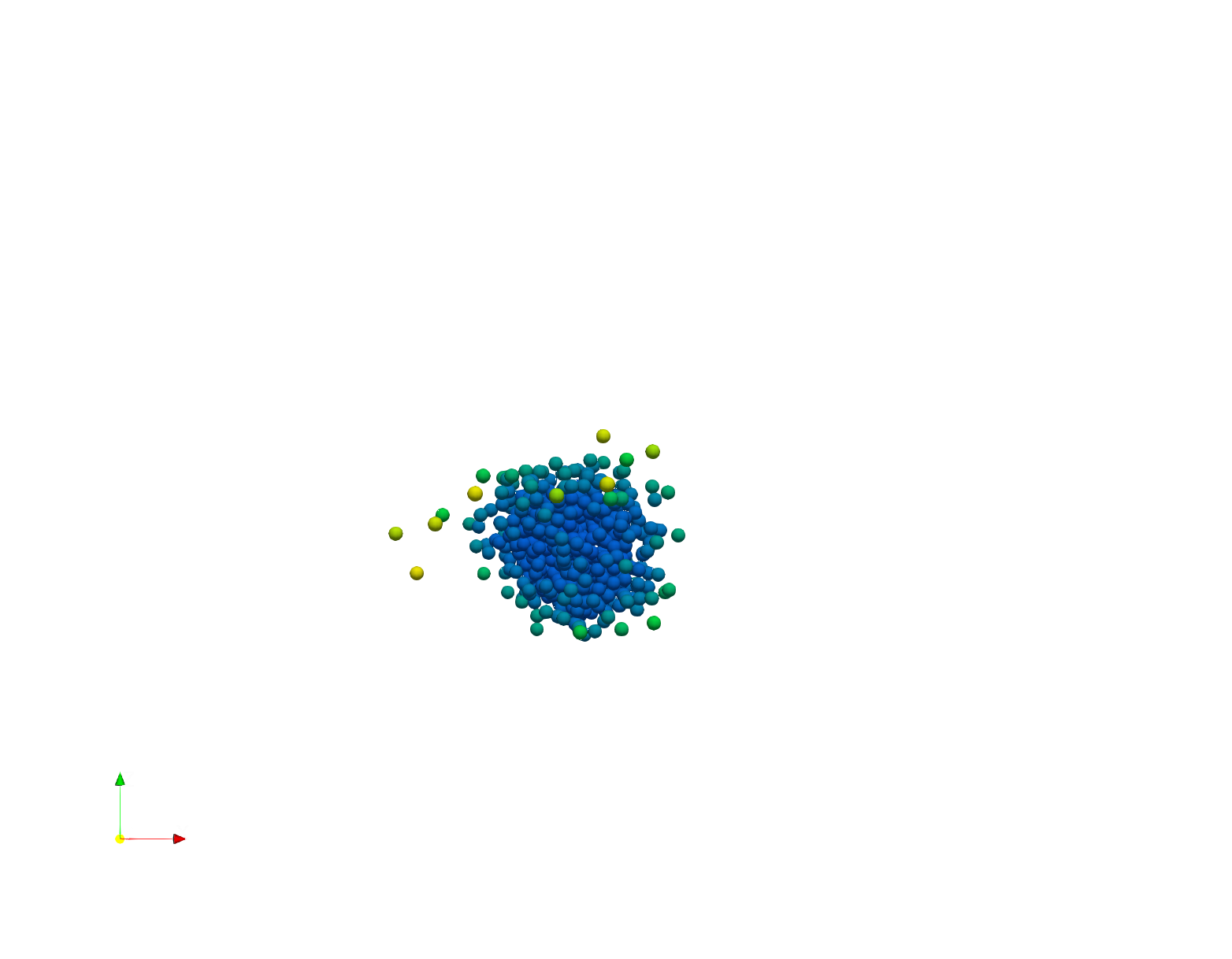}
		\label{subfig:re30_h1_t2}
	}
	\hfill
	\subfloat[$t^*=$572]{%
		\includegraphics[width=0.24\textwidth]{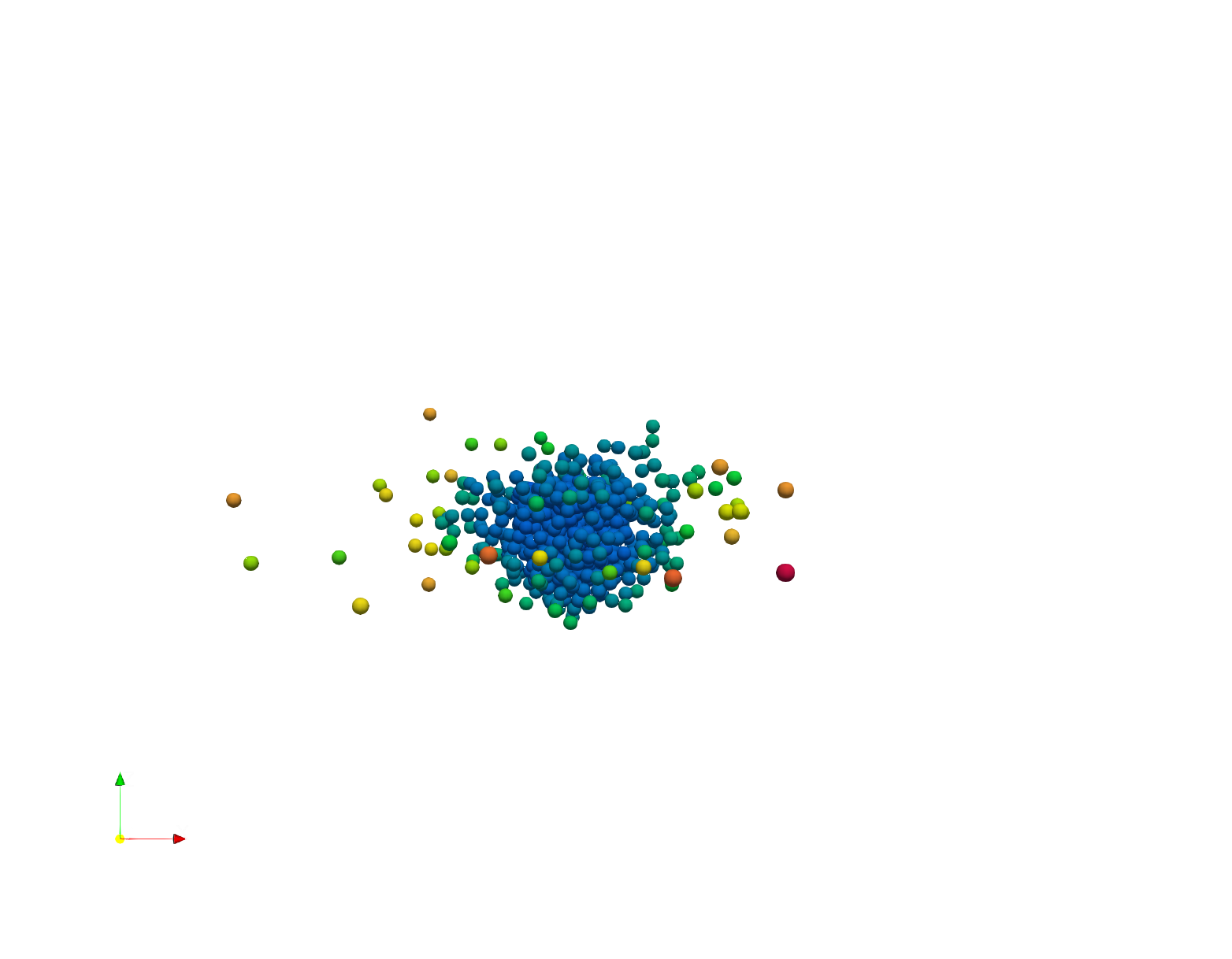}
		\label{subfig:re30_h1_t3}
	}
	\hfill
	\subfloat[$t^*=$792]{%
		\includegraphics[width=0.24\textwidth]{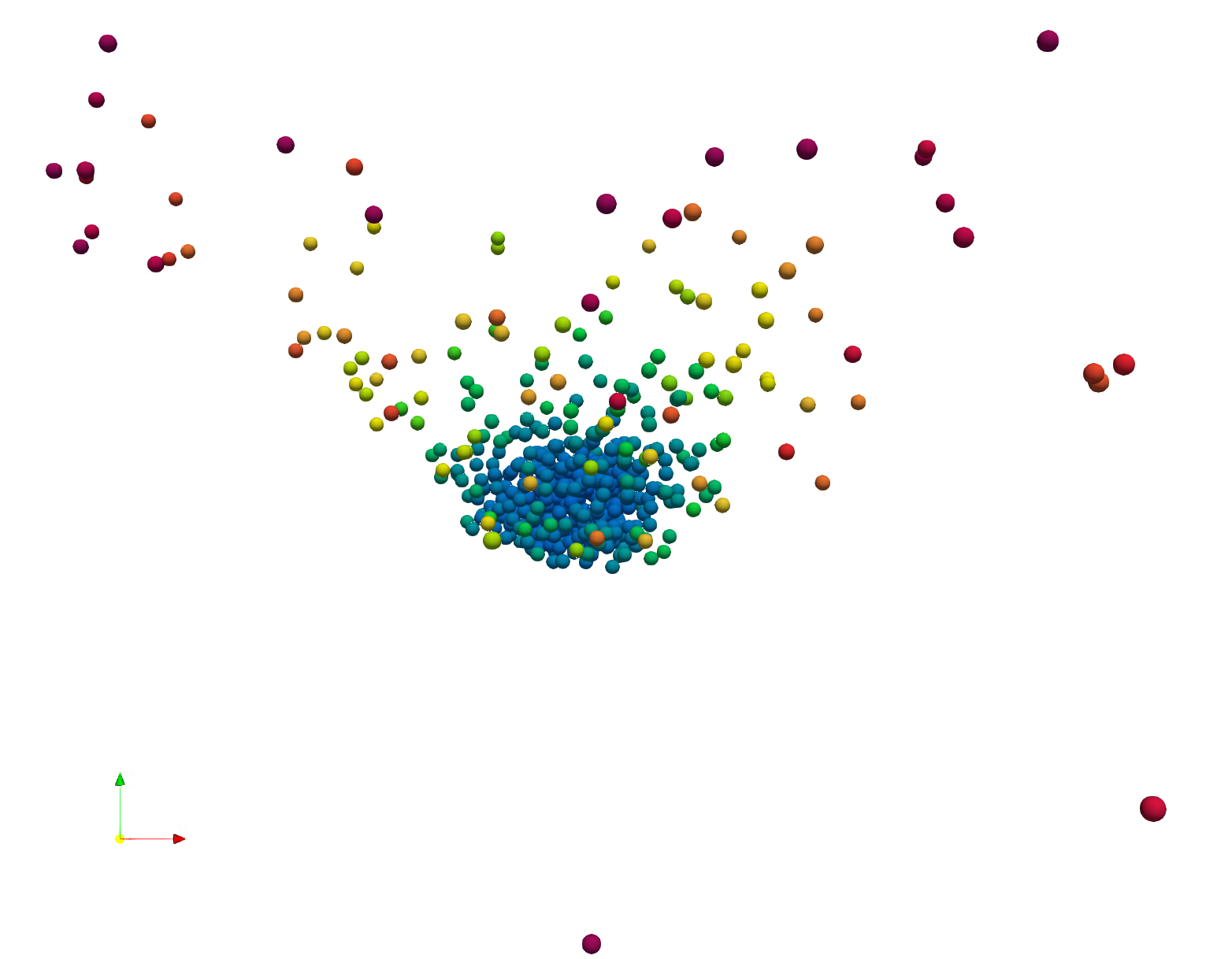}
		\label{subfig:re30_h1_t4}
	}	
	\\\vspace*{-0.8cm}
	\subfloat[{[$Re_{30}, H_{100}$]}, $t^*$=545]{%
		\includegraphics[width=0.24\textwidth]{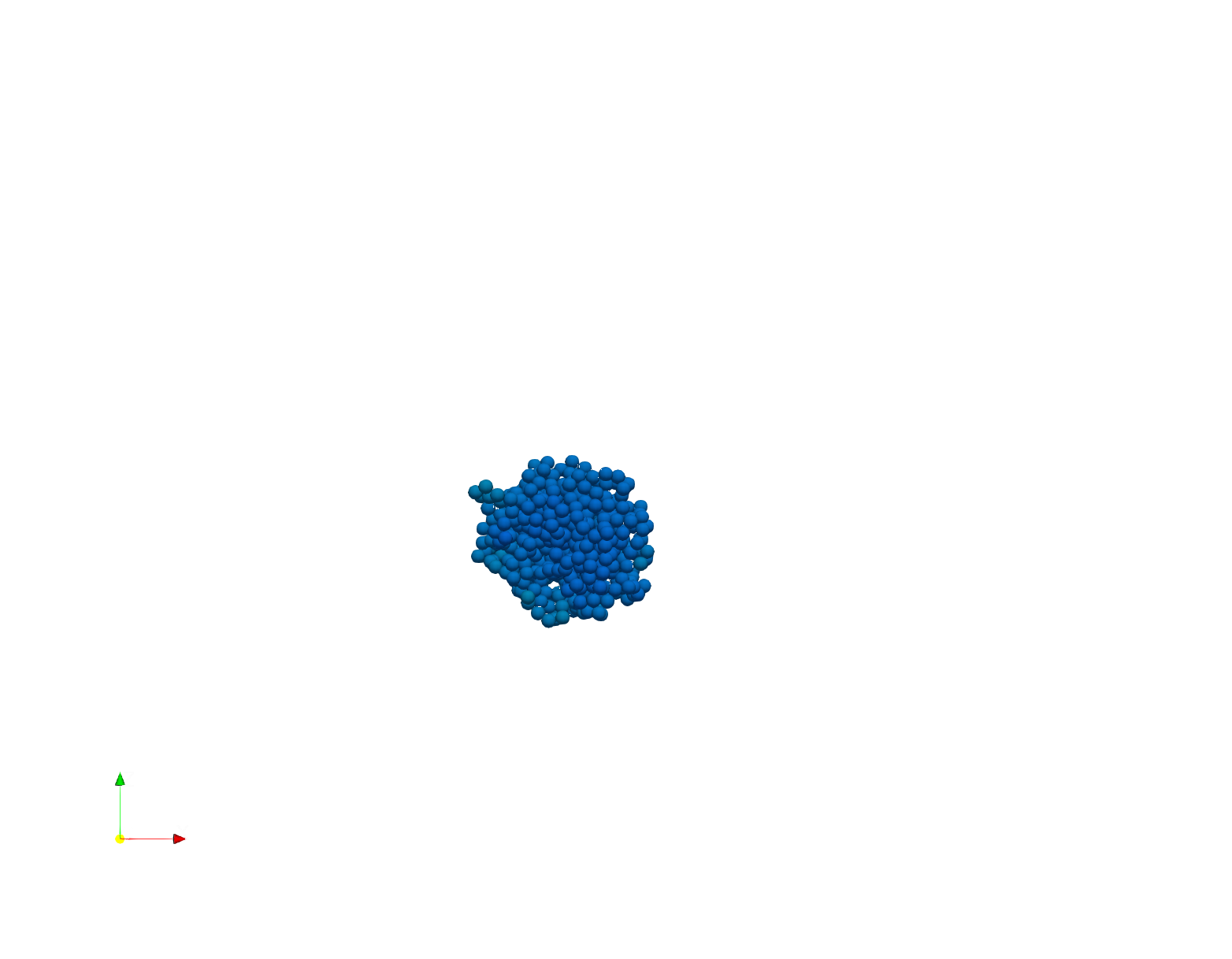}
		\label{subfig:re30_h100_t1}
	}
	\hfill
	\subfloat[$t^*=$803] {%
		\includegraphics[width=0.24\textwidth]{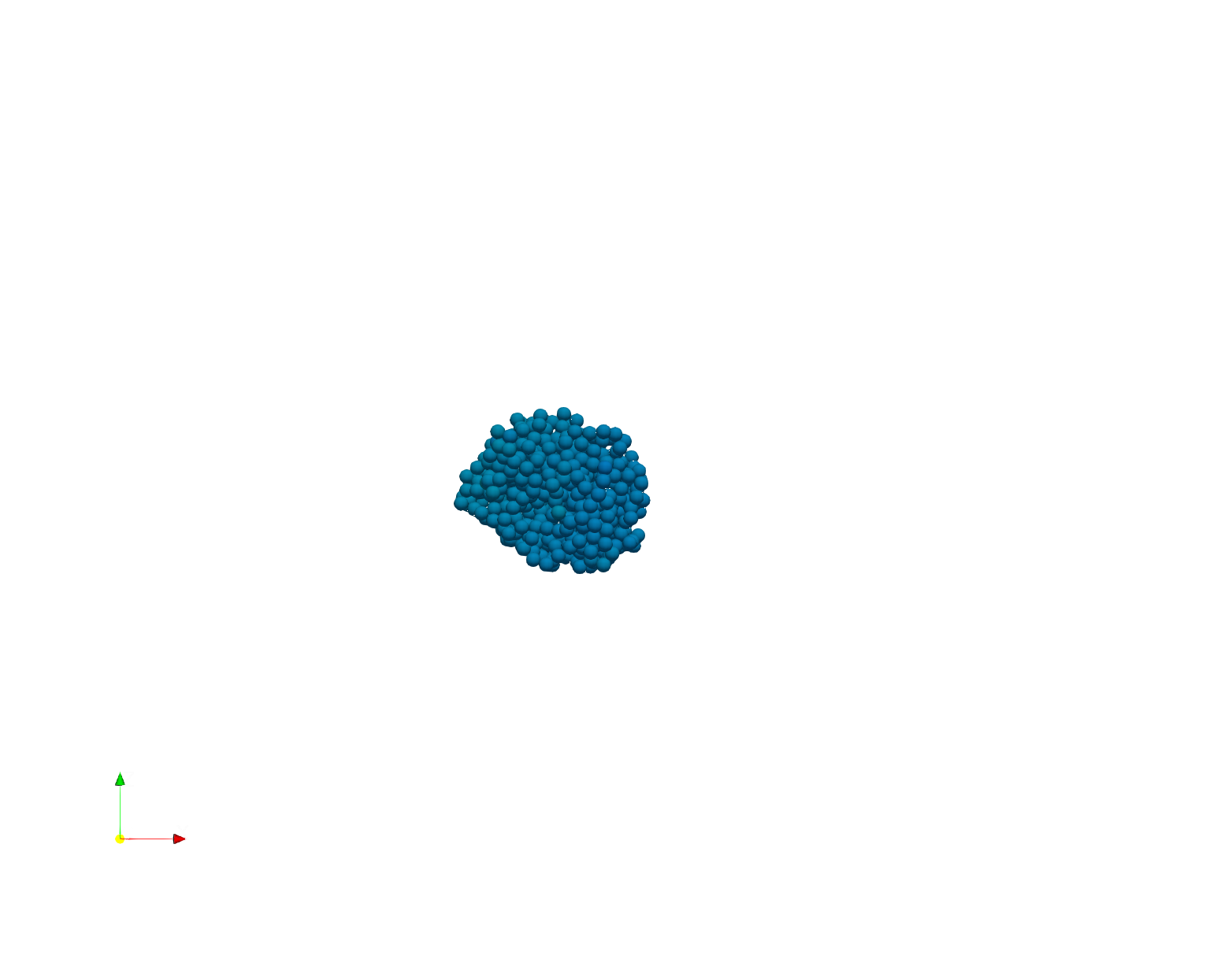}
		\label{subfig:re30_h100_t2}
	}
	\hfill
	\subfloat[$t^*=$1033]{%
		\includegraphics[width=0.24\textwidth]{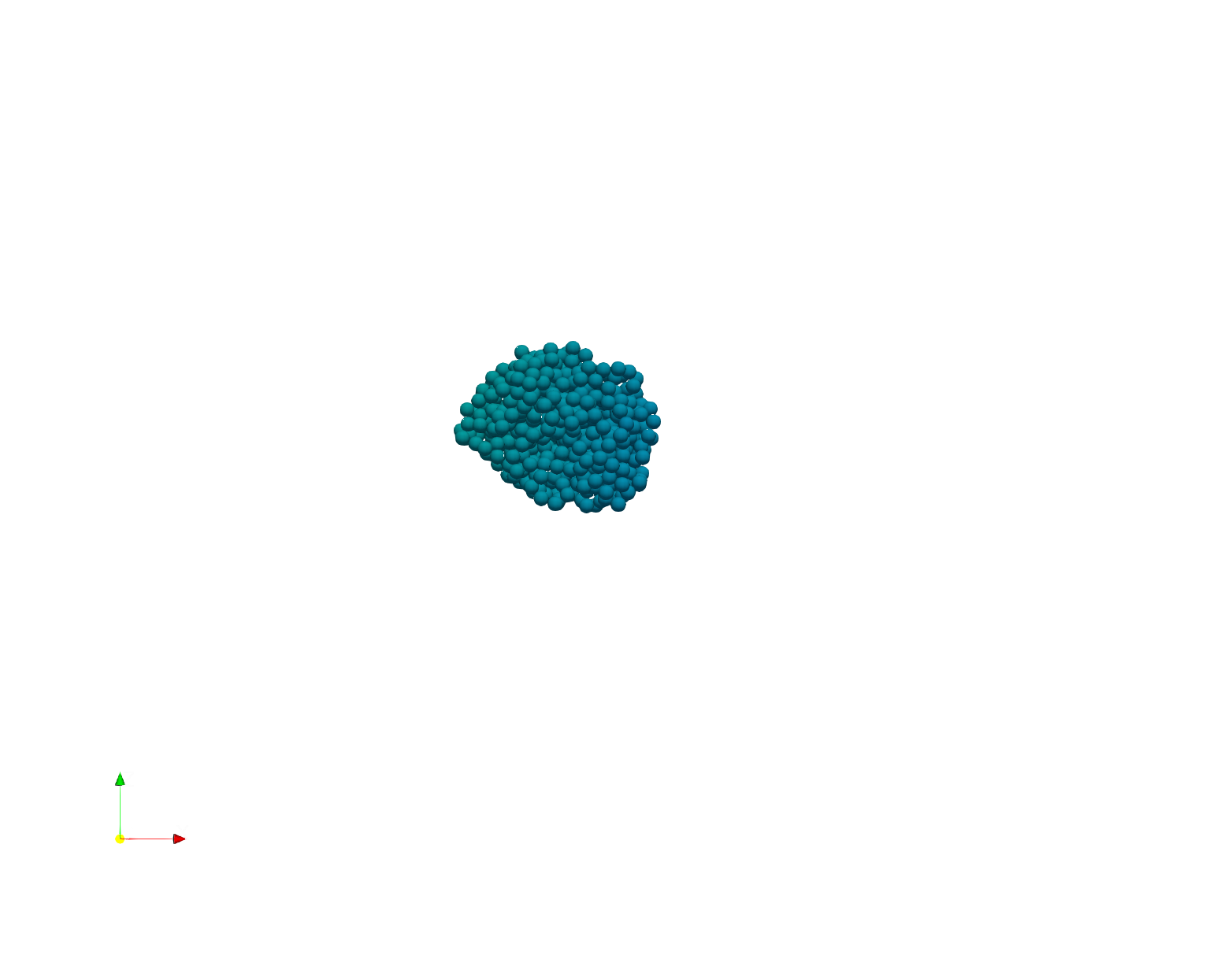}
		\label{subfig:re30_h100_t3}
	}
	\hfill
	\subfloat[$t^*=$1139]{%
		\includegraphics[width=0.24\textwidth]{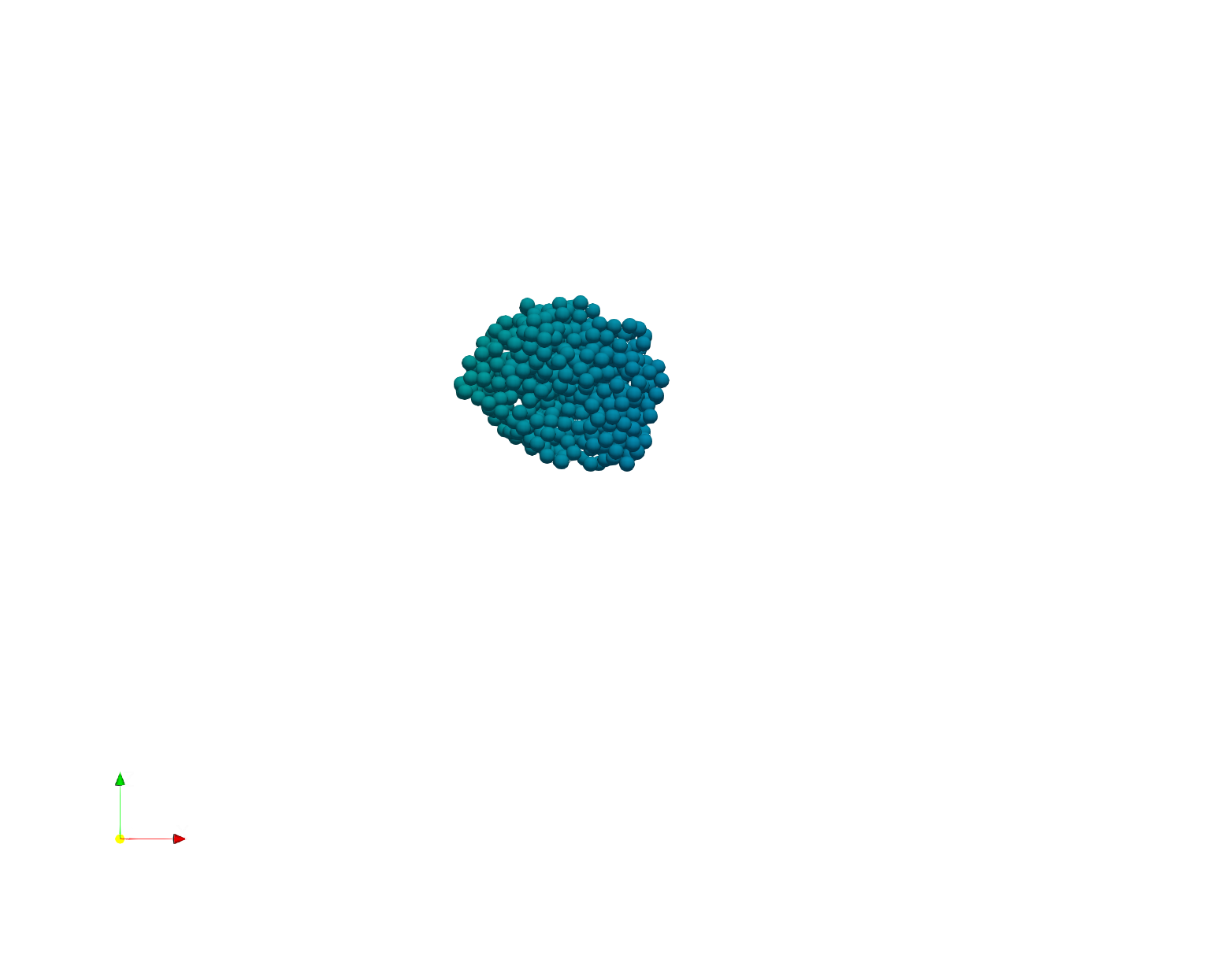}
		\label{subfig:re30_h100_t4}
	}
	\\	\vspace*{-0.8cm}
	\subfloat[{[$Re_{43}, H_{1}$]}, $t^*=$123]{%
		\includegraphics[width=0.24\textwidth]{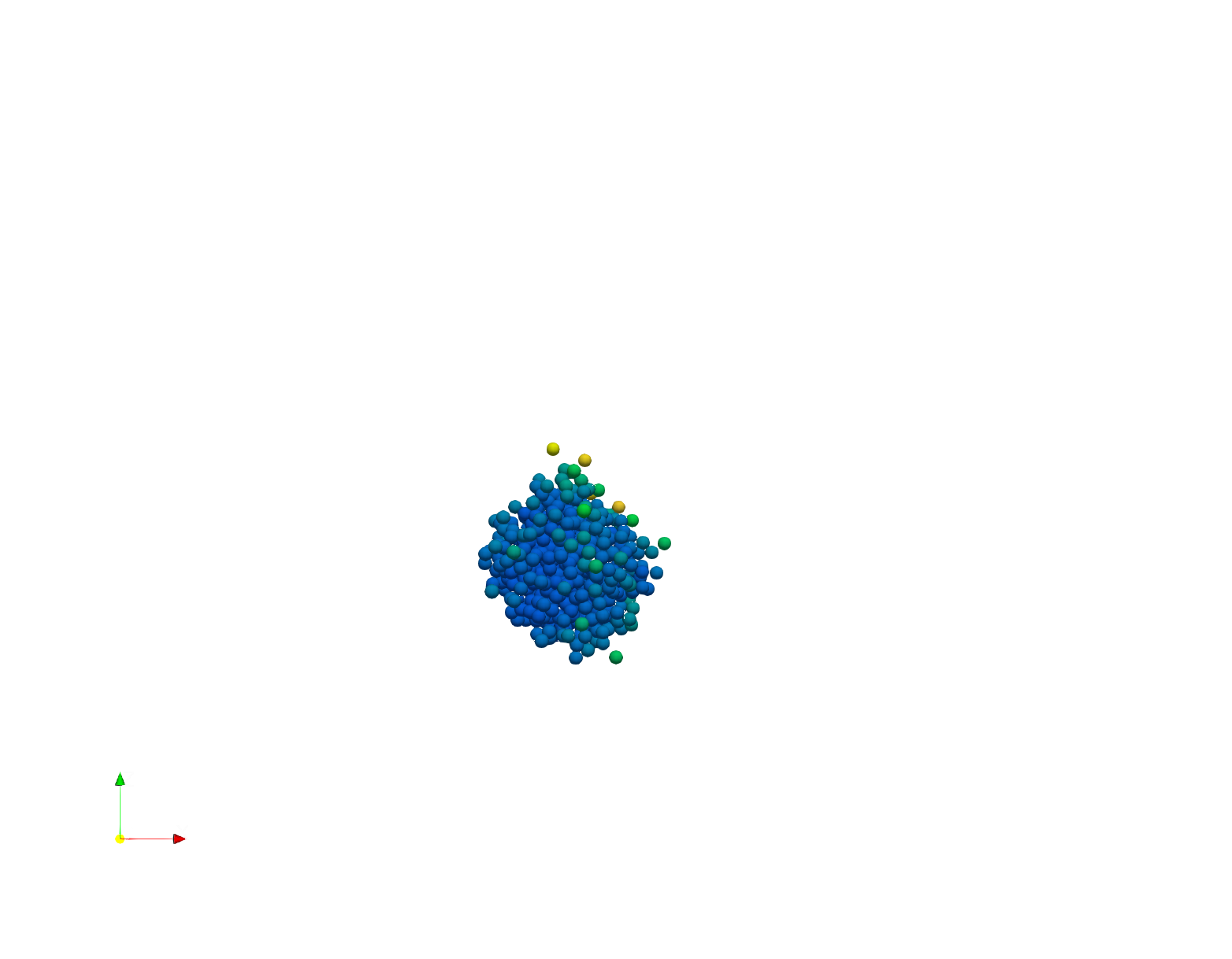}
		\label{subfig:re43_h1_t1}
	}
	\hfill
	\subfloat[$t^*=$232]{%
		\includegraphics[width=0.24\textwidth]{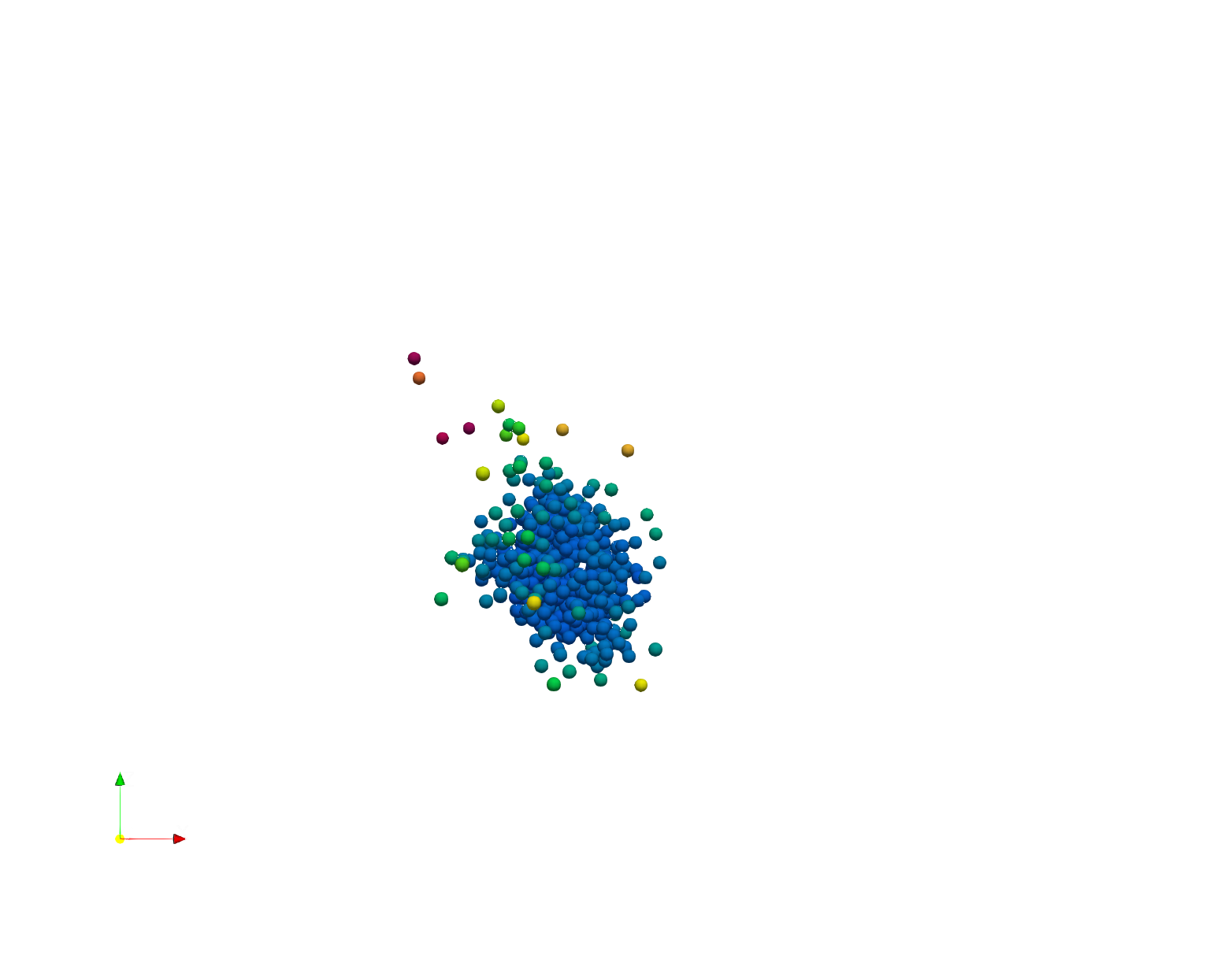}
		\label{subfig:re43_h1_t2}
	}
	\hfill
	\subfloat[$t^*=$423]{%
		\includegraphics[width=0.24\textwidth]{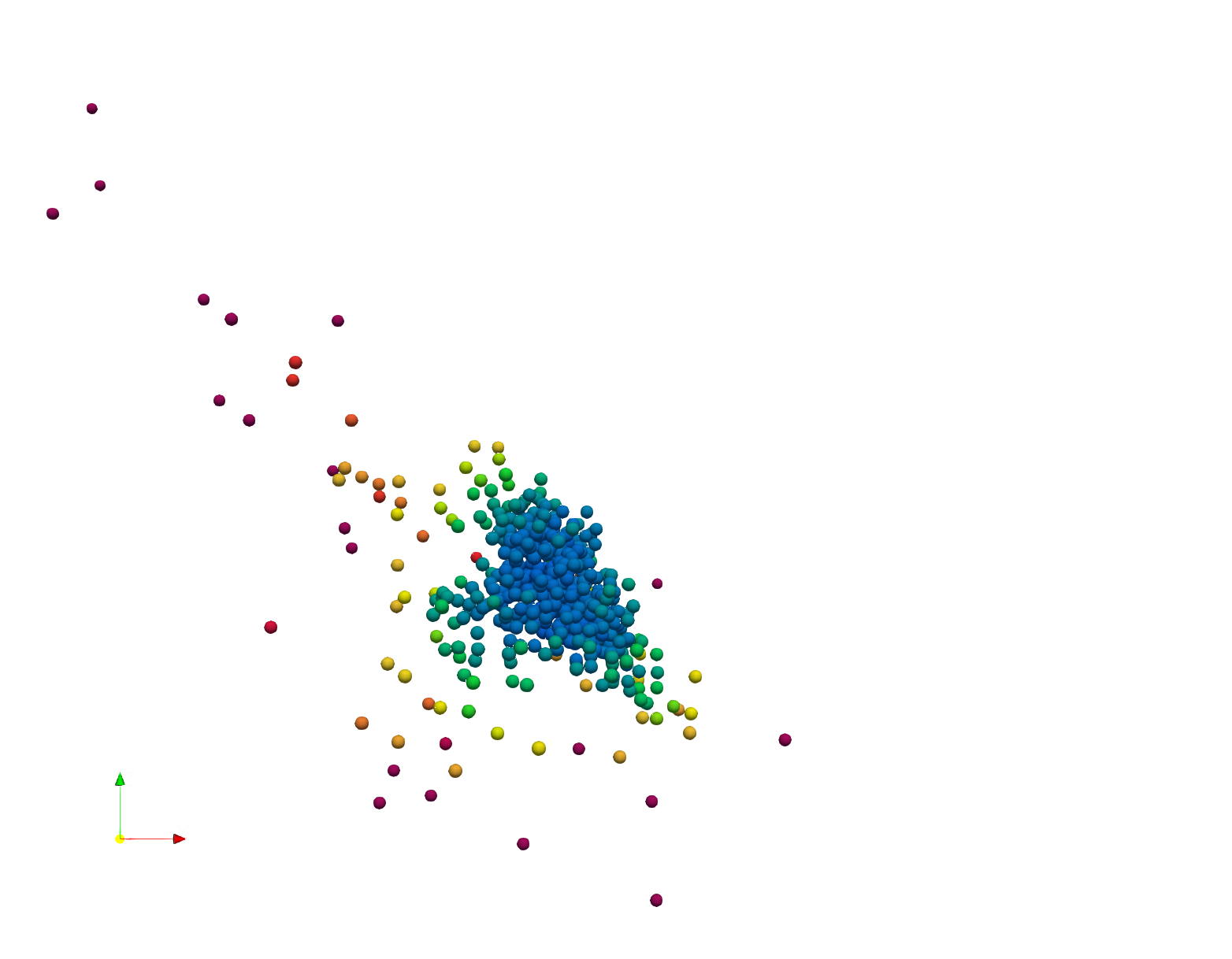}
		\label{subfig:re43_h1_t3}
	}
	\hfill
	\subfloat[$t^*=$682]{%
		\includegraphics[width=0.24\textwidth]{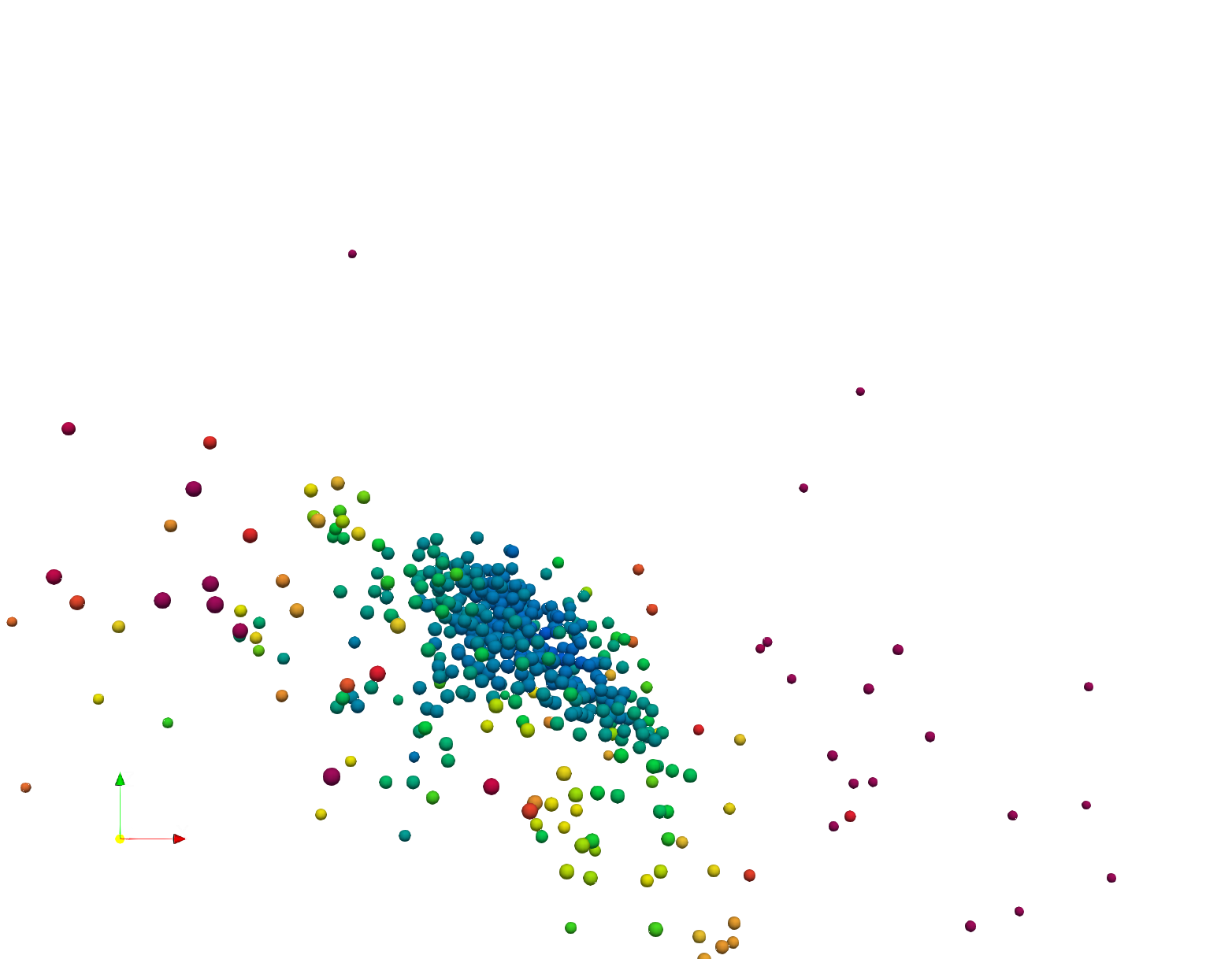}
		\label{subfig:re43_h1_t4}
	}	
	\\\vspace*{-0.8cm}
	\subfloat[{[$Re_{43},H_{100}$]},$t^*$=133]{%
		\includegraphics[width=0.24\textwidth]{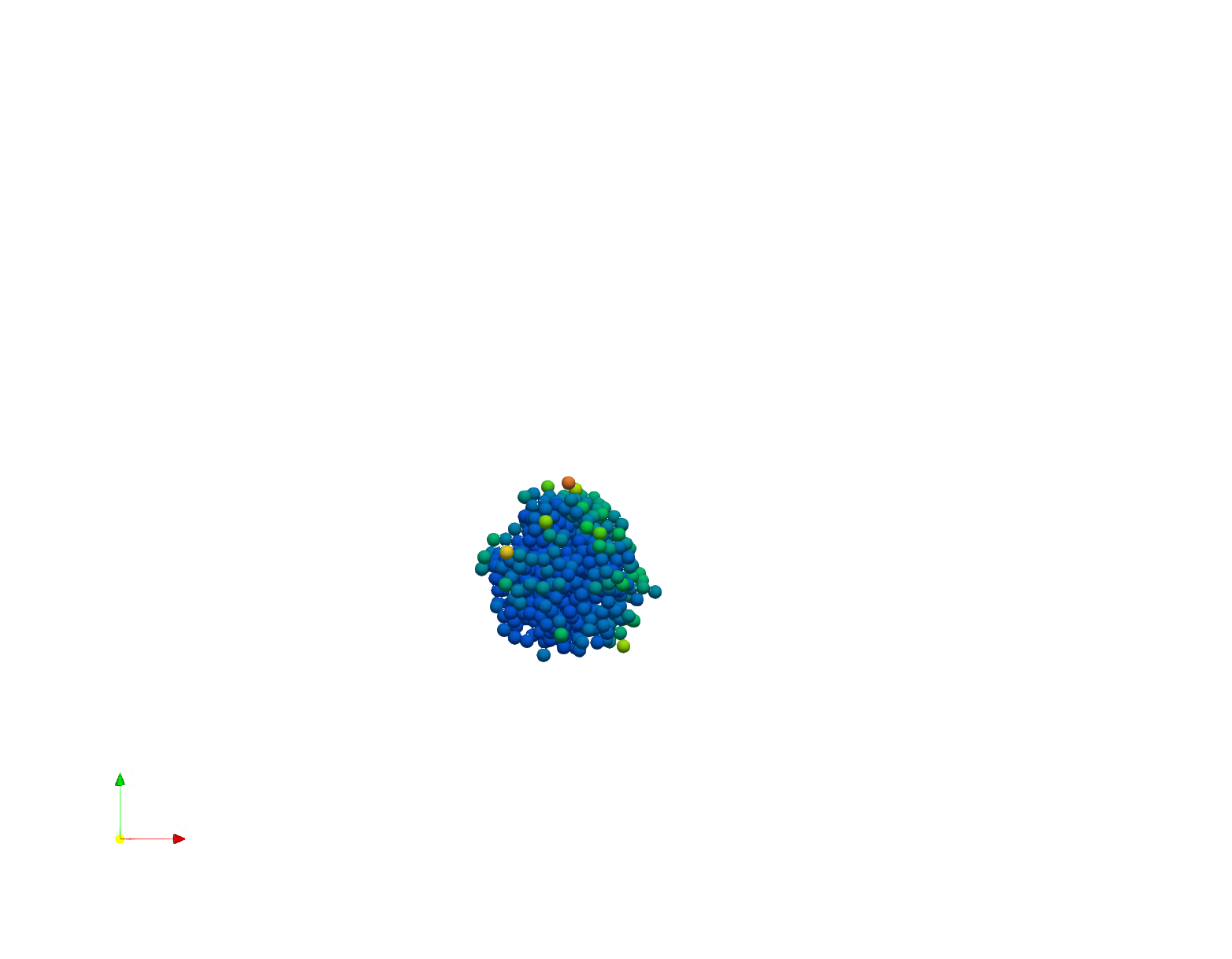}
		\label{subfig:re43_h100_t1}
	}
	\hfill
	\subfloat[$t^*=$277]{%
		\includegraphics[width=0.24\textwidth]{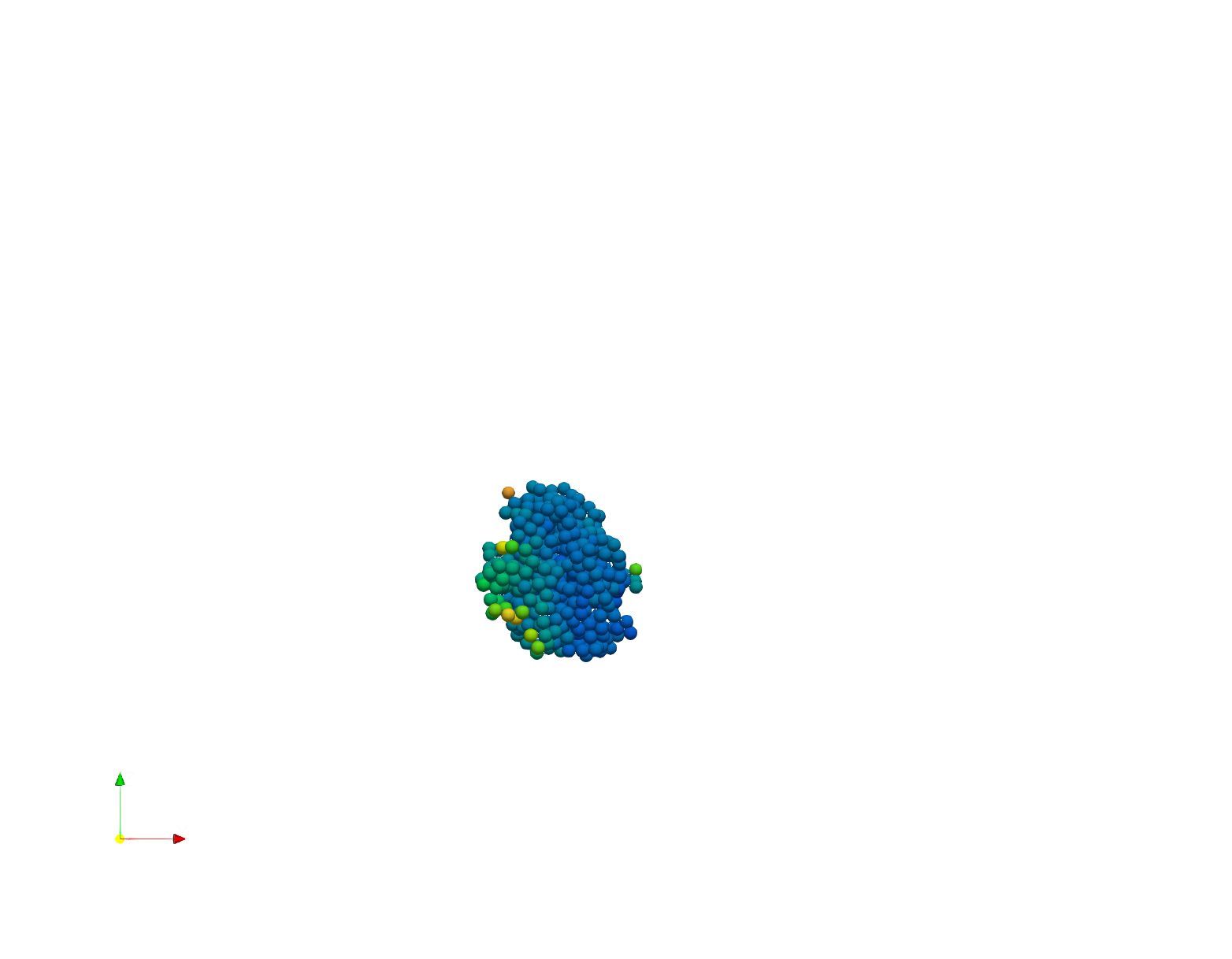}
		\label{subfig:re43_h100_t2}
	}
	\hfill
	\subfloat[$t^*=$326]{%
		\includegraphics[width=0.24\textwidth]{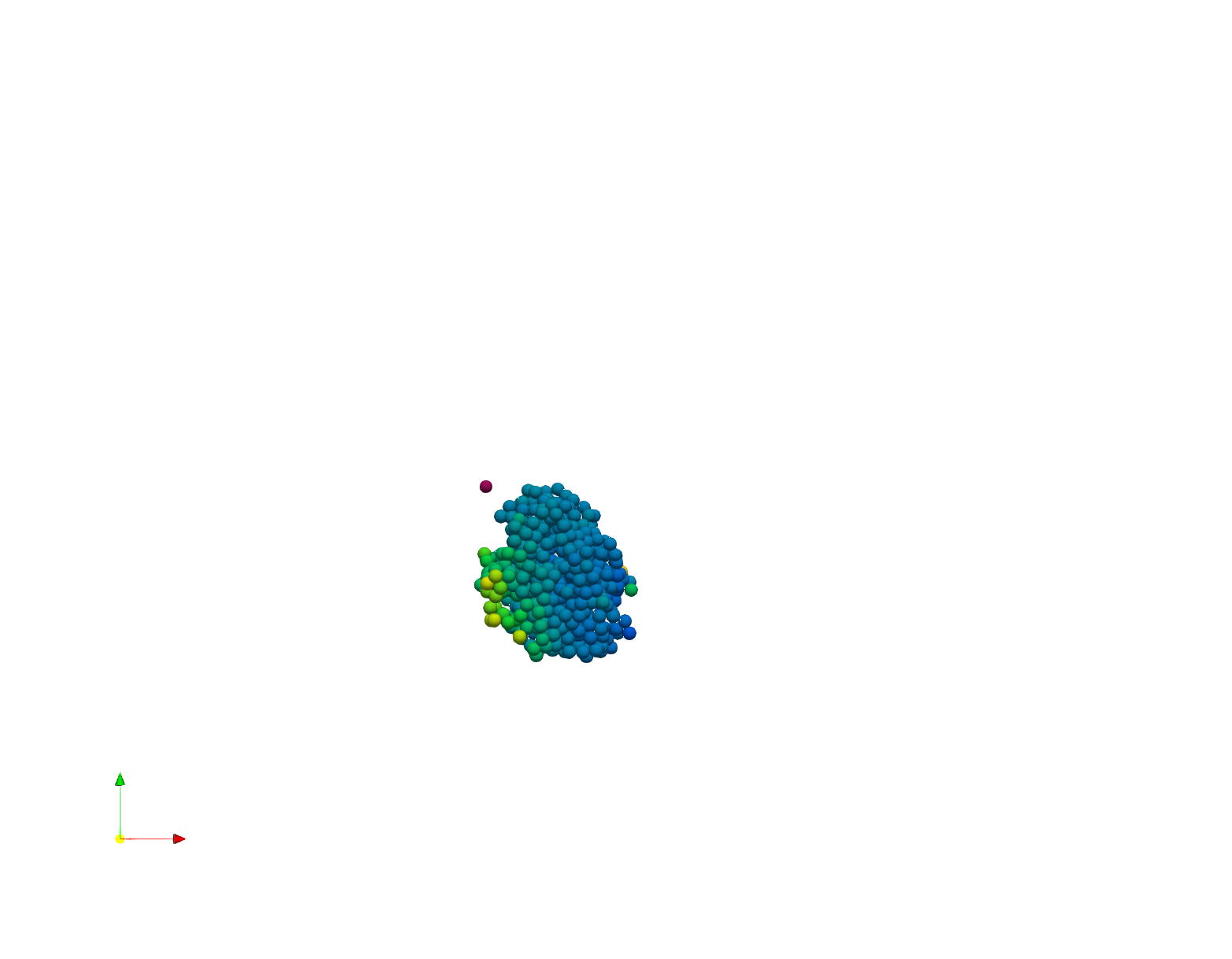}
		\label{subfig:re43_h100_t3}
	}
	\hfill
	\subfloat[$t^*=$1261]{%
		\includegraphics[width=0.24\textwidth]{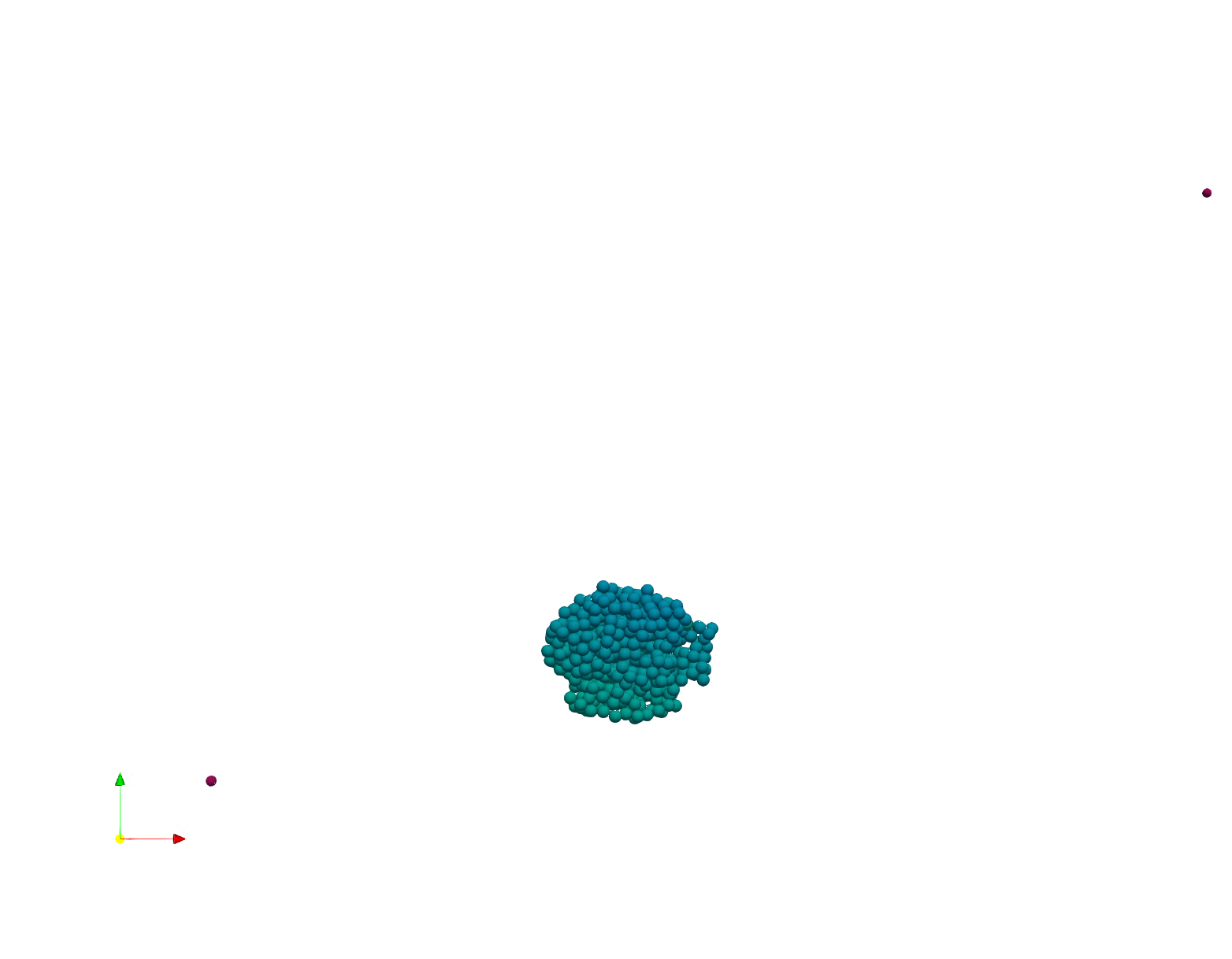}
		\label{subfig:re43_h100_t4}
	}
	\\	\vspace*{-0.5cm}
	\subfloat[{[$Re_{64}, H_{1}$]}, $t^*=$56]{%
		\includegraphics[width=0.24\textwidth]{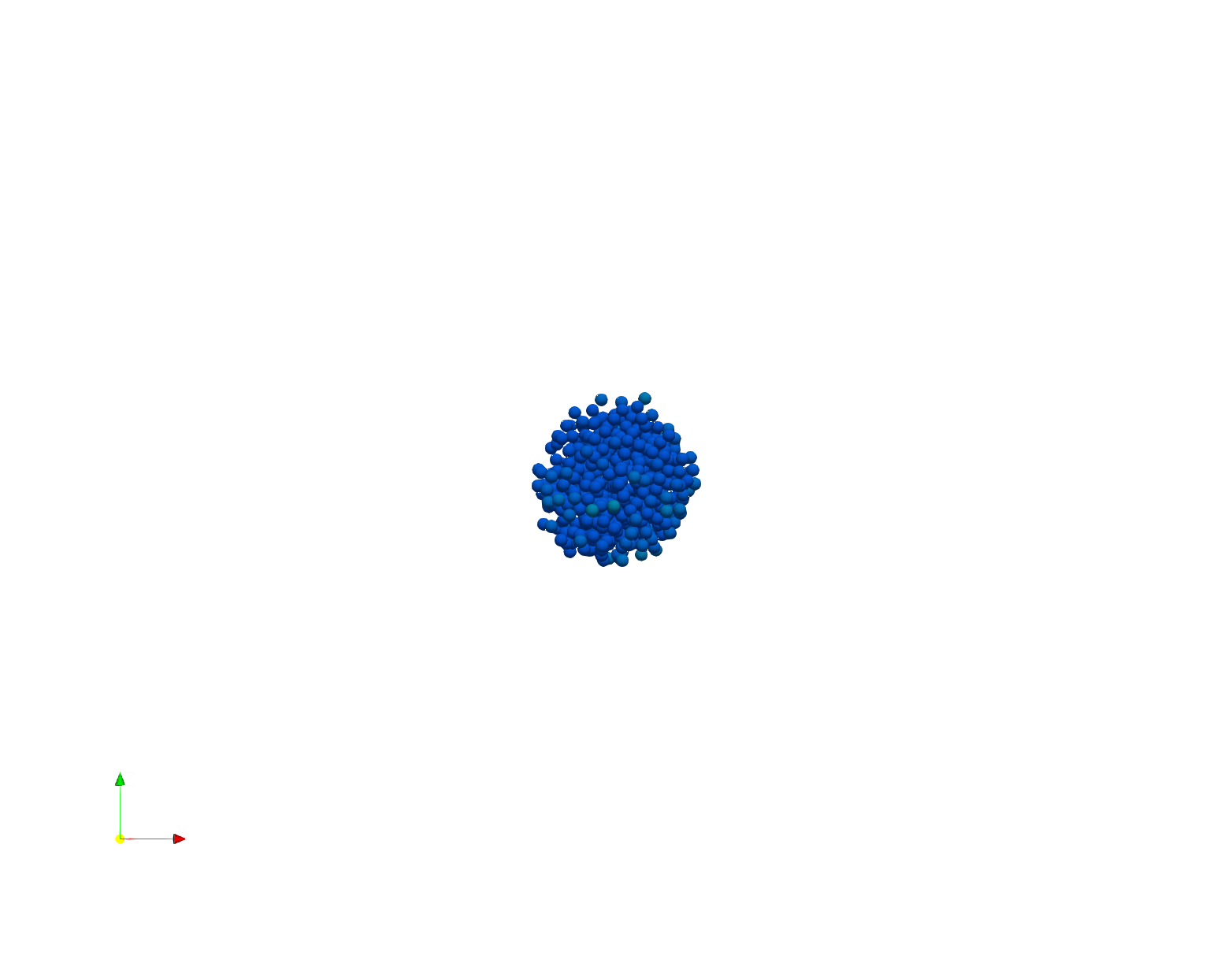}
		\label{subfig:re64_h1_t1}
	}
	\hfill
	\subfloat[$t^*=$81]{%
		\includegraphics[width=0.24\textwidth]{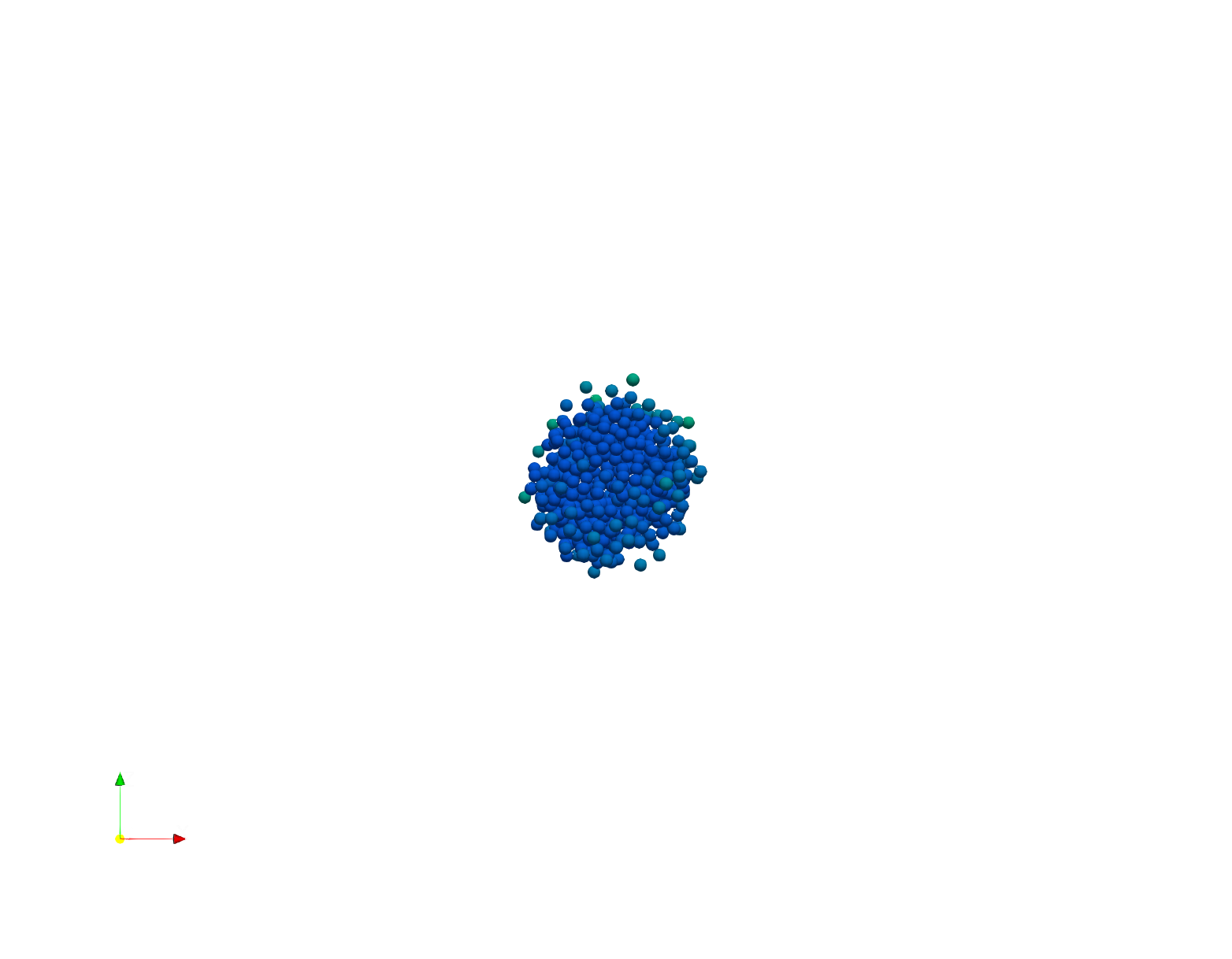}
		\label{subfig:re64_h1_t2}
	}
	\hfill
	\subfloat[$t^*=$129]{%
		\includegraphics[width=0.24\textwidth]{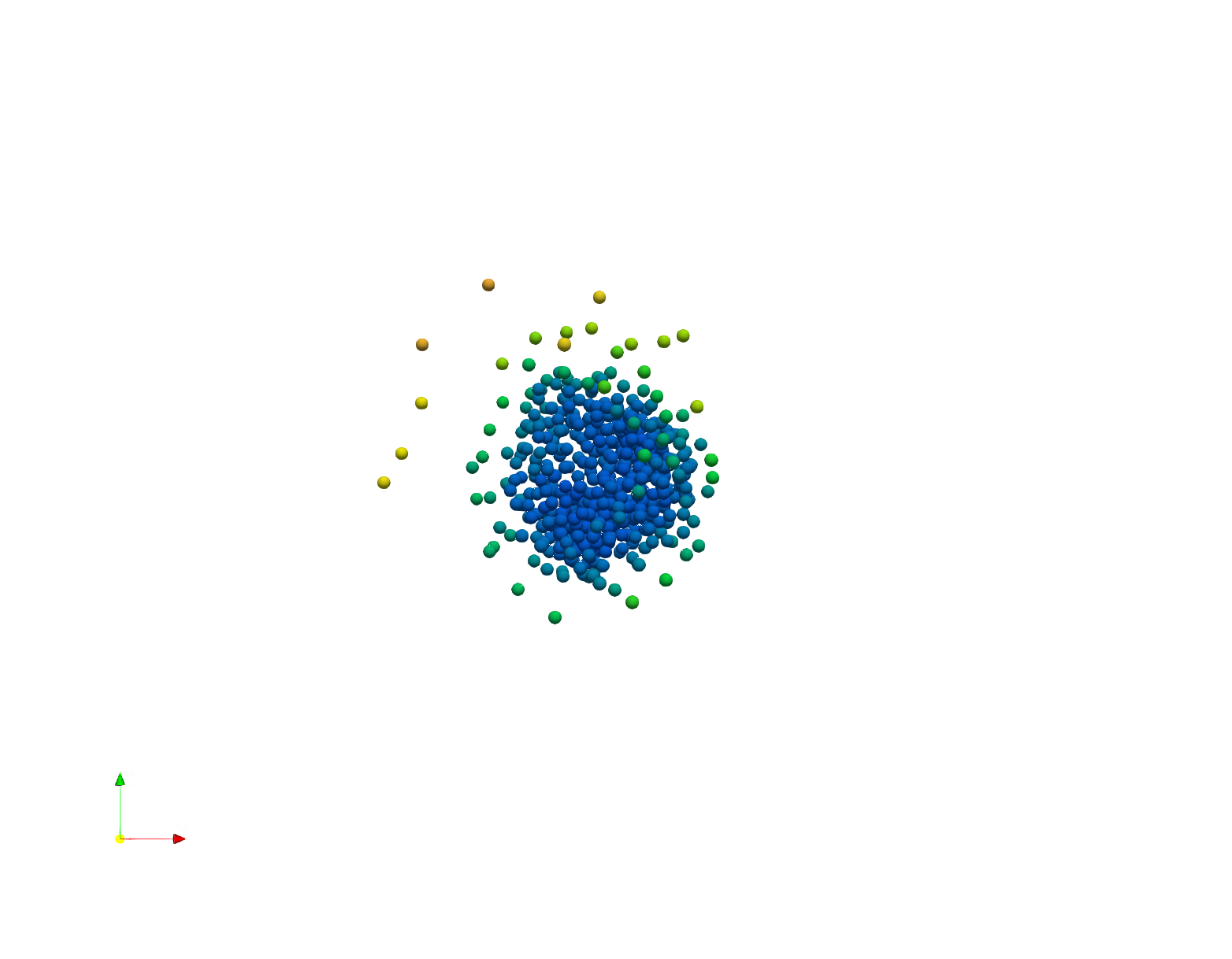}
		\label{subfig:re64_h1_t3}
	}
	\hfill
	\subfloat[$t^*=$364]{%
		\includegraphics[width=0.24\textwidth]{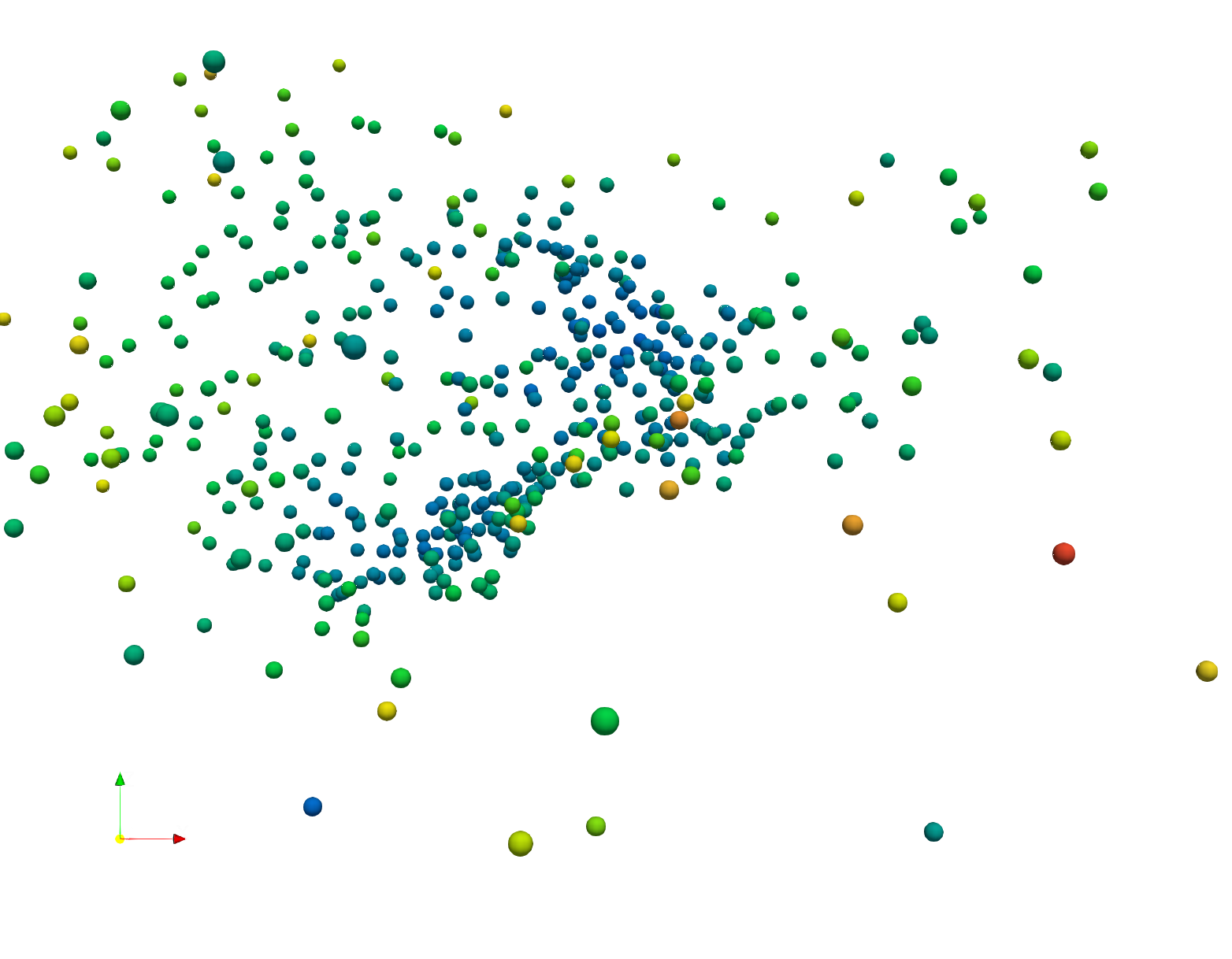}
		\label{subfig:re64_h1_t4}
	}
	\\\vspace*{-0.5cm}
	\subfloat[{[$Re_{64}, H_{100}$]}, $t^*$=195]{%
		\includegraphics[width=0.24\textwidth]{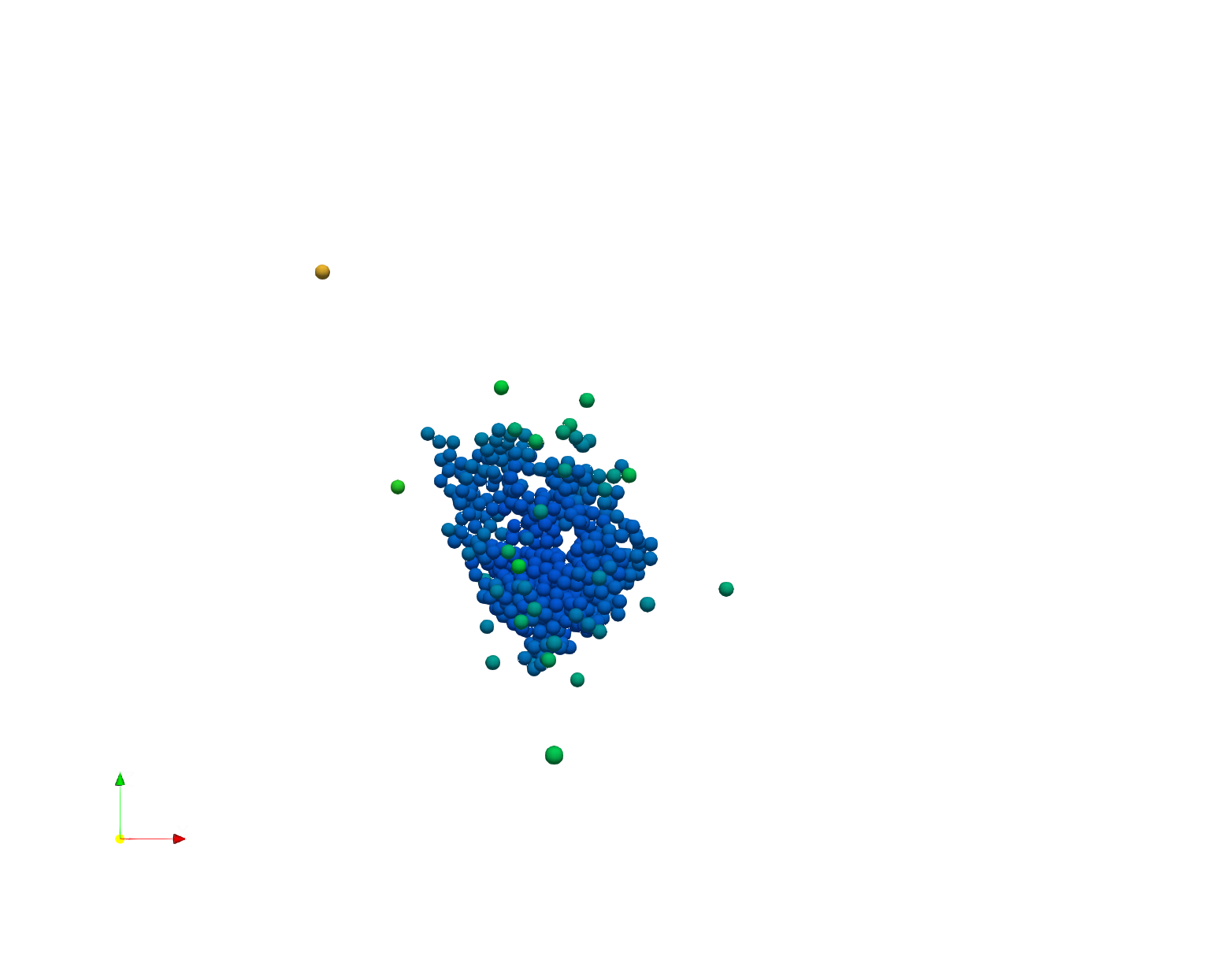}
		\label{subfig:re64_h100_t1}
	}
	\hfill
	\subfloat[$t^*=$364]{%
		\includegraphics[width=0.24\textwidth]{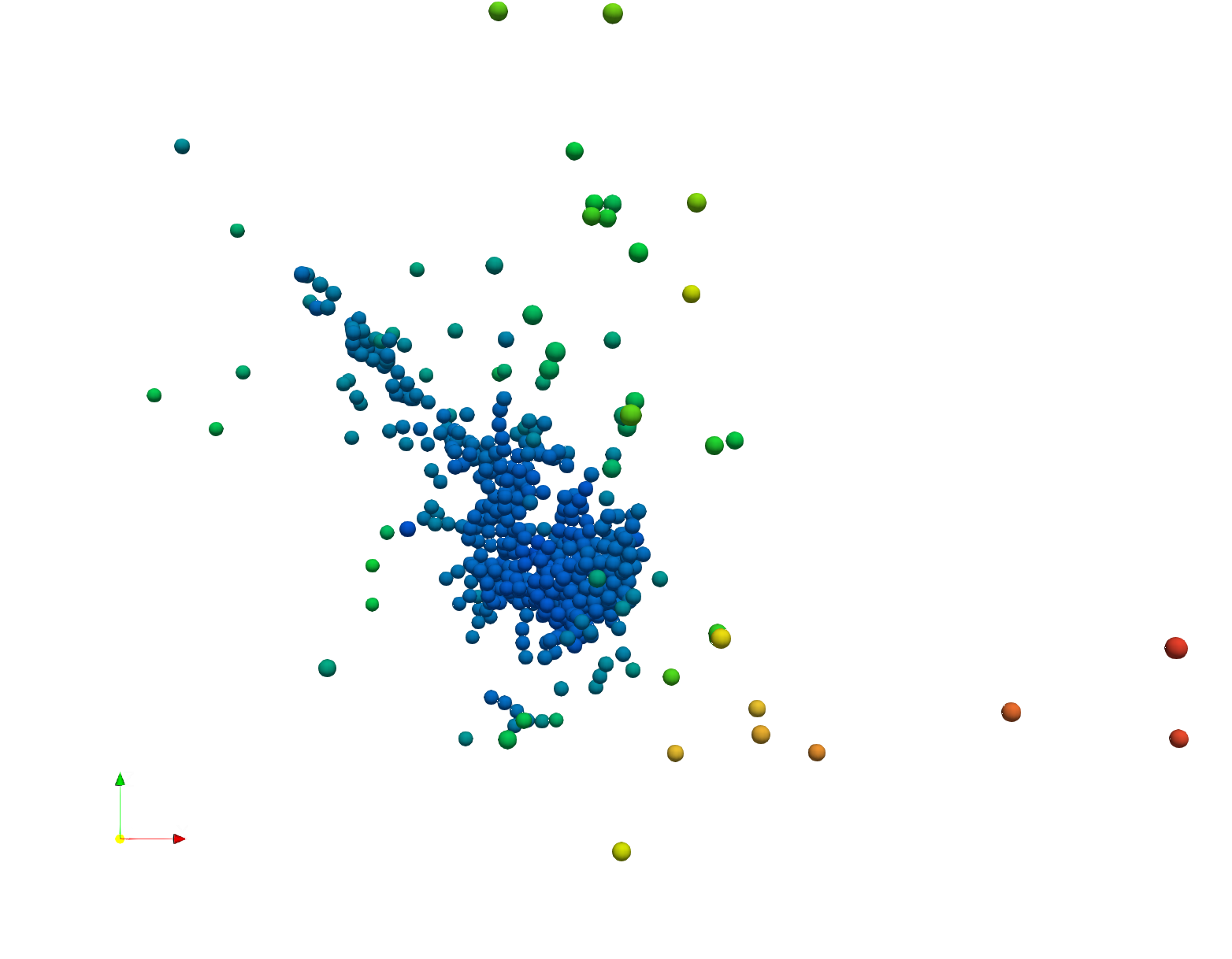}
		\label{subfig:re64_h100_t2}
	}
	\hfill
	\subfloat[$t^*=$484]{%
		\includegraphics[width=0.24\textwidth]{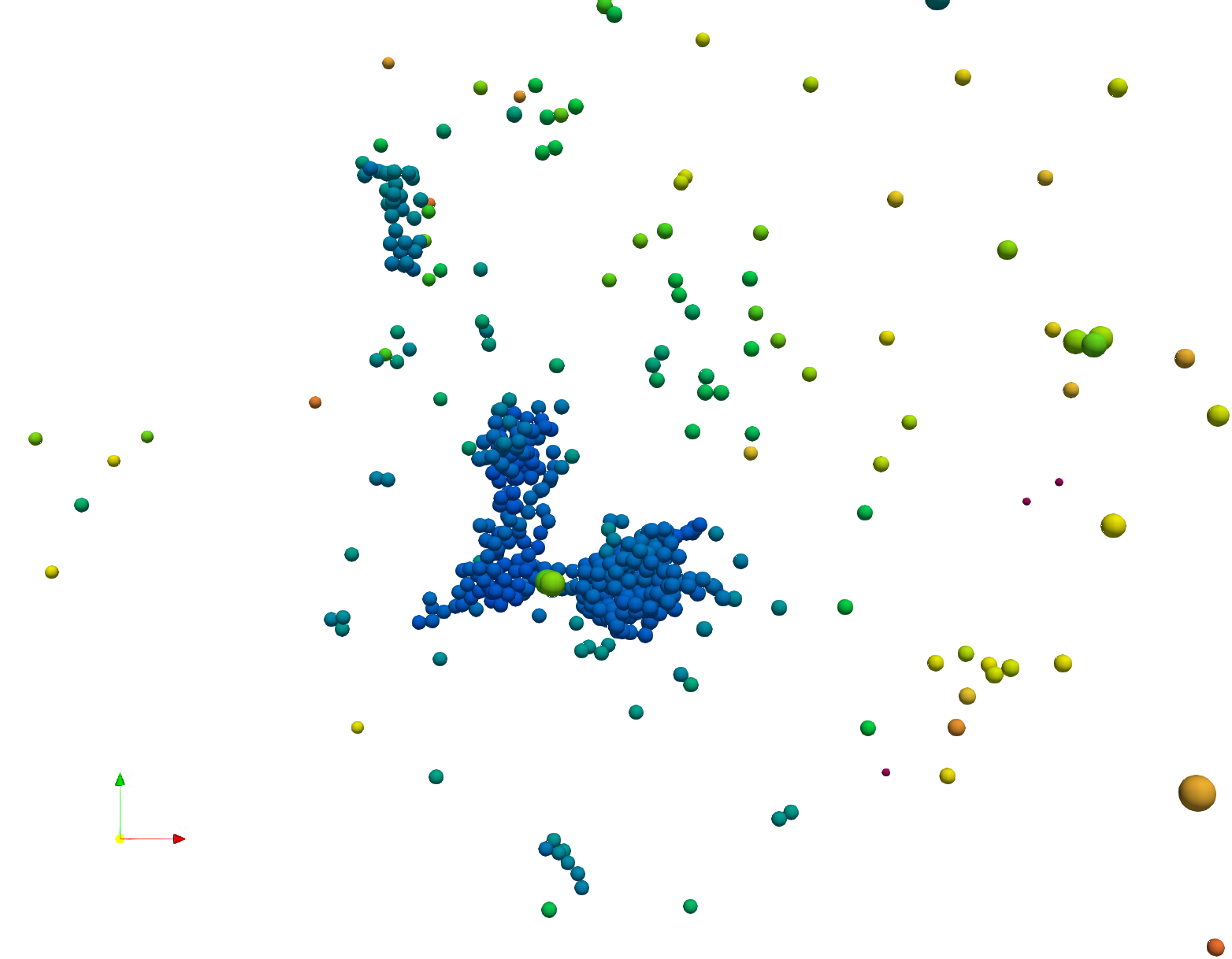}
		\label{subfig:re64_h100_t3}
	}
	\hfill
	\subfloat[$t^*=$597]{%
		\includegraphics[width=0.24\textwidth]{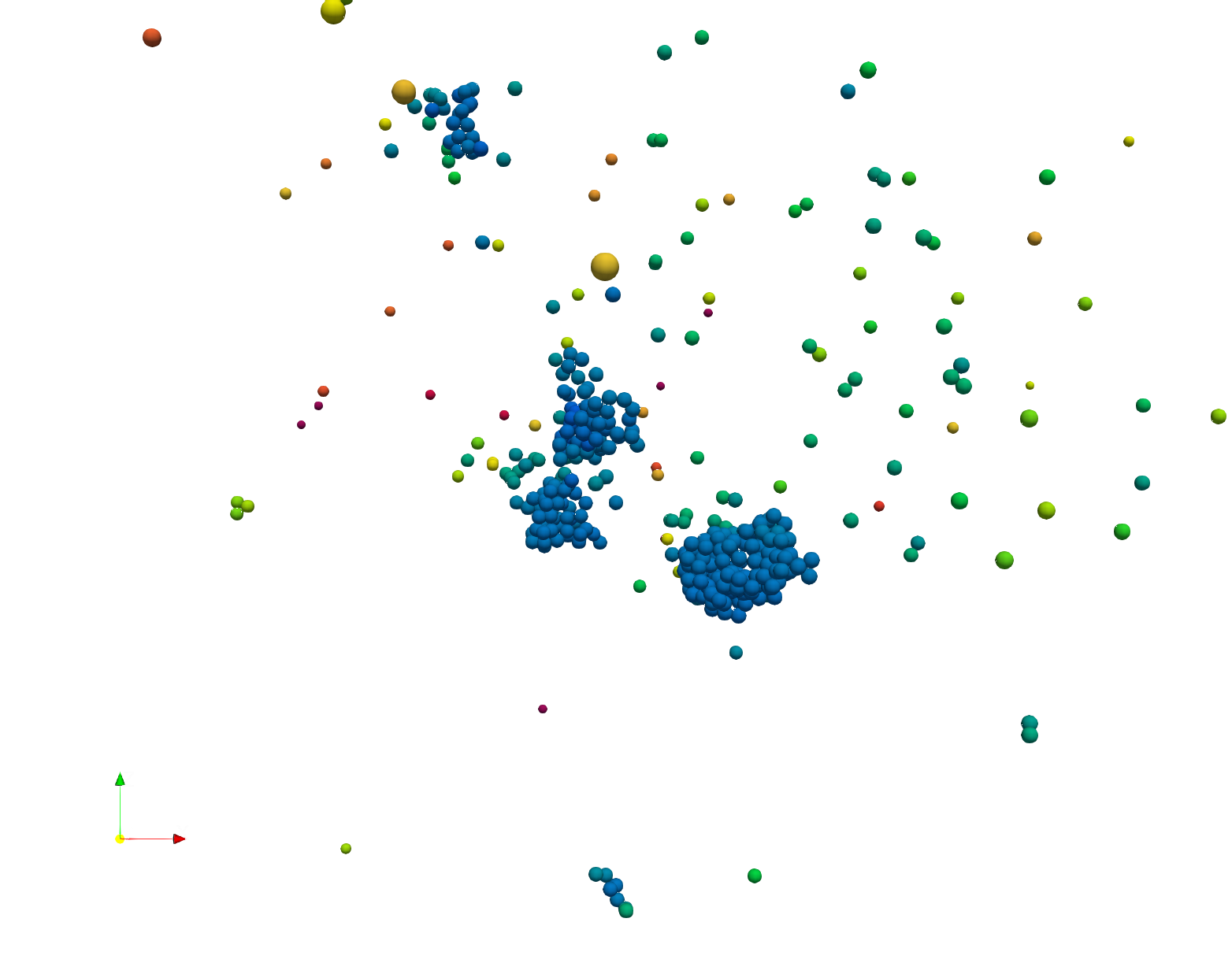}
		\label{subfig:re64_h100_t4}
	}
	\vspace*{-0.4cm}
	\caption{Snapshots of the particles (colored according to velocity 
	magnitude) at four 
	characteristic time instants. Here, the dimensionless time is defined as 
	$t^* = t / \Delta t$.}
	\label{fig:snapshot_par}
\end{figure}

To provide a visual context of the particle configurations at these
time instants, Fig.~\ref{fig:snapshot_par} presents corresponding
snapshots in which particles are colored based on their velocity
magnitude (from blue to red indicating low to high velocities). For
brevity, cases associated with the intermediate Hamaker constant
($H_{10}$) are omitted. Comparing the particle arrangements in these
snapshots under varying flow and cohesion conditions reveals how
cohesive and fluid dynamic forces interact to influence particle
dynamics. Moreover, due to the inherently chaotic nature of the
surrounding turbulent flow and the inertial characteristics of the
particles considered in this study, fragments are observed to disperse
along different axes and in varying directions. This is in contrast to
deagglomeration patterns reported for light particles in simple shear
\citep{ruan2020structural} or even turbulent flows
\citep{saha2013experimental}, where the breakage tends to occur along
more symmetric and flow-aligned paths, often resulting in evenly
distributed fragments that reflect the structure of the surrounding
flow field.

\clearpage
\subsubsection{Preferred direction of fragment detachment}
An additional analysis is conducted to explore potential relationships
between local instantaneous flow structures and the spreading
direction of detaching fragments. Specifically, the alignment between
the fragment's center-of-mass (COM) velocity vector
$\bm{v}^\mathrm{COM}_\mathrm{frag}$ and characteristic directions
derived from the velocity gradient tensor at the fragment's COM is
examined. The velocity gradient tensor field, obtained from the fluid
solver, is interpolated at each fragment's COM using Gaussian-weighted
averaging over the surrounding fluid cells. The Gaussian kernel has a
support radius equal to five times the fragment's radius. This
interpolation uses field values from the last available time step
prior to breakage.

From the interpolated velocity gradient tensor, both the vorticity
vector of the flow $\boldsymbol{\omega}^\mathrm{COM}_\mathrm{f}$ and
the strain-rate tensor $\bm{S}^\mathrm{COM}$ are computed at the
fragment's COM. An eigen-decomposition of $\bm{S}^\mathrm{COM}$ yields
three real eigenvalues, $\lambda_1 > \lambda_2 > \lambda_3$, and the
corresponding orthonormal eigenvectors $\bm{e}_1$, $\bm{e}_2$, and
$\bm{e}_3$, which represent the most extensional, intermediate, and
most compressive directions of the local deformation, respectively.

\begin{figure}[!t]
	\centering
	\subfloat[]{%
		\includegraphics[width=0.49\textwidth]{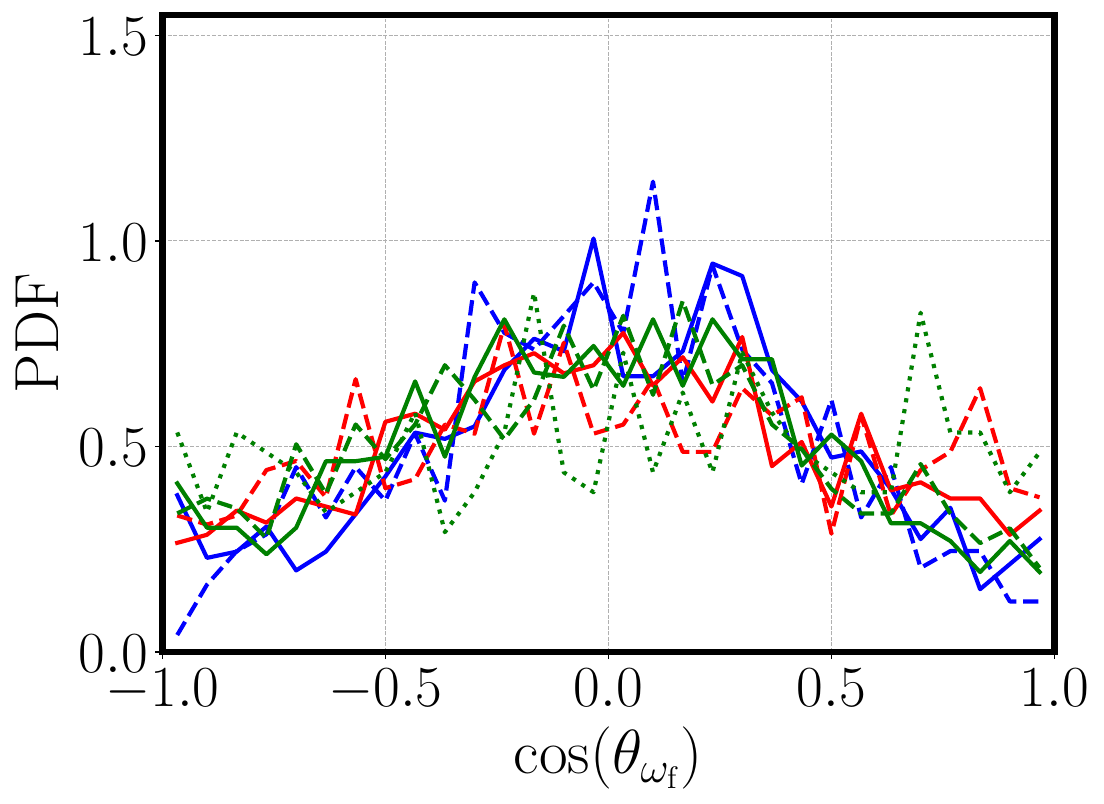}
		\label{subfig:pdf_vort}
	}
	\hfill
	\subfloat[]{%
		\includegraphics[width=0.49\textwidth]{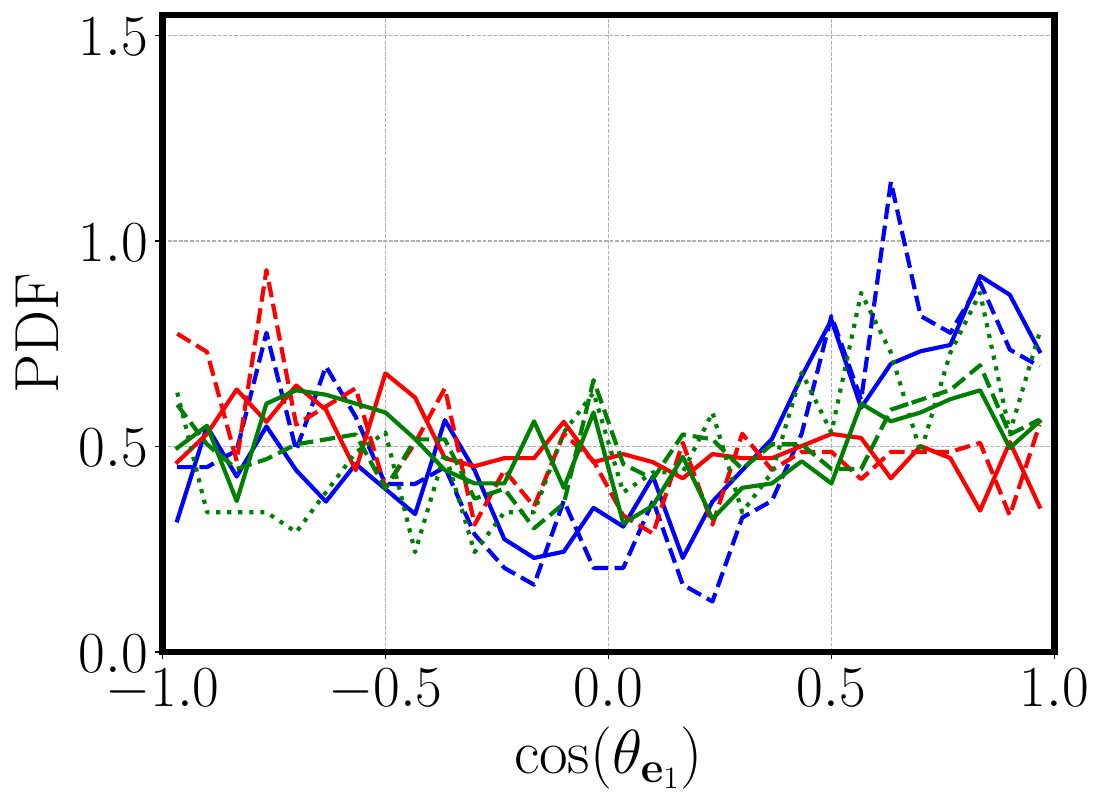}
		\label{subfig:pdf_en1}
	}	
	\\
	\subfloat[]{%
		\includegraphics[width=0.49\textwidth]{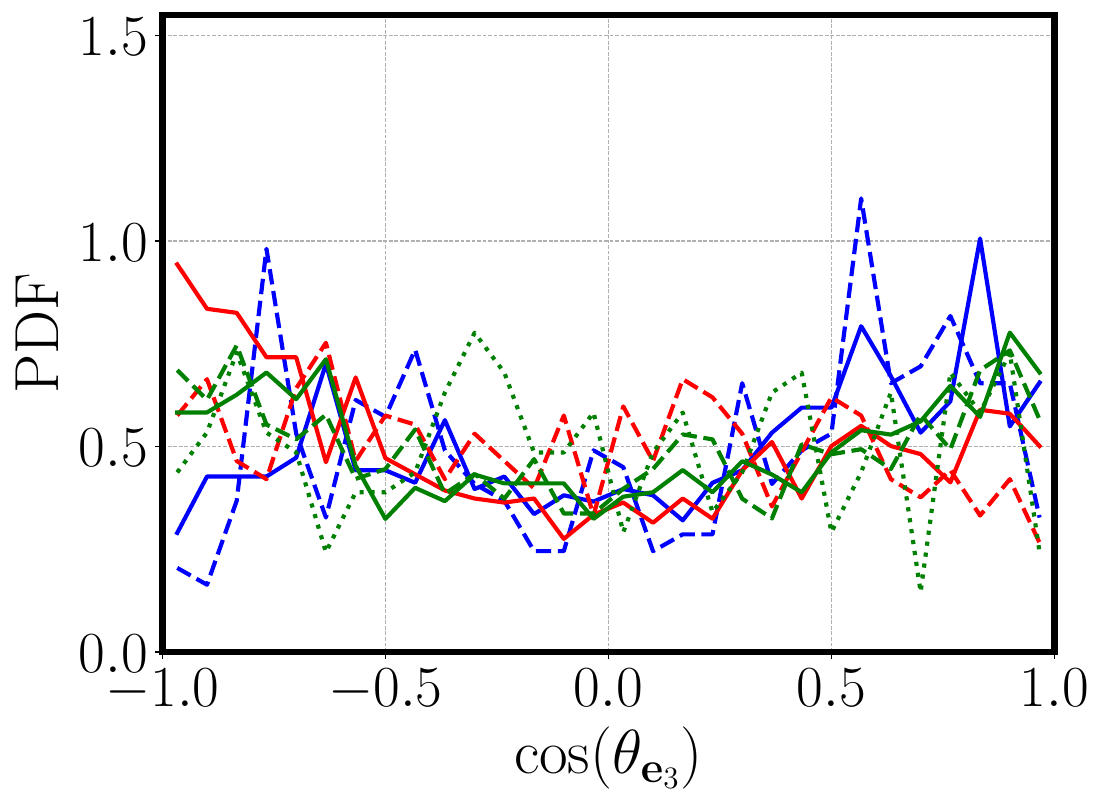}
		\label{subfig:pdf_en3}
	}
	\hfill
	\subfloat[]{%
		\includegraphics[width=0.49\textwidth]{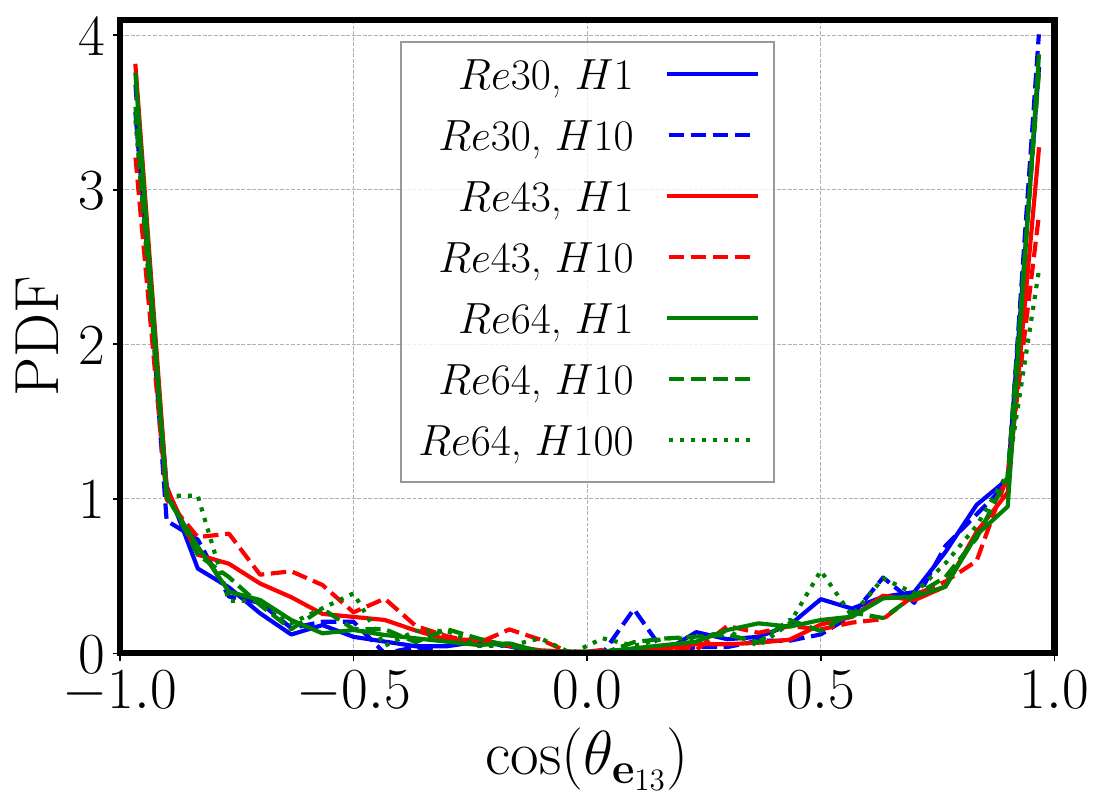}
		\label{subfig:pdf_en13}
	}
	\caption{Probability density function (PDF) of the cosine of
          the angle between the center-of-mass velocity vector
          $\bm{v}^\mathrm{COM}_\mathrm{frag}$ of a detaching fragment
          and (a) the vorticity vector
          $\boldsymbol{\omega}^\mathrm{COM}_\mathrm{f}$, (b) the most
          extensional eigenvector of the strain-rate tensor
          $\bm{e}_1$, (c) the most compressive eigenvector of the
          strain-rate tensor $\bm{e}_3$, and (d) the plane defined by
          $\bm{e}_1$ and $\bm{e}_3$. All vectors are evaluated based
          on the strain-rate tensor interpolated at the fragment's
          center of mass at the moment of breakage.}
	\label{fig:pdf_sepertation_axis}
\end{figure}

In Fig.~\ref{subfig:pdf_vort}, the probability density functions
(PDFs) of the cosine of the angle $\theta_{\omega_\mathrm{f}}$ between
$\bm{v}^\mathrm{COM}_\mathrm{frag}$ and the local vorticity vector
$\boldsymbol{\omega}^\mathrm{COM}_\mathrm{f}$ are shown. The
distributions with seven cases with significant breakup are found to
possess a maximum near $\cos(\theta_{\omega_\mathrm{f}}) = 0$,
indicating that the fragment velocities are predominantly orthogonal
to the local rotation axis of the flow. This observation suggests that
the rotational motion is not a dominant influence on the preferred
detachment direction.

In contrast, the distributions of the angles between
$\bm{v}^\mathrm{COM}_\mathrm{frag}$ and the eigenvectors of the
strain-rate tensor, specifically $\theta_{\bm{e}_1}$ with $\bm{e}_1$
and $\theta_{\bm{e}_3}$ with $\bm{e}_3$, are presented in
Figs.~\ref{subfig:pdf_en1} and \ref{subfig:pdf_en3},
respectively. Peaks around $\cos(\theta) = \pm1$ are observed,
indicating that fragments tend to eject along either the most
extensional or most compressive directions. Although the alignment is
not strongly dominant in all cases, a preferential fragmentation path
aligned with these principal directions of deformation is suggested.

Inspired by these findings, an additional analysis considers the
alignment of $\bm{v}^\mathrm{COM}_\mathrm{frag}$ with the strain plane
defined by $\bm{e}_1$ and $\bm{e}_3$. Since the eigenvectors of a
symmetric tensor form an orthonormal basis, the intermediate
eigenvector $\bm{e}_2$ satisfies:
\begin{equation}
	\bm{e}_2 = \pm (\bm{e}_1 \times \bm{e}_3) \: .
\end{equation}
The projection of 
\(\bm{v}^\mathrm{COM}_\mathrm{frag}\) onto the strain plane is given by:
\begin{equation}
	\bm{v}^\mathrm{COM}_{\mathrm{frag},e_{13}} = 
	\bm{v}^\mathrm{COM}_\mathrm{frag} - \left( 
	\bm{v}^\mathrm{COM}_\mathrm{frag} \cdot \bm{e}_2 \right) 
	\bm{e}_2
\end{equation}
The cosine of the angle $\theta_{\bm{e}_{13}}$ between
$\bm{v}^\mathrm{COM}_\mathrm{frag}$ and
$\bm{v}^\mathrm{COM}_{\mathrm{frag}, \bm{e}_{13}}$ is then computed.
Because this projection-based cosine is always positive by
construction, its sign is inferred by checking the alignment of
$\bm{v}^\mathrm{COM}_\mathrm{frag}$ with $\bm{e}_2$: If the dot
product is positive, the sign remains positive; if negative, the
cosine is assigned a negative sign. This preserves directional
information in the analysis.

Figure~\ref{subfig:pdf_en13} shows that $\cos(\theta_{\bm{e}_{13}})$
exhibits sharp peaks near $\pm 1$, indicating strong alignment of
fragment velocities with the strain plane. These results suggest that
the local strain, rather than rotation, governs the ejection direction
of fragments. The deformation plane, spanned by the extensional and
the compressive direction, appears to be the dominant driver of
fragment motion upon detachment. Moreover, the findings indicate that
breakage results from the interplay between extension and compression
in the flow, acting together on the agglomerate structure within the
dominant strain plane.  This observation aligns with earlier findings
for light and small particles in homogeneous isotropic turbulence,
where fragments have been reported by \citet{saha2013experimental} to
split preferentially along the axis of the most extensional
eigenvector of the strain-rate tensor.

The information on the splitting directions are of great value for
setting up breakup models for less CPU-time intensive Euler--Lagrange
simulations.

\subsubsection{Breakage rate}

A common alternative to modeling the time evolution of the
fragmentation ratio is the characterization of the time-averaged
breakage rate. Owing to its compact formulation, the time-averaged
breakage rate has been frequently used in the development of breakup
kernels for coarse-grained numerical modeling approaches.  The
breakage rate (BR) is defined as the ratio of the change in the number
of fragments to the corresponding time interval, i.e., $\Delta
N_\mathrm{frag}/\Delta t_\mathrm{frag}$.

\begin{figure}[!t]
	\centering
	\subfloat[]{%
		\includegraphics[width=0.49\textwidth]{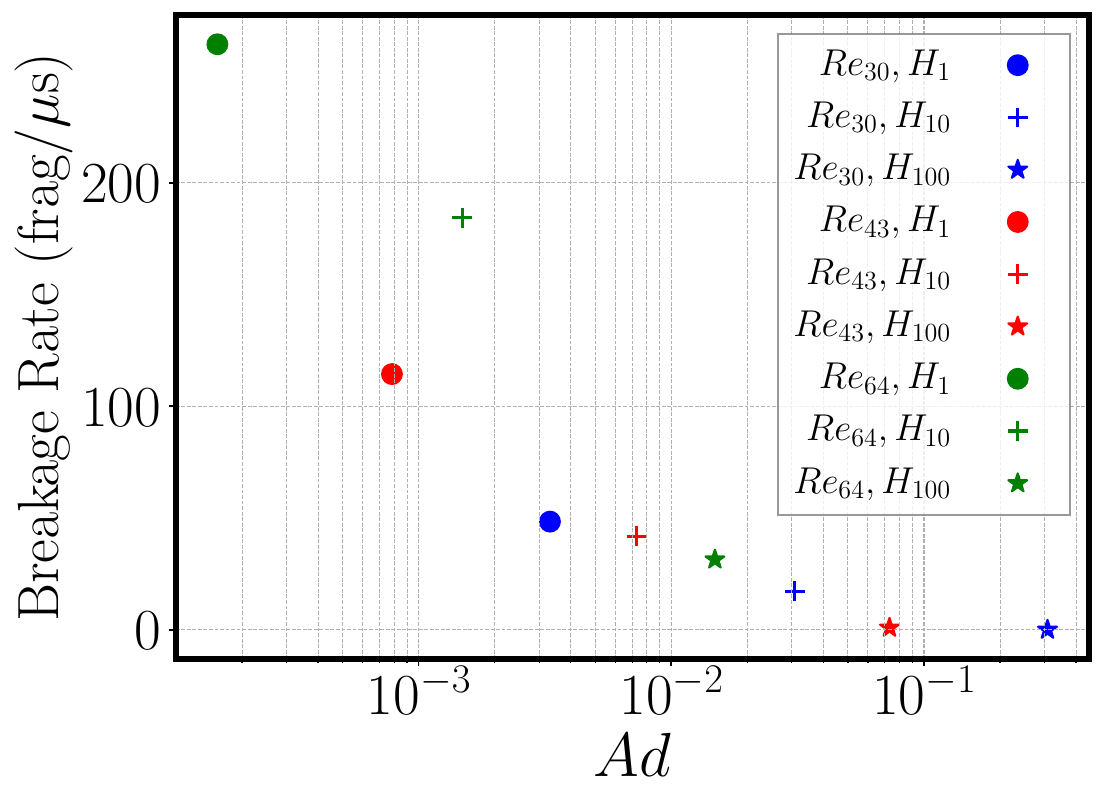}
		\label{subfig:BR_old}
	}
	\hfill
	\subfloat[]{%
		\includegraphics[width=0.49\textwidth]{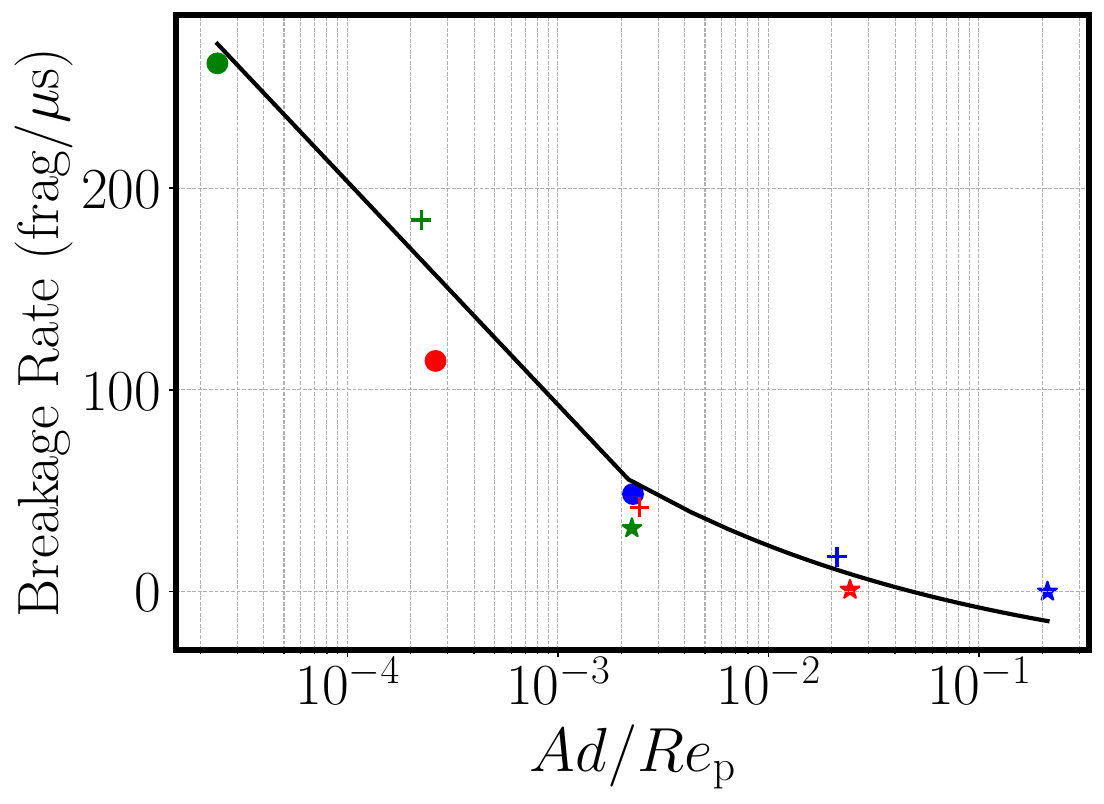}
		\label{subfig:BR_ad}
	}
	\caption{ Breakage 
		rate as function of (a) the adhesion number $Ad$
		vs. (b) the modified adhesion number $Ad/Re_\mathrm{p}$.}
	\label{fig:BR_ad}
\end{figure}

Figure~\ref{subfig:BR_old} displays the time-averaged breakage rate
for each case as a function of the adhesion number, $Ad$ which is
defined as
\begin{equation}
	Ad = \dfrac{H}{12 \: \pi \: \delta_0^2 \: d_\mathrm{p}} \;
        \Big/ \; (\rho_\mathrm{p} u_\mathrm{rms}^2) \; ,
\end{equation}
where the numerator represents the van-der-Waals cohesive stress and
the denominator corresponds to the characteristic turbulent stress
$\rho_\mathrm{p}
u_\mathrm{rms}^2$~(\citet{yao2021deagglomeration}).  Note
that the latter can also be defined with the density of the fluid,
which is not important here since the ratio
$\rho_\mathrm{p}/\rho_\mathrm{f}$ is constant.

It is worth noting that some cases did not reach a fully developed
quasi-steady-state in terms of the number of fragments (see
Fig.~\ref{subfig:fr_t}). To ensure consistency and avoid biases
introduced by plateau regions in the fragmentation curves, the
breakage rate in all cases is evaluated from the start of the
simulation up to the time when a fragmentation ratio of approximately
0.35 is reached. That defines $\Delta t_\mathrm{frag}$ mentioned
above.

An inverse relationship between breakage rate and adhesion strength,
as observed in Fig.~\ref{subfig:BR_ad}, is in general agreement with
trends reported in the literature for deagglomeration in turbulence
\citep{yao2021deagglomeration}.

As expected, a rising adhesion number results in a decreasing breakage
rate, but the results for the different cases are not all
consistent. To improve the correlation, a modified adhesion number
$Ad/Re_\mathrm{p}$ is introduced based on an empirical approach, which
incorporates the particle Reynolds number $Re_\mathrm{p}$ defined in
Eq.~(\ref{eq:reyp}). The motivation for this modified characteristic
number is shown in Figure~\ref{subfig:BR_ad}: By plotting the breakage
rate BR against the modified adhesion number yields a clearer trend
that is well described by an inverse power-law fit of the form:
\begin{equation}
	\mathrm{BR} = a + b \cdot (Ad/Re_\mathrm{p})^c \, ,
\end{equation}
as indicated by the black curve in the figure. The fitting parameters
are given as follows $a = -46.9$, $b = 21.8$, and $c = -0.25$. This
relation can be used within a breakup kernel of coarse-grained
simulations.

%
%

%
%

\section{Conclusions\label{sec:conclusions}}

Based on particle-resolved direct numerical simulations, the breakage
of cohesive agglomerates comprising inertial spherical particles in
homogeneous isotropic turbulence is systematically investigated.  The
simulations resolve both the fluid field and inter-particle
interactions, enabling detailed insights into the complex interplay
between turbulence-induced stresses and cohesive forces.

In contrast to point-particle models, the particle-resolved approach
captures the local flow structures and force distributions responsible
for breakage. Visualizations of $\lambda_2$-criterion isosurfaces and
force fields demonstrates that breakup tends to initiate in regions of
high strain rate and strong fluid-particle interactions. These results
highlight the physical accuracy and descriptive capability of the
particle-resolved simulation framework.

The breakage dynamics analysis reveals that the fragmentation process
is strongly influenced by both the flow Reynolds number and the
Hamaker constant, where the latter governs the strength of
van-der-Waals cohesion. Higher Reynolds numbers and lower Hamaker
constants accelerate and intensify the fragmentation process, in line
with theoretical expectations.

The evolution of fragment size distributions further confirms the
dominant breakup mode, which is predominantly characterized by
erosion.  The analysis shows that in all considered cases, the
agglomerate gradually disintegrates into smaller fragments rather than
splitting into a few similarly sized pieces or shattering into many
small ones simultaneously.

A novel analysis of the fragment ejection direction further reveals
that breakage is not random but guided by the local strain structure
of the turbulent flow. Specifically, the center-of-mass velocity
vectors of the fragments are found to align strongly with the plane
defined by the most extensional and most compressive eigenvectors of
the strain-rate tensor. This indicates that the dominant deformation
plane, not the rotational motion, governs the direction of
detachment. The results suggest that breakup occurs due to the
interplay between local stretching and compression acting on the
agglomerate structure.

The breakage rate follows a power-law decay with increasing values of
the modified adhesion number, generally consistent with trends
reported in previous studies.

Overall, the results demonstrate that particle-resolved simulations
are a highly valuable tool for the mechanistic understanding and
modeling of agglomerate breakup in turbulent flows. The insights
gained here contribute to the development of more accurate breakup
models for less CPU-time intensive but also less detailed simulation
methodologies.

%
%



\section*{Acknowledgments}
Computational resources (HPC cluster HSUper) have been provided by the
project hpc.bw, funded by dtec.bw -- Digitalization and Technology
Research Center of the Bundeswehr. dtec.bw is funded by the European
Union -- NextGenerationEU.

\section*{Author declarations}

\subsection*{Conflict of interest}

The authors have no competing interests.

\subsection*{Data availability}

The data that support the findings of this study are
available from the corresponding author upon
reasonable request.
\clearpage
\appendix
\section{Properties of the primary particles}
%
\begin{table}[!h]
	\centering
	\caption{Physical properties and other characteristic parameters 
		of 
		the 
		considered silica particles (SiO$_2$) 
		\citep{weiler2008generierung,khalifa2020data,khalifa2021efficient}.}
		\begin{threeparttable}
			\begin{tabular}{lcc}
				\toprule[1.5pt]
				\textbf{Parameter}  & \textbf{Unit} & \textbf{Value} \\
				\midrule[1.5pt]				
				Primary particle diameter $d_\mathrm{p}$ & m & 0.97
				$\cdot 10^{-6}$ \\
				Primary particle density $\rho_\mathrm{p}$ & 
				kg$\cdot$m$^{-3}$ 
				& 
				2000  \\
				Poisson's ratio  $\nu$ & - &  0.17 $^e$   \\
				Modulus of elasticity  $E$ & N/m$^2$  & $7.2 \cdot 
				10^{10}$ $^e$  \\
				Hamaker constant $H$ & J &  $2.148 \cdot 10^{-20} \; 
				^{a},  \: 2 \cdot 10^{-19}, \; 2 \cdot 10^{-18}$ \\  
				Min.  inter-particle distance $\delta_0$ & m &   
				$4.0 \cdot 10^{-10} \; ^b $ \\ 
				Restitution coefficient $e$ & - & 0.97 
				$^c$ \\
				Static friction coefficient $\mu_{s}$  
				& - &  0.94 $^c$   
				\\
				Rolling friction coefficient $\mu_{r}$   & - &  
				$2 \cdot  10^{-3}$ $^d$  \\
				$\bm{F}^{\text{vdW}}$ cut-off distance 
				$l_{\text{max}}$& m
				&  $5\cdot10^{-2} \; d_p $ 
				\\
				\bottomrule[1.5pt]
				\multicolumn{3}{l}{%
					\begin{minipage}{1\textwidth} ~\\%
						\footnotesize 
						$^a$\citet{schubert2003handbuch}, 
						$^b$\citet{krupp1967particle}, 
						$^c$\citet{foerster94} for soda-lime-silica 
						glass,
						$^d$\citet{yang2008agglomeration} for a 
						material 
						with comparable properties, 
						$^e$\citet{azomaterials2018}.
				\end{minipage}}\\
			\end{tabular}
		\end{threeparttable}
		\label{tab:sio2prop}
	\end{table}
	
	
	\clearpage
	
	\bibliography{mybib,particle,agglo,allgemein,breakup,diss}
	
\end{document}